\documentclass[letterpaper,twocolumn,10pt]{article}
\usepackage{usenix-2020-09}  

\usepackage{tikz}
\usepackage{amsmath}
\usepackage{amssymb}
\usepackage{amsthm}
\usepackage{microtype}
\usepackage{subcaption}
\usepackage{booktabs}
\usepackage{tabularx}
\microtypecontext{spacing=nonfrench}

\usepackage{xspace}
\usepackage{algorithm}
\usepackage[noend]{algpseudocode}
\DeclareRobustCommand{\circled}[1]{\tikz[baseline=(char.base)]{\node[shape=circle,draw,inner sep=1pt,font=\small] (char) {#1};}}
\newcommand{\ours}{Koala\xspace}
\newcommand{\stitle}[1]{\vspace{1ex}\noindent\textbf{#1}}

\theoremstyle{definition}
\newtheorem{definition}{Definition}
\newtheorem{property}{Property}

\usepackage{enumitem}
\setlist{noitemsep,topsep=1pt,parsep=1pt,partopsep=0pt, leftmargin=*}

\newcounter{VasiaNOC}
\stepcounter{VasiaNOC}

\newcounter{LeiNOC}
\stepcounter{LeiNOC}

\newcounter{JohnNOC}
\stepcounter{JohnNOC}

\newcounter{SakshiNOC}
\stepcounter{SakshiNOC}

\usepackage{balance}
\usepackage{stfloats}

\graphicspath{{usenixbadges/}}
\usepackage[available,functional,reproduced]{usenixbadges/usenixbadges}

\begin{document}


\date{}

\title{Koala: Non-disruptive Reconfiguration for Long-running Stateful Dataflows}


\author{
\begin{tabular}{c@{\hspace{1.8em}}c@{\hspace{1.8em}}c@{\hspace{1.8em}}c}
{\rm Lei Huang} & {\rm Sakshi Sharma} & {\rm Chenxu Yu} & {\rm Zikun Wang}$^\dagger$\thanks{Work done while at Boston University.} \\
{\rm Yuanli Wang} & {\rm Zichen Zhu} & {\rm John Liagouris} & {\rm Vasiliki Kalavri}
\end{tabular}\\[1em]
Boston University \qquad $^\dagger$Northeastern University
}

\maketitle

\begin{abstract}
Stateful dataflow systems must decide where state lives relative to the tasks that process it. For long-running applications that require frequent reconfiguration due to changes in resource allocation, partitioning, or task placement, this design choice presents a core trade-off between adaptability and performance. Co-locating state with computation ensures low latency at steady state but makes reconfiguration disruptive, as any change forces state migration. On the other hand, keeping state in an external store makes reconfiguration seamless, but incurs overhead on every access.

We present \ours, a reconfiguration protocol for distributed dataflow systems that retains the operational performance of embedded-state architectures while matching the disruption-free adaptability of external-state designs. Unlike prior protocols which relocate all state regardless of whether it will soon be accessed, \ours migrates state in a fine-grained, lazy manner. To achieve that, it employs a lightweight ownership transfer mechanism that safely decouples event routing from state location. We implement \ours in a new streaming dataflow system and evaluate it on real-world and synthetic benchmarks. When compared with the state-of-the-art protocol DRRS, \ours consumes backlog up to $23\times$ faster and transfers $8.7\times$ less state on average during reconfiguration.
\end{abstract}

\section{Introduction}
A core architectural choice in distributed stateful dataflow systems is whether to co-locate and co-partition state with computation tasks or maintain it in a separate remote store.  This design decision is particularly crucial for long-running applications, like streaming analytics, which may require frequent adaptation to variable workloads~\cite{xu2021move,venkataraman2017drizzle,mei2020turbine,streamingsurvey_vldb24}.
At one end of the design space, embedded-state systems~\cite{carbone2017state,zhu2025impeller,lin2016streamscope,murray2013naiad,sparkstructuredstreaming} deliver low latency at steady state but tightly couple logical task assignment to the physical location of data. When the system must adapt, whether to accommodate additional resources, respond to load imbalance, or change task placement, state must be relocated, interrupting processing. 
In contrast, systems with external state management, such as MillWheel~\cite{MillWheel} and Flink 2.0~\cite{flink2},  offer non-disruptive reconfigurations, since no state needs to be migrated upon change, but incur higher latency and resource consumption at steady state.

In this paper, we argue that this trade-off is not fundamental and we propose \ours, a reconfiguration protocol that retains the nominal performance of embedded-state architectures while matching the disruption-free adaptability of remote-state designs.
Koala exploits the fact that real-world streaming workloads exhibit \emph{temporal locality} and \emph{skew}~\cite{barber2021bladerunner,asyabi2022new,rappaz2021twitchdataset,newyorktaxi,AzureVMData,GoogleClusterData}: at any point in time, only a small, slowly evolving working set is hot while the vast majority of keys sit idle.
Yet, existing reconfiguration protocols~\cite{Meces,megaphone,spacker,Falcon,DRRS,carbone2017state,castro2013integrating} migrate all state, regardless of whether it will soon be accessed.  \ours exploits this property to lazily migrate only state that is necessary for computation, leaving cold partitions in place. 

\begin{figure}[t]
    \centering
    \includegraphics[width=0.44\textwidth]{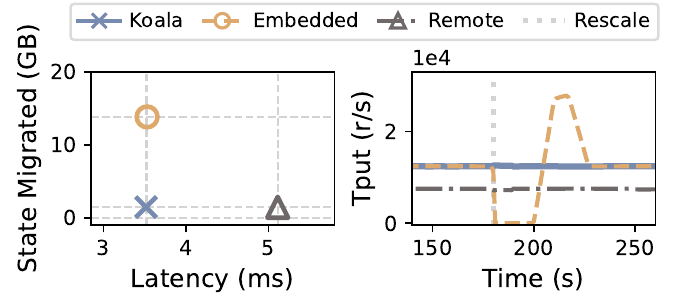}
    \caption{Reconfiguration disruption vs. latency.}
    \label{fig:intro}\vspace{-4mm}
\end{figure}

Figure~\ref{fig:intro} illustrates how \ours combines the strengths of both ends of the design space in a rescaling experiment where the parallelism of a stateful operator is updated after 3 minutes (setting in \S\ref{sec:exp-methodology}). The embedded-state system provides low latency but experiences downtime, as it must migrate GBs of state.  The remote-state system avoids downtime since state is already external, but incurs $1.5\times$ higher processing latency due to remote state access. \ours achieves the best of both: it migrates $6\times$ less state by fetching hot keys on demand, exhibits no disruption during scaling, and its performance is on par with that of the embedded-state approach.


To achieve non-disruptive reconfiguration without externalizing state, \ours employs a lightweight \emph{ownership transfer} mechanism that safely decouples updates to key space partitioning and updates to state location.  Unlike existing protocols, which consider a reconfiguration complete only once data partitioning and state location are fully synchronized, a \ours reconfiguration completes as soon as minimal ownership metadata has been transferred and upstream routing has been updated, enabling fast and frequent reconfigurations without disruption.

 

\subsection{Limitations of existing \emph{live} approaches}\label{sec:limitations}
Despite a substantial body of research on live reconfiguration protocols~\cite{Meces,megaphone,spacker,Falcon,DRRS,mao2021trisk,wang2023fries}, production stream processing systems continue to rely on \emph{stop-and-restart} approaches~\cite{gulisano2012streamcloud,Mai2018Chi,castro2013integrating,gedik2014elastic,state-survey,streamingsurvey_vldb24}. 
We attribute this gap between research and practice to the following limitations of existing live reconfiguration protocols that make simpler, more disruptive approaches preferable in production settings.  



\stitle{Unnecessary state transfer.} A fundamental limitation shared by all existing live reconfiguration protocols is that they offer no advantage over stop-and-restart in terms of how much state is transferred. Even approaches that distinguish between hot and cold keys~\cite{Meces,DRRS} eventually migrate \emph{entire} state partitions, regardless of how many keys are accessed after the reconfiguration.  This is particularly problematic under diurnal or bursty workloads, where the system repeatedly alternates between two configurations. In such cases, large amounts of cold state are shuffled across workers with every transition, only to be moved back again shortly after. Even when migration is performed asynchronously or progressively~\cite{DRRS, Meces, megaphone}, it still consumes resources and can disrupt performance.



\stitle{Incomplete descriptions and implementations.} The literature on live reconfiguration is fragmented, with individual protocols tackling specific aspects of the problem while leaving others unaddressed. For example, correct reconfiguration must preserve event-time semantics~\cite{akidau2021watermarks,akidau2015dataflow}, ensuring that time-based windows do not fire too early, too late, or more than once. These aspects are often overlooked or underspecified in the context of state migration~\cite{Meces,spacker,DRRS},  complicating adoption in real systems.  In addition, some protocols support only a restricted set of operators~\cite{DRRS}, while others target specific scenarios, such as scaling out~\cite{Meces},  and do not discuss how to handle scale-in or other reconfiguration use cases.




\stitle{Benign experimental settings.} Another important limitation is simplistic or favorable experimental conditions: (i) small operator state in the range of MBs to GBs~\cite{DRRS,spacker,Sponge,Meces,Falcon}, while production workloads involve an order-of-magnitude larger state~\cite{flink2}, (ii) backpressure-free deployments with stable input rates~\cite{DRRS,Rhino,spacker,Falcon}, despite reconfiguration being typically triggered in response to overloaded tasks, making backlog accumulation and recovery essential to evaluate, and (iii) single-operator reconfiguration scenarios~\cite{DRRS,Meces,megaphone,spacker,Falcon,Sponge}, even though policy controllers can trigger multiple concurrent changes to the same dataflow~\cite{ds2,mei2020turbine,capsys}.\vspace{-2mm}



\subsection{Technical challenges and contributions}
\ours addresses the limitations of prior work and delivers disruption-free adaptability without sacrificing performance. Realizing its lazy, fetch-on-demand approach in practice requires solving three core technical challenges:

\noindent \textbf{C1: Correctness under divergent routing and state location.} 
All existing protocols fundamentally rely on synchronizing state location with data routing before resuming processing to ensure no two tasks concurrently access the same key.  \ours must provide the same guarantee while allowing access to state that may reside on remote workers.  

\noindent \textbf{C2: Efficient state location tracking.} When reconfigurations occur repeatedly over the lifetime of a long-running job,  \ours must track where each key's state resides,  and it must do so without introducing overhead.  This problem does not arise in prior work where state always moves with its partition. 

\noindent \textbf{C3: Forcing full state migration.} The proposed lazy migration approach is not suitable for certain reconfiguration scenarios, such as scaling in or task relocation, which inherently require full state transfer. Nevertheless, \ours must handle such cases without causing performance disruption. \vspace{1mm}

\noindent We make the following contributions:

\begin{itemize}

\item We present \ours, a lazy reconfiguration protocol for embedded-state dataflow systems. \ours introduces a lightweight ownership transfer mechanism that safely decouples key space partitioning from state location (\S\ref{subsec-procedure}).  Ownership transfer guarantees correctness (\textbf{C1}) and allows on-demand, concurrent state access without coordination. \ours supports a broad range of reconfiguration scenarios, including scaling, skew mitigation, and task migration.  We also describe a watermark inheritance mechanism that preserves event-time semantics across reconfigurations (\S\ref{subsec-watermark}).

\item We describe how \ours can accurately track state location across repeated reconfigurations by maintaining and transferring compact metadata across tasks (\textbf{C2}).  \ours tasks only store location information for their own assigned partitions and always fetch remote state in a single hop (\S\ref{subsec-multiple}).   

\item To support reconfiguration scenarios that require full state transfer (\textbf{C3}), we extend \ours's lazy mechanism to enable piggybacking of additional state on regular remote requests, gradually migrating cold state without disruption (\S\ref{subsec-full}).

\item We present a new distributed dataflow runtime that may be of independent interest (\S\ref{sec:implementation}). The system provides a set of control-plane utilities that enable easy integration of different reconfiguration strategies.  In addition to \ours, we provide implementations of three baseline protocols: (i) an optimized version of stop-and-restart, (ii) an external-state approach, and (iii) the state-of-the-art DRRS protocol, which we extend with watermark propagation support.

\end{itemize}

\begin{figure*}[t]
  \begin{center}
    \includegraphics[width=0.9\textwidth]{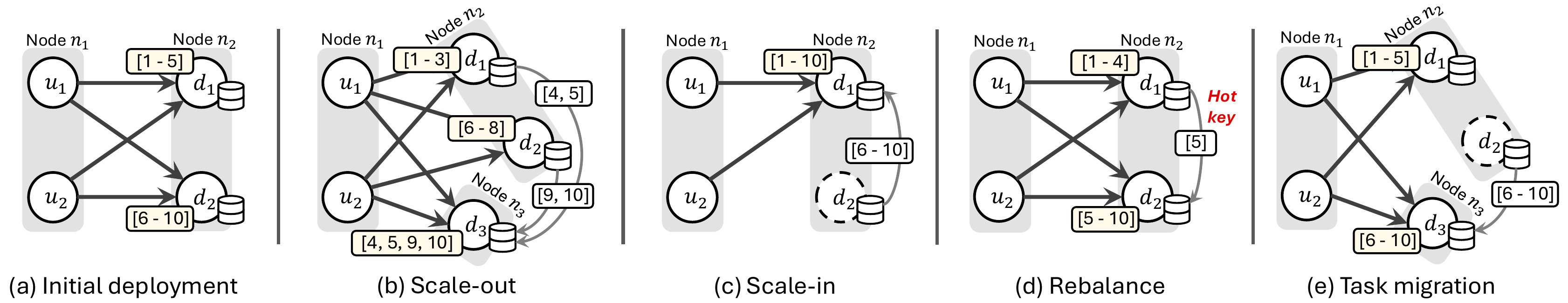}
  \end{center}
  \vspace{-3mm}
  \caption{Dataflow reconfiguration scenarios supported by \ours.}
  \label{fig:reconfig-types}
  \vspace{-4mm}
\end{figure*}

\stitle{Evaluation summary.} We evaluate \ours on real-world and synthetic benchmarks under diverse reconfiguration scenarios (\S\ref{sec:eval}).  We design a challenging experimental setting that accumulates state and backlog before triggering reconfiguration to evaluate \ours and baselines under realistic conditions.  We further include experiments with state sizes of up to $320$GB and present extensive microbenchmarks to study how individual parameters affect \ours's behavior.
Our results show that \ours matches the performance of embedded-state approaches at steady state while ensuring no downtime during reconfiguration. It recovers from backlog up to $23\times$ faster than the baselines and transfers $8.7\times$ less state on average.  Perhaps counter-intuitively, we also show that our carefully implemented stop-and-restart protocol can sometimes consume backlog faster than the live reconfiguration baseline.

\ours is open source and available at \url{https://github.com/CASP-Systems-BU/koala}.
\vspace{-1mm}

\section{Stateful dataflow reconfiguration}\label{sec:background}
We introduce core concepts of the stateful dataflow model and describe the reconfiguration scenarios that \ours supports.

\vspace{-1mm}
\subsection{The stateful dataflow model}\label{sec:dataflow-model}

\stitle{Operators and tasks.}
The dataflow model~\cite{akidau2015dataflow,streamingsurvey_vldb24} represents an application as a directed acyclic graph, where vertices are \emph{operators} that transform an input data stream into an output stream, and edges are \emph{channels} that carry records between operators. Each record consists of a \emph{key}, a \emph{value}, and an \emph{event timestamp}. A \emph{stateless} operator (e.g., filter, projection) processes each record independently of all others, while a \emph{stateful} operator (e.g., windowed aggregation, join) produces outputs that may depend on prior inputs.  Upon deployment, each operator is instantiated as a set of data-parallel \emph{tasks} distributed across a set of \emph{workers}. The \emph{parallelism} of an operator is the number of tasks executing it.

\stitle{State partitioning.}
The key space of a stateful operator is partitioned into \emph{disjoint subsets}, with each task assigned exactly one partition. In embedded-state dataflow systems, state is represented as key-value pairs and is \emph{co-located} with the task responsible for the corresponding key partition. This design enforces \emph{per-key read/write isolation}: at any point in time, each key is read and written by exactly one task~\cite{state-survey,carbone2017state,kalavri2020support}.

\stitle{Event time.}
To correctly process out-of-order records from distributed data sources, the dataflow model adopts \emph{event-time semantics}~\cite{akidau2015dataflow,akidau2021watermarks}. 
Tasks periodically emit \emph{watermarks} that denote their event time progress. A watermark $W(t_w)$ asserts that no future record with event timestamp $t \leq t_w$ will arrive. 
Maintaining the monotonicity of watermarks is essential for the correctness of time-dependent computations.

\subsection{Reconfiguration scenarios}\label{sec:reconfig-scenarios}

\emph{Reconfiguration} actions change the key assignment or physical location of tasks (or both) and are triggered by a policy controller~\cite{floratou2017dhalion,Mai2018Chi,bhardwaj2023cilantro,kalim2018henge,ds2,Sponge,mei2020turbine} in response to changes in workload conditions or resource availability. 
We refer to the operator whose deployment (key assignment or physical location) changes as the \emph{target} operator.
Figure~\ref{fig:reconfig-types} shows four reconfiguration scenarios \ours supports. In the initial deployment (Figure~\ref{fig:reconfig-types}a), upstream tasks $u_1$ and $u_2$ are co-located on node $n_1$ and ingest events over the key range $[1, 10]$. Events are partitioned by key and forwarded to downstream tasks $d_1$ and $d_2$ on node $n_2$: $d_1$ is responsible for keys $1$--$5$ and $d_2$ for keys $6$--$10$, each storing its associated state locally.

\stitle{Scale-out and scale-in} change the parallelism of the target operator by adding or removing tasks, causing repartitioning of its key space.  Figure~\ref{fig:reconfig-types}b shows a scale-out action where a new task, $d_3$, takes over keys $\{4, 5, 9, 10\}$ from $d_1$ and $d_2$. Figure~\ref{fig:reconfig-types}c shows a case of scale-in, where task $d_2$ is removed and its key partition and state are assigned to $d_1$.

\stitle{Key rebalance} reassigns keys among existing tasks by modifying the key assignment but without changing the operator parallelism.
Figure~\ref{fig:reconfig-types}d shows an example, where rebalancing is performed to mitigate skew. The hot key, $5$, is reassigned to task $d_2$ to balance the load. 

\stitle{Task migration} relocates a task from one node to another. Such an action may be necessary to balance resource utilization or move computation to new hardware~\cite{flink2,capsys}.
Figure~\ref{fig:reconfig-types}e shows task $d_2$ being migrated from node $n_2$ to $n_3$. Unlike the scenarios above, task migration does not alter the key partitioning but it changes the physical location of tasks.

\section{The \ours reconfiguration protocol}\label{sec:protocol}

\begin{figure*}[t]
    \centering
    \includegraphics[width=0.92\textwidth]{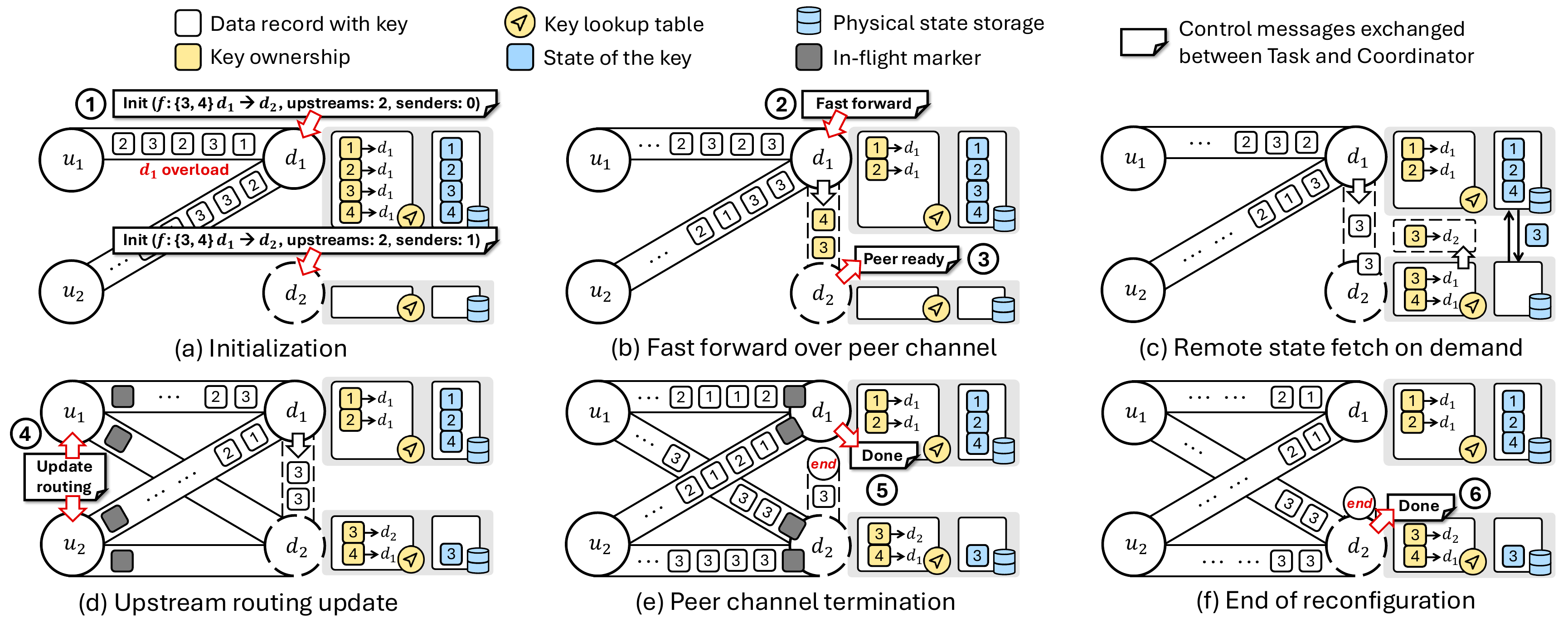}
    \caption{\ours protocol steps in a scale-out scenario, where the target operator $d$ is reconfigured to use one more data-parallel task $d_2$. Keys $\{3,4\}$ from the global key space $\{1, 2, 3, 4\}$ are reassigned from task $d_1$ to $d_2$. Markers are injected by upstream tasks $u_1$ and $u_2$ to notify downstream tasks $d_1$ and $d_2$ that records for migrating keys $\{3,4\}$ are now being routed to~$d_2$.}
    \label{fig:protocol-steps}\vspace{-2mm}
\end{figure*}


\ours is a new reconfiguration protocol for long-running distributed dataflows that matches the performance of embedded-state architectures while providing disruption-free adaptability.
Unlike existing protocols, \ours decouples state ownership---\emph{which task is responsible for processing a key}---from physical state location---\emph{which task stores the key's state}---enabling key space repartitioning to take effect independently of the migration of associated key-value pairs. Records are forwarded to new owners through \emph{peer channels} established between tasks of the target operator and relevant state is fetched \emph{per-key}, on demand. This enables fast ``work stealing'': new owners quickly take over computation from existing, often overloaded, tasks, while previous owners serve as remote state backends.


\stitle{Protocol assumptions.} \ours makes minimal assumptions about the underlying dataflow system: (i) state in the form of key-value pairs, (ii) support for event-time semantics along with a mechanism for tracking computation progress (\S\ref{sec:dataflow-model}), and (iii) FIFO communication channels between tasks, in practice, TCP connections. These features are common in all modern dataflow engines~\cite{akidau2015dataflow, carbone2015apache, akidau2021watermarks, sparkstructuredstreaming}. \ours also relies on a highly available coordinator that issues reconfiguration commands according to a policy. The coordinator is external to the dataflow runtime and can be implemented on top of a Paxos-based service. \ours is agnostic to the controller's policy but requires that reconfiguration commands are processed sequentially.
Like all existing protocols, \ours assumes no failures during reconfiguration. In practice,  a failure during reconfiguration prompts recovery from a checkpoint (see \S\ref{sec:implementation}).

\stitle{\ours ownership metadata.}
Decoupling key ownership from physical state location eliminates the constraint that state must reside on the task owning the key. A task can therefore claim logical ownership of a key without first acquiring the underlying state. To keep track of state location, \ours introduces a Key Lookup Table (KLT) that maps keys to the tasks where the state resides.
Each task maintains its own KLT; when accessing a key, the task first consults this table and redirects the request to a remote task if the state is not locally available. 


A KLT is updated in three cases: (i) the first time a key appears, (ii) upon transferring ownership of a key, and (iii) upon fetching remote state.
Ownership transfers are triggered by reconfiguration commands from the coordinator: the old owner removes all KLT entries  of the reassigned keys and sends them to the new owner, which inserts them into its own KLT. No state is migrated at this point.  The actual key-value pairs are fetched lazily by the new owner when processing input records that require access to remote state.  In this case, the owner first retrieves the state and then updates its KLT to reflect that the state is now local. Keys that are never accessed incur no KLT updates and no migration overhead.

In addition to the KLT, each task also maintains a Routing Table (RT), which is used to route output events to downstream tasks based on the event key. Upon a reconfiguration, \ours updates one or more KLTs of the target operator's tasks (to reflect the new key ownership) and one or more RTs of the upstream operator's tasks (to reflect the new routing scheme). To ensure correctness, the order in which these updates are applied requires careful coordination, as we explain below.\vspace{-2mm}

\subsection{Protocol overview} \label{subsec-procedure}

We illustrate \ours with an example and analyze its correctness properties in Appendix~\ref{appendix:analysis}. Figure~\ref{fig:protocol-steps} shows a scale-out scenario, where the parallelism of the target operator $d$ changes from 1 to 2. Task $d_1$ originally owns the entire key space
$\mathcal{K}=\{1, 2, 3, 4\}$, reflected in its KLT, and holds all associated state locally. 
Upon scaling out, the key space is repartitioned across the two data-parallel tasks $d_1$ and $d_2$. We refer to the task that takes ownership of the reassigned keys as the \emph{receiver} ($d_2$) and to the task that previously owned those keys as the \emph{sender} ($d_1$). \ours's coordinator exchanges control messages with the tasks and orchestrates protocol execution.

\stitle{Initialization.}
The coordinator prepares the reconfiguration by deploying the new task $d_2$ and repartitioning the key space $\mathcal{K}$ to generate the new key ownership scheme $f$. In Figure~\ref{fig:protocol-steps}a, the set of keys $\mathcal{K}^{'} = \{3, 4\}$ is reassigned from $d_1$ to $d_2$. \ours processes the scheme $f$ sent by the coordinator to derive two parameters for each task: (i) the number of expected upstream connections (\texttt{upstreams}), and (ii) the number of expected inbound peer connections (\texttt{senders}). 

\stitle{Reconfiguration steps.}
In step \circled{1}, the coordinator initiates the protocol by sending \texttt{Init} control messages to all tasks of the target operator $d$. Each \texttt{Init} message carries the key ownership scheme $f$ and the task-specific parameters \texttt{upstreams} and \texttt{senders}. In step \circled{2}, shown in Figure~\ref{fig:protocol-steps}b, the coordinator sends \texttt{Fast forward} messages to sender tasks to establish peer channels with their corresponding receiver tasks. 

Peer channels enable fast forwarding of in-flight records to receiver tasks according to $f$. Sender tasks first transfer key ownership by sending the respective KLT entries over the peer channels, in this case, entries $E= \{3\rightarrow d_1,~4\rightarrow d_1\}$,  and then begin redirecting records for those keys to receiver tasks instead of processing them locally.
As a result, existing overloaded tasks can push work to new tasks immediately, without having to wait for upstream tasks to update their routing. Ownership transfers are atomic to ensure that no two tasks update the same state.  

Once a receiver task establishes peer channels with all its \texttt{senders} and receives the corresponding KLT entries, it sends a \texttt{Peer ready} message to the coordinator (step \circled{3} in Figure~\ref{fig:protocol-steps}b).
Note that receiver tasks do not wait for fast forwarding to finish before sending \texttt{Peer ready} to the coordinator; they only wait to receive the KLT entries.

State corresponding to reassigned keys stays at the old owners ($d_1$) until the first access by the new owners ($d_2$). As shown in Figure~\ref{fig:protocol-steps}c, when $d_2$ receives its first record for key $3$, it consults its KLT to identify the state location, fetches the state from $d_1$, and updates its KLT entry to reflect that the state is now local. 
Because $d_1$'s and $d_2$'s key partitions are disjoint and the coordinator preserves this property when generating $f$, concurrent access to $d_1$'s state requires no coordination, allowing $d_2$ to process $d_1$'s backlog.
Subsequent records for key $3$ are served locally by $d_2$, incurring no further remote accesses. State corresponding to keys that are never accessed (e.g., key 4) remains in $d_1$, avoiding unnecessary migrations.

Upon receiving \texttt{Peer ready} messages from all receiver tasks, the coordinator proceeds to step \circled{4} shown in Figure~\ref{fig:protocol-steps}d. It instructs all upstream tasks ($u_1$ and $u_2$) to update their RTs according to the new ownership scheme. Specifically, each upstream task establishes output connections to newly deployed downstream tasks (in this case, $d_2$), injects markers into all output channels, and begins forwarding records according to its updated RT. Markers are used to notify the downstream tasks that all in-flight records for migrating keys are now being routed according to $f$.
Step \circled{4} must happen after the ownership transfer (steps \circled{2}--\circled{3}) to guarantee a single owner task per migrating key (\S\ref{subsec-semantics}).
At this point, all records are routed according to the new configuration: in-flight records for reassigned keys $\mathcal{K}^{'} = \{3,4\}$ already sent to $d_1$ are rerouted to $d_2$ via the peer channel, while any new records for $\mathcal{K}^{'}$ generated by $u_1$ and $u_2$ are sent directly to $d_2$.

As shown in Figures~\ref{fig:protocol-steps}e-f, once $d_1$ has received markers from all upstream channels confirming that routing has changed, it closes the peer channel 
and sends a \texttt{Done} message to the coordinator at step \circled{5}. Similarly, once $d_2$ has received markers from all upstream channels and, in addition, the peer channel is closed by $d_1$ (confirming that all in-flight records have been consumed), it sends a \texttt{Done} message to the coordinator at step \circled{6}. 

In \ours, a task can act both as a sender and a receiver simultaneously. In this case, it can safely terminate all outbound peer channels once all expected markers arrive, and it can afterwards send a \texttt{Done} message once all inbound peer channels have closed. When the coordinator receives \texttt{Done} messages from all tasks of the target operator, the reconfiguration completes. 



\begin{figure}[t]
    \centering
    \includegraphics[width=0.46\textwidth]{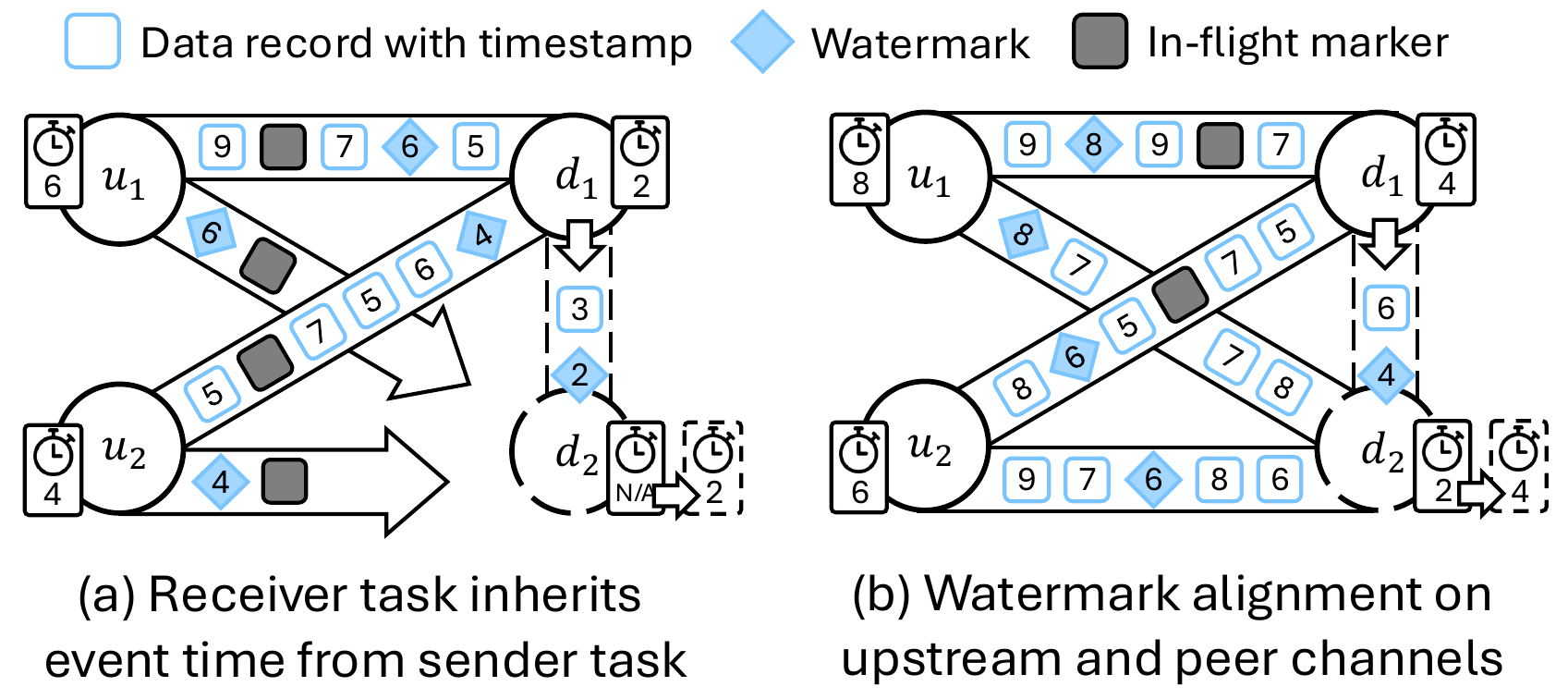}
    \caption{Watermark propagation in \ours.}
    \label{fig:watermark}
\end{figure}

\subsection{Watermark propagation}\label{subsec-watermark}

So far, we have purposely omitted watermark handling in \ours to keep the presentation simple. We now describe how the protocol handles event time to ensure correct execution during reconfiguration.
In the dataflow model (\S\ref{sec:dataflow-model}), a task advances its event time by computing the minimum of all watermarks received across its upstream channels. \ours, however, introduces additional peer channels that must be taken into account to avoid premature event-time progress. Moreover, \ours needs to decide when and how to set the event time of any new tasks introduced by a reconfiguration. 

A naive but correct approach is to block until all peer channels drain in-flight records and markers from all upstream channels arrive at the new task. However, this would prevent the new task from emitting any results in the meantime. Instead, \ours leverages watermark monotonicity to let tasks immediately \emph{inherit} their senders' event time:
since watermarks are propagated from the upstream to the downstream tasks of a dataflow, \emph{the event time of a receiver's upstream tasks is guaranteed to be higher than or equal to its sender's event time}.
Figure~\ref{fig:watermark}a shows an example. Upon establishing the peer channel, sender $d_1$ forwards its watermark 2 to the receiver, $d_2$, directly initializing $d_2$'s event time to 2 before upstream channels are connected. It is worth noting that, even though $d_2$ must wait for watermark messages from all peers to set its event time, \emph{it does not need to pause execution} and can keep processing input records. Setting its event time only enables it to safely emit time-based results downstream, such as aggregates of closed windows.




The upstream tasks also send their current watermarks to the receiver task as the first message after injecting the in-flight markers (Figure~\ref{fig:watermark}a). 
While a reconfiguration is underway, \ours treats peer channels as additional upstream channels for watermark alignment; that is, receiver tasks advance their event time by computing the minimum watermark across all input channels, i.e., both upstream and peer channels.
This is shown in Figure~\ref{fig:watermark}b: $d_2$ advances to event time 4 after receiving the watermark 4 from both upstream and peer channels. When the sender closes the peer channel, the receiver task recalculates its watermark, potentially advancing its time.
We provide a proof sketch of correctness for watermark propagation in Appendix~\ref{subsec-semantics}.

\subsection{Supporting multiple reconfigurations}\label{subsec-multiple}

\begin{figure}[t]
    \centering
    \includegraphics[width=0.45\textwidth]{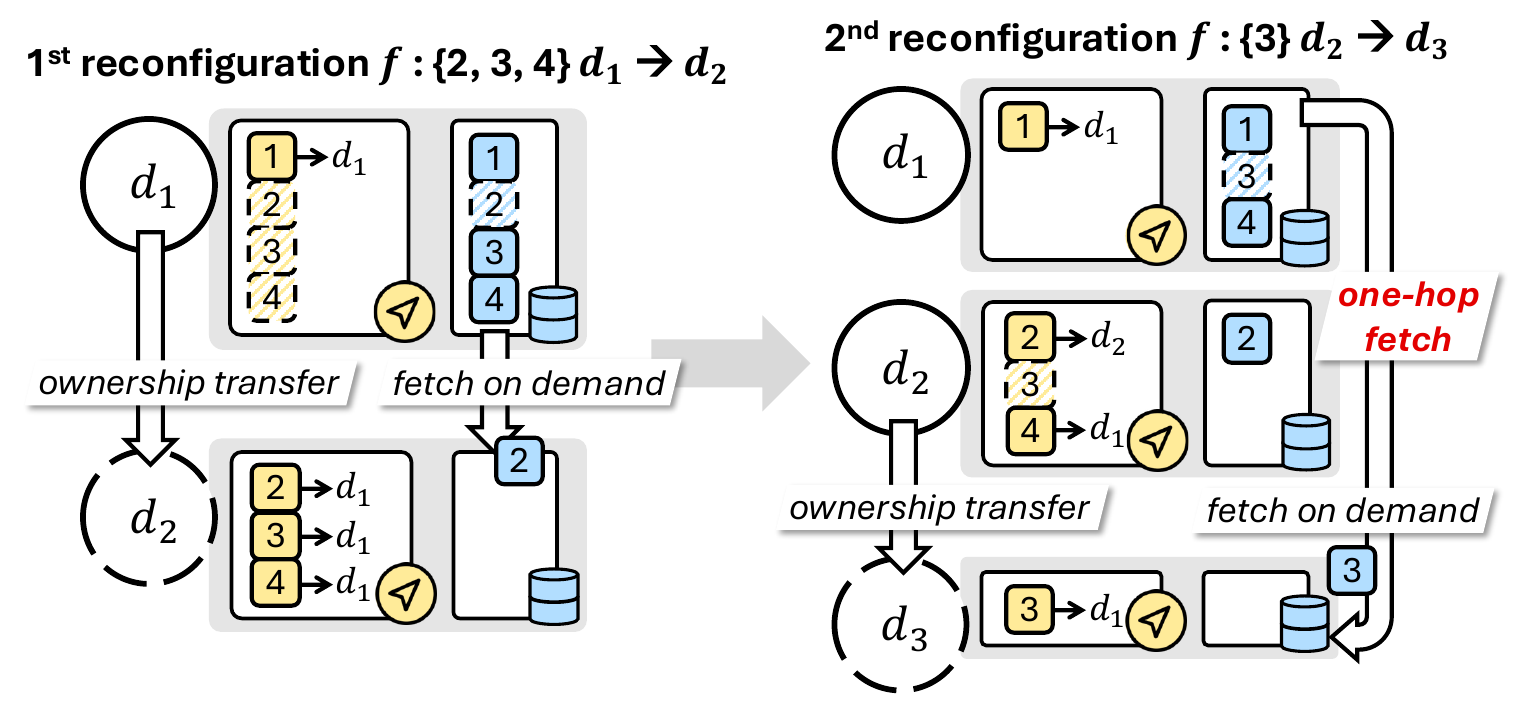}
    \caption{By propagating KLT entries to receiver tasks, \ours guarantees that remote state is fetched in a single call, even in case of repeated reconfigurations of the same operator.}
    \label{fig:multi-reconfig}
\end{figure}

\ours expects that reconfiguration commands are executed sequentially, in some order defined by the coordinator. 
Yet, because a \ours reconfiguration is considered complete as soon as ownership transfer and rerouting take effect, multiple target operators can be reconfigured in short succession and their on-demand state migrations can proceed in parallel without conflicts.
A core challenge, however, lies in tracking where a key's state resides after repeated reconfigurations, which can be common in long-running jobs due to inaccurate predictions, noisy performance metrics, or oscillating workloads.


With \ours's fetch-on-demand strategy, key ownership and physical state location are likely to be out-of-sync upon repeated reconfigurations. Consider the example of Figure~\ref{fig:multi-reconfig}, where we increase the parallelism of operator $d$ from 1 to 2 tasks, with keys $\mathcal{K}_2 = \{2, 3, 4\}$ reassigned from $d_1$ to $d_2$, followed by a second scale-out from 2 to 3 tasks, with keys  $\mathcal{K}_3=\{3\}$ reassigned from $d_2$ to $d_3$. When the second reconfiguration starts, $d_2$ owns key 3 but its state is still on $d_1$. In this case, a strawman approach would require 2 hops to locate the state associated with key $3$: one from $d_3$ to $d_2$ and a second one to $d_1$. In general, the number of hops grows linearly with the number of reconfigurations, increasing tail latency of state accesses and  degrading overall performance.

To address this problem, \ours provides a \emph{one-hop state fetch} guarantee regardless of prior reconfigurations. 
Recall that KLT entries, which carry the ground truth of state location, (i) are passed to the receiver task with every key ownership transfer, and (ii) are only updated by the owner on a remote state fetch (to indicate that the state is now local). 
As a result, the KLT entries held by the owner task always contain the correct state location. In the example of Figure~\ref{fig:multi-reconfig}, when $d_3$ needs to access state for key 3, it consults its KLT entry for key 3 and fetches the state from $d_1$ in a single call. This entry was passed to $d_3$ by $d_2$ during the second reconfiguration, yet still points to $d_1$, as the state location for key $3$ had not changed since the first reconfiguration (and had it changed, $d_2$ would have passed the updated entry to $d_3$). The property holds by induction for any number of reconfigurations. 

In addition to reducing tail latency of remote state accesses, \ours's one-hop guarantee also reduces the amount of state transferred when canceling tasks. For example, if $d_2$ is removed in a subsequent reconfiguration, only state for key 2 needs to be migrated, as state for key 4 remains in $d_1$ until it is fetched on demand by $d_3$, or $d_1$ is also canceled. We evaluate such a scenario in \S\ref{sec:protocol-generality:repeated-operator-reconfiguration}.






\subsection{Forcing a full state transfer}\label{subsec-full}

When a reconfiguration cancels tasks (e.g., due to scale-in or task migration), their entire key space must be reassigned to other tasks and therefore full state migration is inevitable. To handle such scenarios, \ours adopts a gradual migration strategy that reuses the existing fetch-on-demand mechanism, avoiding additional synchronization overhead for bulk state transfer.
Within each remote fetch request, which potentially includes multiple remote keys, \ours piggybacks $N$ extra keys for state migration if (i) the target task is canceled, and (ii) the requesting task owns keys whose state still resides in the canceled task. In case a task does not need any keys from a canceled task in a given batch, \ours still initiates a fetch request carrying only the extra keys and continues doing so until there is no state left in the canceled task. The technique works with any number of canceled tasks.



State transfer for extra keys proceeds in the background, without disrupting regular record processing. The number of extra keys $N$ is a tunable parameter that trades off performance disruption against migration duration. A large $N$ migrates state aggressively, which is preferable when canceled tasks must be terminated as soon as possible. In certain cases, such as periodic workloads that oscillate between a fixed set of configurations, a fetch-on-demand approach that does not migrate any extra keys may be desirable to avoid repeatedly moving cold state back and forth. We empirically evaluate the impact of $N$ in \S\ref{sec:protocol-generality:task-migration}.

While similar in spirit to Megaphone's fluid strategy~\cite{megaphone} and DRRS's subscales~\cite{DRRS}, a critical difference is that \ours's gradual migration does not divide a single reconfiguration into multiple smaller ones. Piggybacking gradually fetches cold state and does not prolong the duration of ownership transfer.

\subsection{Protocol generality}


\ours generalizes to all common stateful operators in dataflow systems, including continuous aggregations, windows, and joins.  For operators that are not time-dependent, such as running aggregates, the protocol simply skips the watermark forwarding and event time computation steps. For two-input operators such as joins,  \ours works as described, with one peer channel per logical input.  

\ours supports any reconfiguration that can be expressed as a key ownership scheme $f$: (i) if $f$ reassigns a subset of keys from existing tasks to new tasks, it corresponds to \emph{scale-out}, (ii) if $f$ moves all keys from existing tasks to other existing tasks, it corresponds to \emph{scale-in}, (iii) if $f$ reassigns keys without changing resource allocation, it corresponds to \emph{rebalance}, which can be used for skew mitigation, and (iv) if $f$ moves all keys from an existing task to a new task, it corresponds to \emph{task migration}.
\ours is also agnostic to the key space partitioning policy and works with consistent hashing (our default), uniform hashing, or any custom scheme.

Finally, \ours is workload-agnostic. While \ours primarily benefits from realistic workloads with temporal locality and skew, where a small and slowly-evolving working set is migrated and cold state is left in place, it operates correctly under any state access pattern. We evaluate \ours's worst-case scenario---uniform state access across the entire key space---in Appendix~\ref{appendix:uniform-access}.

\section{Implementation}\label{sec:implementation}

We implement \ours as part of a new distributed dataflow system we developed from scratch in Go. The system is designed to support pluggable reconfiguration protocols, enabling a fair experimental comparison of different approaches on top of a common runtime. We describe the system architecture and practical aspects of the \ours implementation.


\stitle{Execution model.} Our dataflow runtime follows a standard coordinator-worker architecture. The coordinator manages resource allocation, task scheduling, periodic checkpoints, job monitoring, and reconfiguration actions. We use sticky task scheduling, where each task is assigned to a worker at deployment time and runs continuously unless a reconfiguration occurs. Workers communicate via FIFO dataflow channels that are implemented on top of standing TCP connections.

The runtime exposes a familiar data-parallel programming model and supports standard dataflow operators, including event-time windows, custom aggregations, and joins. Each task executes a continuous processing loop that pulls a batch of records from upstream channels, applies user-defined logic, and forwards results downstream. 

\stitle{State management and fault tolerance.}
Stateful tasks access state through a unified backend API that can be configured as either an embedded or an external key-value store. Both options are currently backed by PebbleDB~\cite{pebble}. Upon receiving an input batch, a task fetches the state of all unique keys it contains, processes all records, and flushes updates back to the backend. Per-batch deduplication ensures that state for a key is fetched at most once per batch, regardless of how many times it appears. This optimization benefits all reconfiguration protocols implemented on top of the runtime and matters even for local state,  as it reduces (de-)serialization overhead. 

To support large key spaces in realistic deployments, keys are grouped into buckets, where each bucket is a logical partition of the key space and serves as the basic unit of state ownership assignment. The runtime is agnostic to the state partitioning strategy, which can be configured per deployment.  By default, we use consistent hashing~\cite{dynamodb} with 128 buckets.

To provide fault tolerance, we have implemented a standard distributed checkpointing mechanism~\cite{carbone2017state, chandy1985distributed} with the addition that KLT entries are included in task snapshots.
If a failure occurs, tasks recover from the last checkpoint. 


\stitle{Reconfiguration utilities.}
To support pluggable reconfiguration protocols, the runtime provides a set of shared control-plane utilities. These include APIs for modifying key space partitioning, the number of workers, and task placement, as well as mechanisms for establishing control-plane connections between workers and the coordinator. The runtime also exposes extensible APIs for defining custom control messages and their handlers, and built-in coordination mechanisms for marker injection,  routing updates,  and collecting acknowledgments.  We implement both \ours and the three baseline protocols evaluated in \S\ref{sec:exp-baselines} on top of these utilities.  We describe the \ours implementation below and discuss the implementation of other protocols in \S\ref{sec:exp-baselines-intro}.

\stitle{\ours implementation.} In \ours, a task may act as a sender, a receiver, or both.  A sender has transferred ownership of certain buckets but may continue to receive records for keys in them while the reconfiguration is in progress.  To handle forwarding efficiently, each sender spawns one goroutine per peer connection alongside its main processing goroutine.  A receiver acquires ownership of new keys and fetches state on demand from the corresponding peer senders.  For each input batch,  the main goroutine first consults the KLT to classify event keys as having local or remote state. 
Remote state is fetched from all relevant peers in parallel, and processing begins as soon as the required state arrives, while preserving the order in which keys appeared in the input batch.  Local and remote state accesses can safely happen concurrently without locking, because ownership transfer guarantees per-key isolation (see Appendix~\ref{subsec-semantics}). State backend flushing and KLT updates are performed in the background and synchronization is only enforced at batch boundaries.

To keep the KLT compact, \ours maintains it at bucket granularity. Each bucket owned by a task holds an in-memory mapping from keys to their current state location, with a new entry inserted upon a key's first appearance.  When a bucket is reassigned during reconfiguration, its lookup entries are transferred to the receiving task as a unit.

%
%

\section{Evaluation}\label{sec:eval}


Our evaluation is organized into three parts: In \S\ref{sec:exp-baselines}, we compare \ours with prior protocols under a challenging setting that accumulates large state and backlog before reconfiguration.  In \S\ref{sec:protocol-generality}, we demonstrate \ours's broad applicability and generality, evaluating (i) a production-scale experiment with large state,
(ii) repeated reconfigurations under variable workloads, (iii) the reconfiguration scenarios of \S\ref{sec:reconfig-scenarios}, and (iv) a multi-operator reconfiguration. Finally,  in \S\ref{sec:exp-microbenchmarks}, we present a set of controlled experiments to understand how \ours behaves under varying state size,  workload characteristics, and progressive state migration strategies.

\subsection{Experimental methodology}\label{sec:exp-methodology}

We implement all protocols and workloads on top of our distributed dataflow framework (\S\ref{sec:implementation}). Sharing a common runtime ensures that the evaluation reflects genuine differences in protocol design and state management architecture rather than implementation artifacts. In Appendix~\ref{sec:appendix-drrs}, we also validate that our dataflow runtime outperforms Apache Flink.

\stitle{Experimental setting.}
We conduct experiments on two testbeds: (i) \textbf{E1-AWS} consists of  \texttt{c5d.4xlarge} EC2 instances, each with 16 vCPUs, 32 GB RAM, and 400 GB NVMe SSD, and (ii) \textbf{E2-CloudLab} uses \texttt{c6620} nodes in the CloudLab Utah cluster~\cite{Cloudlab}, each with 24 vCPUs, 96 GB RAM, and 1.6 TB NVMe SSD.  Unless otherwise specified,  we deploy a 12-node cluster: 4 nodes for Kafka brokers and producers,  4 nodes for the PebbleDB store used by the Remote baseline (see \S\ref{sec:exp-baselines-intro}), and 4 nodes for the dataflow runtime, with a configurable number of worker tasks.  The dataflow coordinator is colocated with one of the Kafka brokers.

%

\subsubsection{Workloads}
We use two types of workloads:

\noindent \textbf{(1) Real-world queries} are constructed from publicly-available data streams that inherently exhibit temporal locality and skew: (i) \textbf{Twitch}, a live-streaming analytics pipeline of audience engagement events~\cite{rappaz2021twitchdataset,rappaz2021recommendation} on the Twitch platform; (ii) \textbf{Taxi}, a trending routes query~\cite{DEBS2015} based on New York City taxi trip records~\cite{newyorktaxi}; (iii) \textbf{Azure}, a CPU utilization analytics workload based on Microsoft Azure traces~\cite{AzureVMData}; and (iv) \textbf{Borg}, a resource monitoring pipeline based on Google cluster traces~\cite{GoogleClusterData}. We describe the queries in detail in Appendix~\ref{appendix:workload}.

\noindent \textbf{(2) Nexmark}~\cite{nexmark} is the de facto stream processing benchmark. We implement queries \textbf{Q3}--\textbf{Q8} (Q1, Q2 are stateless) and two variants, \textbf{Q4*} and \textbf{Q6*}, introduced in prior work~\cite{Meces}.

\subsubsection{Baselines}\label{sec:exp-baselines-intro}

We compare \ours with three baselines:

\noindent \textbf{(1) Optimized stop-and-restart (S\&R)} is our implementation of the stop-and-restart protocol~\cite{wang2023fries,castro2013integrating} that pauses only the affected tasks during reconfiguration.  This is significantly less disruptive than Apache Flink's strategy, which restarts the entire execution graph from a checkpoint~\cite{carbone2017state}.  We further leverage PebbleDB's range queries over migrating buckets and parallelize transfers across all sender-receiver pairs to reduce the migration duration.  S\&R uses a consistent hashing partitioning strategy, like \ours.

\noindent \textbf{(2) DRRS}~\cite{DRRS} is the state-of-the-art live migration protocol for embedded state.  We choose DRRS as a baseline because it incorporates techniques from all prior live migration approaches: reconfiguration signal ingestion~\cite{megaphone, wang2023fries}, fluid migration~\cite{megaphone, Meces}, and prioritized migration~\cite{Meces}. 
In addition, DRRS reroutes in-flight data between tasks of the target operator and leverages out-of-order processing within and across input buffers to reduce periods of blocking.  
Our implementation reuses the reconfiguration utilities and peer channels built for \ours and extends the original protocol with proper watermark propagation to guarantee event-time correctness.  We use uniform key partitioning, according to the protocol description.  In Appendix~\ref{sec:appendix-drrs}, we show that our implementation outperforms the original DRRS on Apache Flink~\cite{drrs_github}.


   
   
\noindent \textbf{(3) Remote} is an external-state baseline that persists state in a separate cluster of PebbleDB instances. We deliberately built our own distributed store rather than relying on an existing system like TiKV~\cite{Tikv},  to avoid overheads from features unrelated to the state management architecture. We pre-assign key partitions to PebbleDB instances, routing requests directly to their designated instance over a persistent connection and we disable data replication. Our batch-oriented state access mechanism (\S\ref{sec:implementation}) further benefits this baseline by grouping keys and issuing per-destination requests in parallel. Our implementation results in seamless reconfiguration with no downtime, as we show in the rest of this section.  As a point of reference, Flink 2.0 with external state reports reconfiguration downtime in the order of tens of seconds~\cite{flink2}.

\begin{figure*}[t]
\centering
\begin{subfigure}{0.9\textwidth}
\centering
\includegraphics[width=\linewidth]{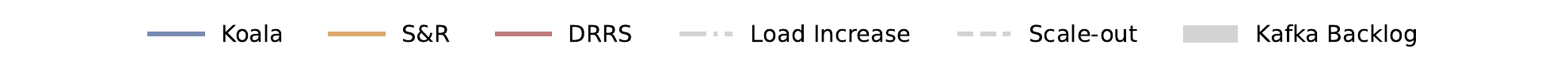}
\vspace{-6mm}
\end{subfigure}
\hfill
\begin{subfigure}{0.49\textwidth}
\centering
\includegraphics[width=\linewidth]{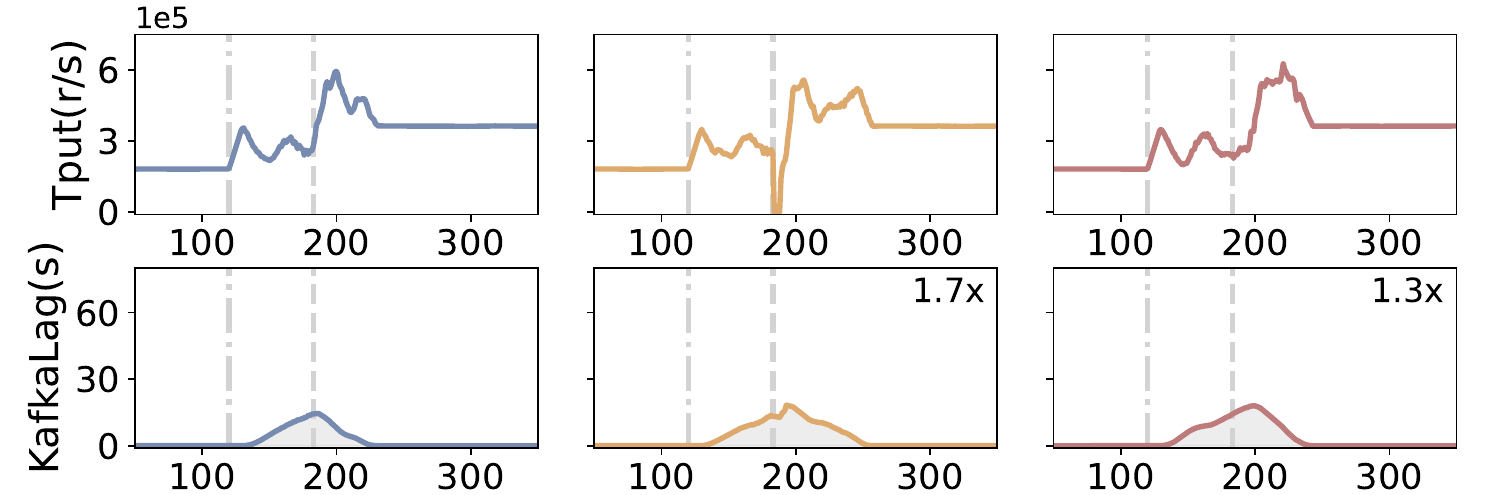}
\vspace{-6mm}
\caption{Twitch}
\label{fig:twitchBaselineComparison}
\end{subfigure}
\hfill
\begin{subfigure}{0.49\textwidth}
\centering
\includegraphics[width=\linewidth]{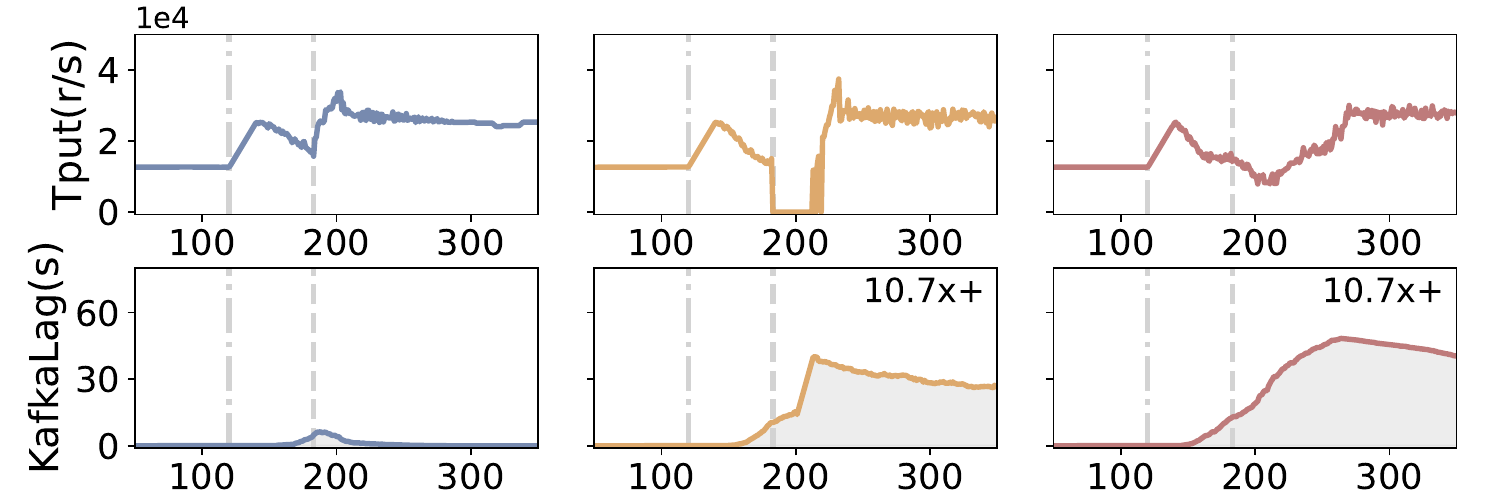}
\vspace{-6mm}
\caption{Taxi}
\label{fig:taxiBaselineComparison}
\end{subfigure}
\hfill
\begin{subfigure}{0.49\textwidth}
\centering
\vspace{-1mm}
\includegraphics[width=\linewidth]{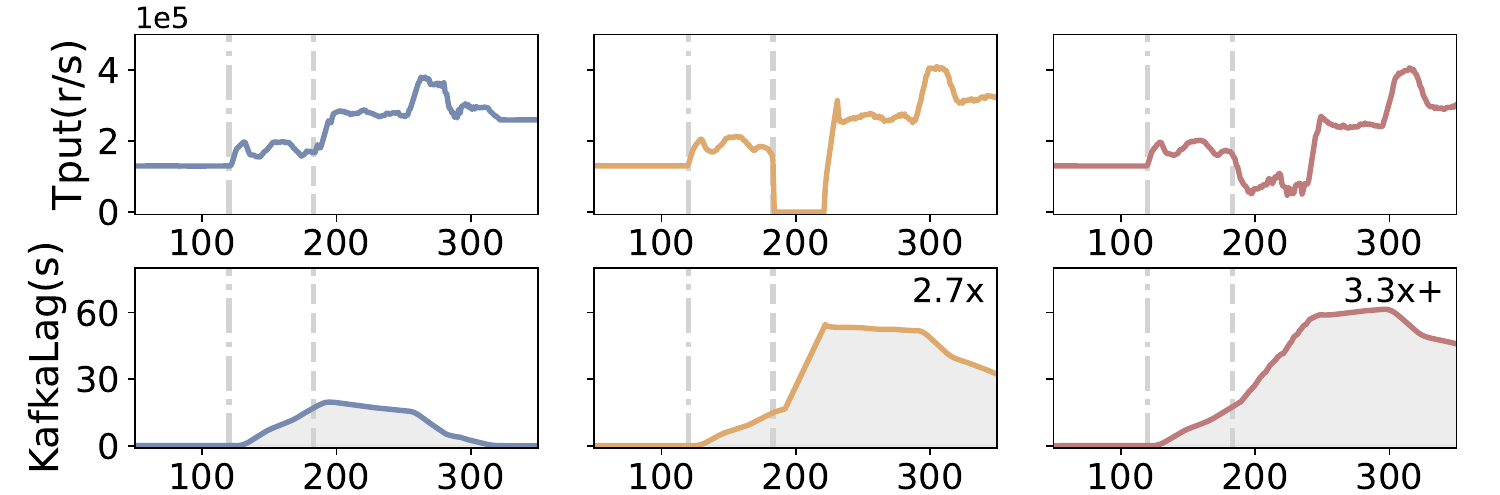}
\vspace{-6mm}
\caption{Borg}
\label{fig:borgBaselineComparison}
\end{subfigure}
\hfill
\begin{subfigure}{0.49\textwidth}
\centering
\vspace{-1mm}
\includegraphics[width=\linewidth]{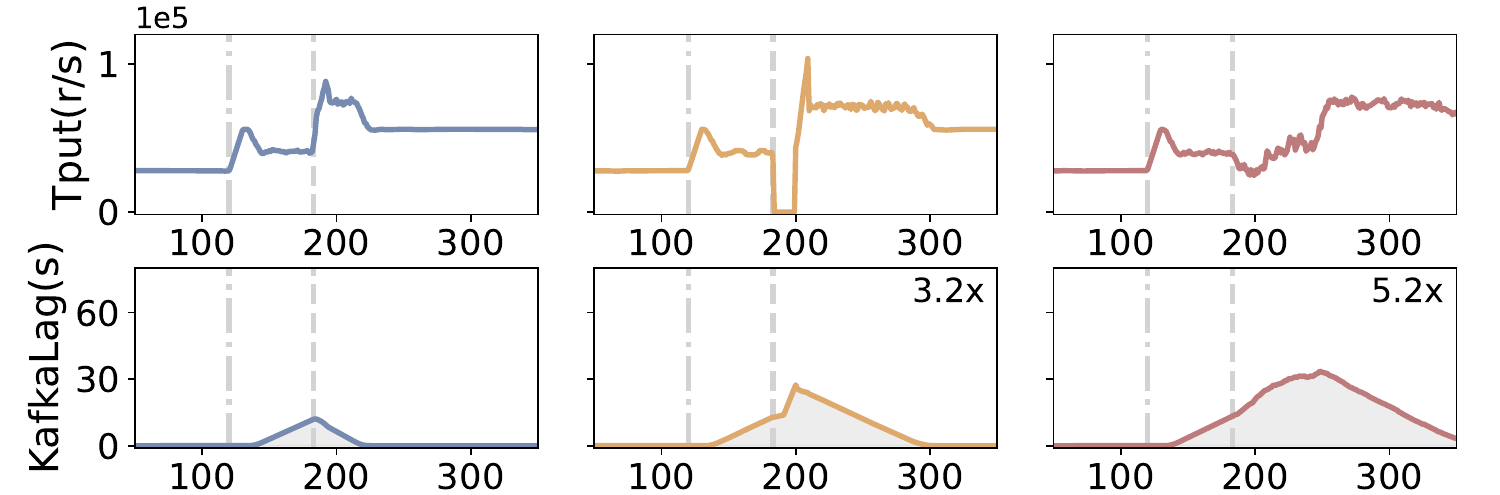}
\vspace{-6mm}
\caption{Azure}
\label{fig:azureBaselineComparison}
\end{subfigure}
\hfill
\begin{subfigure}{0.49\textwidth}
\centering
\vspace{-1mm}
\includegraphics[width=\linewidth]{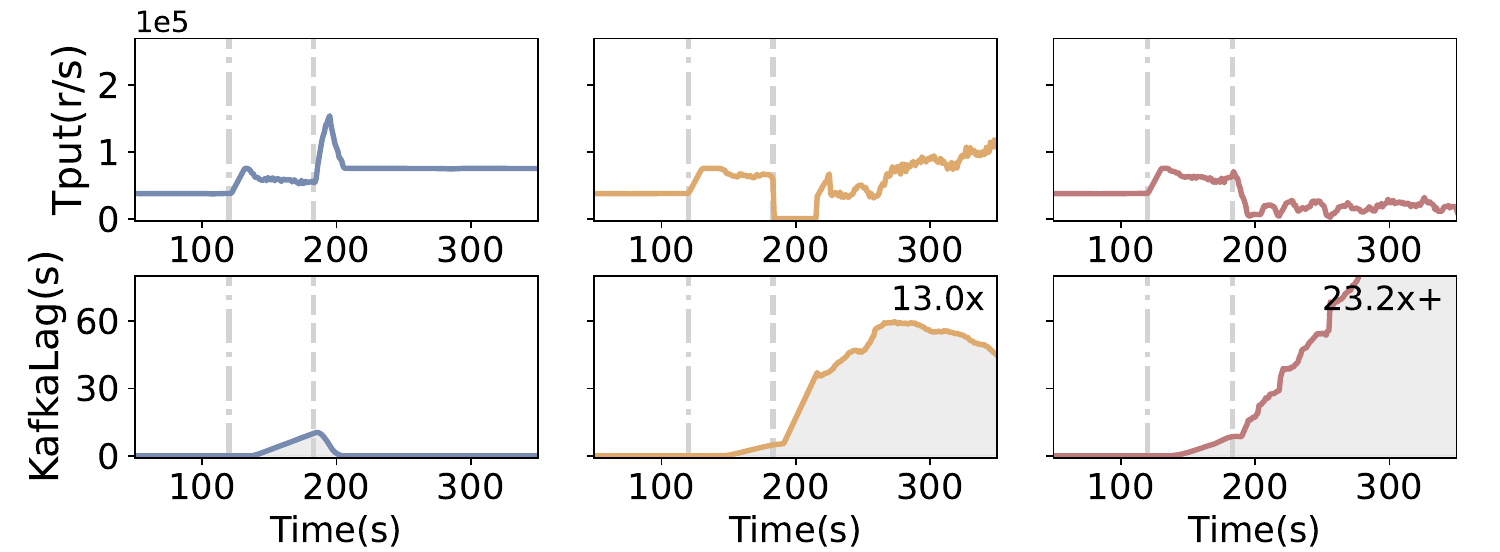}
\vspace{-6mm}
\caption{Q3}
\label{fig:q3BaselineComparison}
\end{subfigure}
\hfill
\begin{subfigure}{0.49\textwidth}
\centering
\vspace{-1mm}
\includegraphics[width=\linewidth]{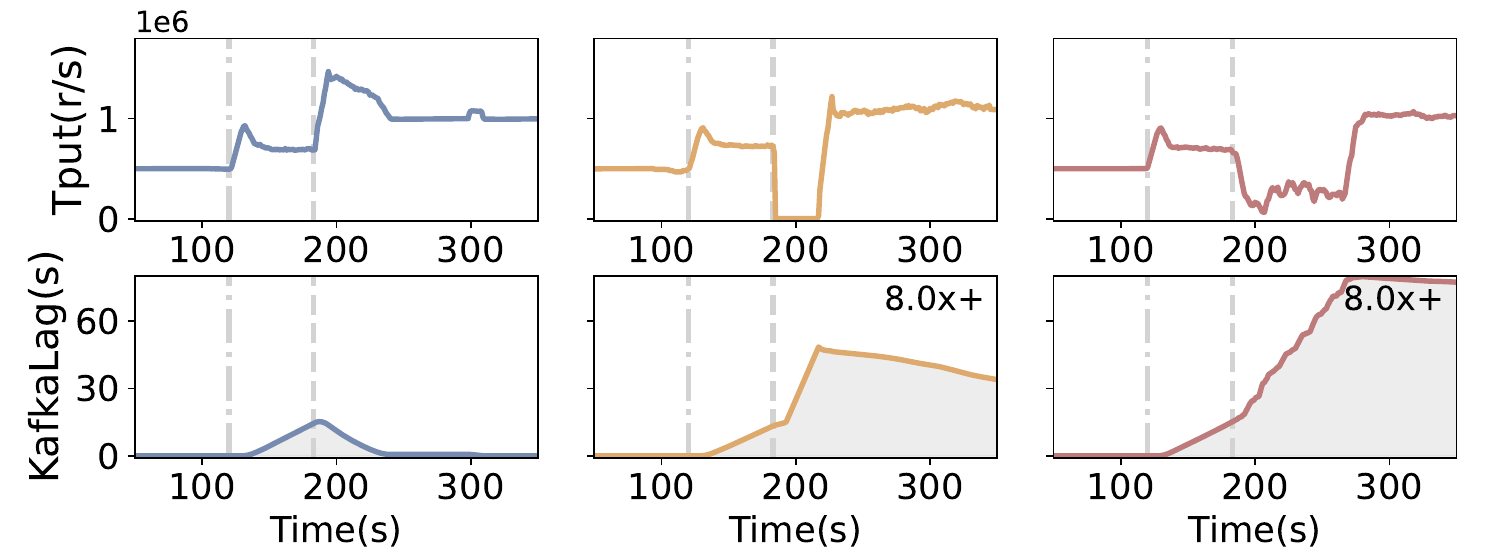}
\vspace{-6mm}
\caption{Q6*}
\label{fig:q6ModBaselineComparison}
\end{subfigure}
\caption{Comparison of \ours against S\&R and DRRS with embedded state. The Kafka-lag speedup (upper-right corner) reports the baseline's backlog recovery time relative to \ours; a trailing `+' indicates recovery exceeded the 10-minute runtime.}
\vspace{-3mm}
\label{fig:baseline_comparison}
\end{figure*}

\subsubsection{Evaluation Metrics}\label{sec-eval-metrics}
We collect metrics at $1s$ granularity and plot them after applying a $10s$ moving-average window for visual clarity. 
For each experiment, we report a subset of the metrics below that are most relevant to the scenario under study and we provide the remaining metrics in Appendix~\ref{appendix:full-results}.

\stitle{Throughput and Kafka lag.} To reflect realistic deployment conditions, we deploy parallel Kafka producers that generate records at a configurable input rate, with the dataflow consuming from Kafka as fast as possible.  When the input rate stays below the processing capacity, no backlog accumulates. When input rate increases beyond capacity, backlog builds up in Kafka.  As a result,  after a scale-out reconfiguration, the dataflow can temporarily exceed the input rate to drain the backlog before returning to steady state.

We report \textbf{throughput} as the aggregated output rate at the source operator, reflecting the end-to-end processing capacity of the dataflow. We also report \textbf{Kafka lag},  as the average time records spend waiting in the Kafka queue before being consumed. Kafka lag is a more informative metric than latency for evaluating reconfiguration,  because it demonstrates how quickly a system can recover from backlog accumulation. We also derive \textbf{backlog recovery time} as the time elapsed from the start of reconfiguration until lag returns to zero.

\stitle{Per-batch latency.} This metric measures the average time to process a batch of records, including state pre-fetching, computation, and state flushing.  Per-batch latency is useful because it captures the latency impact of embedded versus remote state access. We note that this metric is not meaningful for DRRS because its out-of-order processing design allows batches to be partially processed while remaining records are deferred until state migration completes.

We do not report end-to-end latency: operators may drop, transform, or generate records, so output records cannot be attributed to individual input records. Existing approaches approximate the end-to-end latency by injecting latency markers at the source~\cite{flink2}, which, however, bypass operator computation and only capture queuing and transmission delays.

\stitle{State transfer metrics.} We report \textbf{total migrated state}, as the total volume of state transferred during reconfiguration, and \textbf{state migration rate}, as the rate at which state is transferred.  

\begin{table}[t]
\centering
\footnotesize
\setlength{\tabcolsep}{2pt}
\begin{tabularx}{\linewidth}{>{\raggedright\arraybackslash}p{1.8cm} *{6}{>{\centering\arraybackslash}X}}
\toprule 
\textbf{Workload} & Twitch & Taxi & Borg & Azure & Q3 & Q6* \\
\midrule
\textbf{State Size (GB)} & 11.2 & 26.8 & 40 & 21.6 & 44 & 40\\
\bottomrule
\end{tabularx}
\vspace{1mm}
\caption{Accumulated state size of the target stateful operator in each workload after warmup.}
\label{tab:state_size:total_state_accumulated}
\end{table}

\begin{figure}
    \centering
    \includegraphics[width=0.465\textwidth]{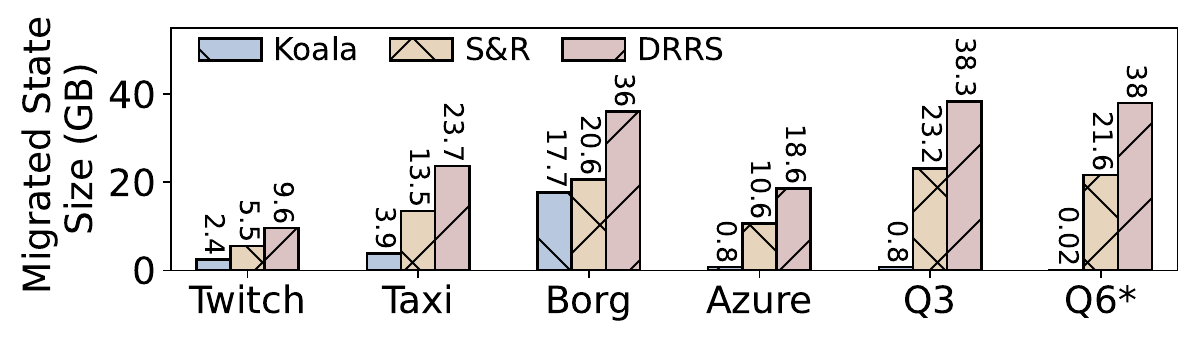}
    \caption{Total state migration size during reconfiguration.}\vspace{-3mm}
    \label{fig:state_size_transferred}
\end{figure}

\subsection{Comparison with baselines on large state}\label{sec:exp-baselines}

In this set of experiments, we evaluate all protocols on a scale-out scenario using six queries that can accumulate large state: the four \textbf{real-world queries} and Nexmark \textbf{Q3}, \textbf{Q6*}. The remaining queries have state in the order of MBs, and we include results for them in Appendix~\ref{appendix:full-nexmark-results}.

\stitle{Setup.}
We run this experiment on \textbf{E1-AWS} with the target operator parallelism set to 4 and the input rate at 70\% of processing capacity for the first 2 minutes (Phase 1). We then double the input rate to simulate a workload surge (Phase 2), during which backlog starts building up in Kafka. One minute later, we scale out from 4 to 8 tasks and continue for 7 more minutes (Phase 3). To accumulate large state before scale-out, we warm up each query for 30 minutes prior to Phase 1, resulting in the state sizes shown in Table~\ref{tab:state_size:total_state_accumulated}.


\begin{figure*}[t]
\centering
\begin{subfigure}{0.9\textwidth}
\centering
\includegraphics[width=\linewidth]{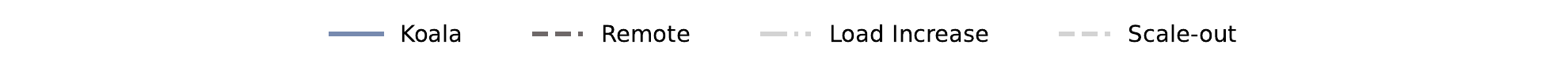}
\vspace{-6mm}
\end{subfigure}
\hfill
\begin{subfigure}{0.171\textwidth}
\centering
\includegraphics[width=\linewidth]{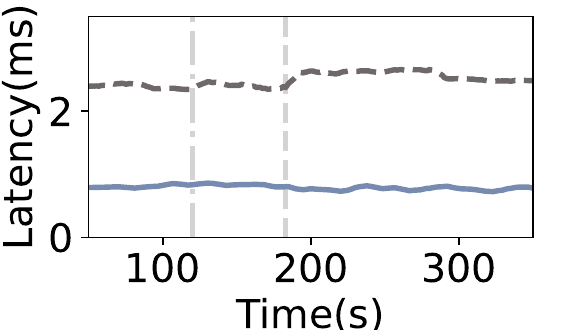}
\vspace{-5mm}
\caption{Twitch}
\label{fig:latency_twitch}
\end{subfigure}
\hfill
\begin{subfigure}{0.158\textwidth}
\centering
\includegraphics[width=\linewidth]{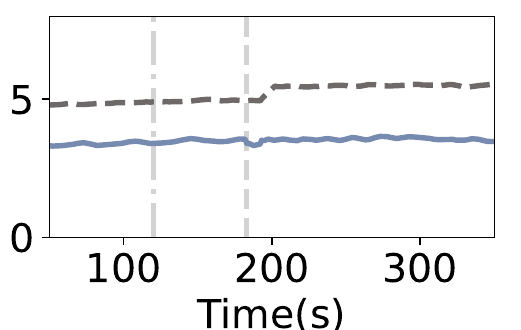}
\vspace{-5mm}
\caption{Taxi}
\label{fig:latency_taxi}
\end{subfigure}
\hfill
\begin{subfigure}{0.158\textwidth}
\centering
\includegraphics[width=\linewidth]{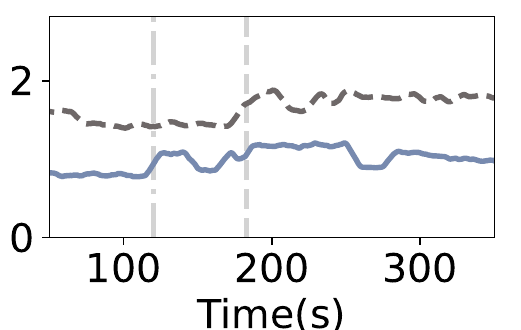}
\vspace{-5mm}
\caption{Borg}
\label{fig:latency_borg}
\end{subfigure}
\hfill
\begin{subfigure}{0.158\textwidth}
\centering
\includegraphics[width=\linewidth]{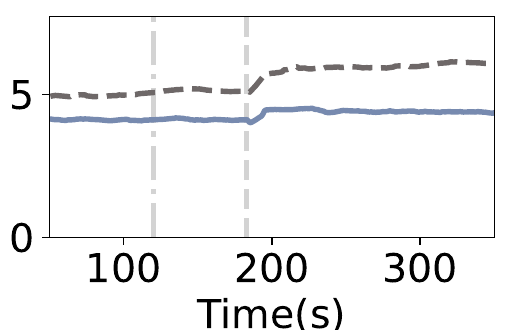}
\vspace{-5mm}
\caption{Azure}
\label{fig:latency_azure}
\end{subfigure}
\hfill
\begin{subfigure}{0.158\textwidth}
\centering
\includegraphics[width=\linewidth]{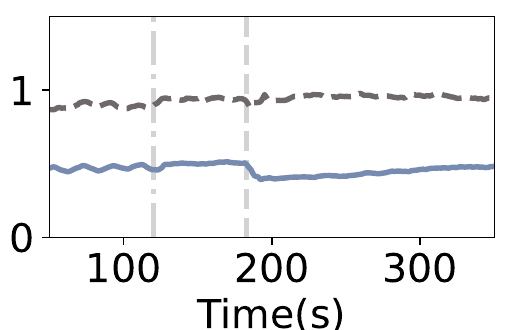}
\vspace{-5mm}
\caption{Q3}
\label{fig:latency_q3}
\end{subfigure}
\hfill
\begin{subfigure}{0.158\textwidth}
\centering
\includegraphics[width=\linewidth]{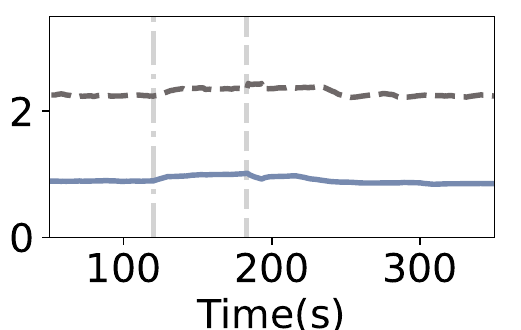}
\vspace{-5mm}
\caption{Q6*}
\label{fig:latency_q6mod}
\end{subfigure}
\caption{Comparison of \ours against Remote baseline with external state.}
\vspace{-4mm}
\label{fig:latency_comparison}
\end{figure*}

\stitle{Comparison with embedded-state protocols.}
Figure~\ref{fig:baseline_comparison} shows the throughput and Kafka lag for \ours, S\&R and DRRS.  After the load increases in Phase 2,  queries cannot keep up and Kafka lag grows. Upon reconfiguration in Phase 3, only \ours can instantly start consuming backlog across all queries, quickly stabilizing throughput.

S\&R experiences downtime during reconfiguration, causing the backlog to grow further until it resumes processing. By that point, significantly more Kafka backlog has accumulated, prolonging recovery time.  By contrast, DRRS sustains processing during reconfiguration through its live migration protocol. However, the large volume of transferred state triggers extensive database compaction and resource contention, reducing throughput and delaying backlog recovery.

\ours consumes the backlog within 18s--52s across queries, with \textbf{Borg} as an outlier at 127s due to its larger state migration size (Figure~\ref{fig:state_size_transferred}). S\&R and DRRS take multiple minutes to recover, and DRRS fails to consume the backlog within the 10-minute experiment duration for four of the queries, exhibiting up to $23.2\times$ longer backlog recovery time than \ours.


Figure~\ref{fig:state_size_transferred} shows the corresponding total state migration size and explains why \ours is non-disruptive: it transfers only a small fraction of the total state, while S\&R and DRRS migrate more than half of the total state. Specifically, \ours transfers $5\times$ less state than S\&R and $8.7\times$ less than DRRS on average across real-world queries.
Note that DRRS migrates more state than S\&R due to its uniform partitioning strategy.

\begin{figure}[t]
    \centering
    \includegraphics[width=0.46\textwidth]{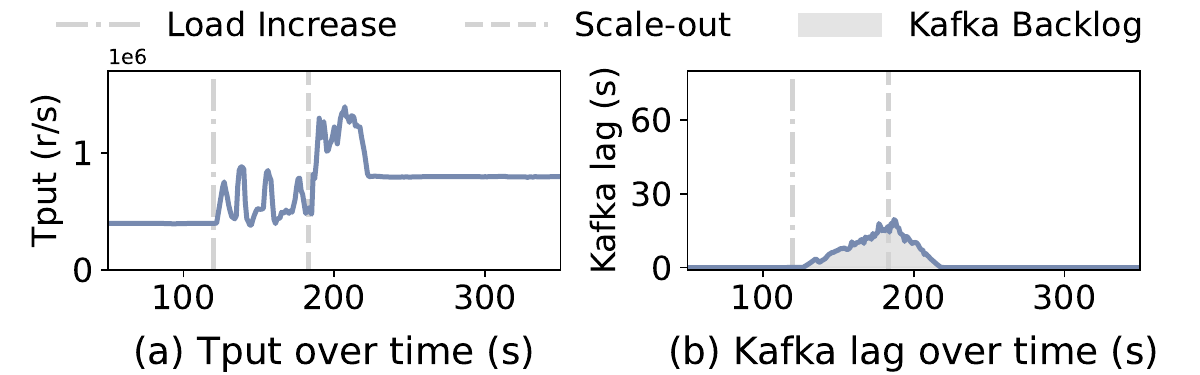}
    \caption{Scaling to 32 tasks with 320 GB total state size.}
    \vspace{-4mm}
    \label{fig:scalability}
\end{figure}

\stitle{Comparison with remote state.}
We also compare \ours with Remote, which migrates no state during reconfiguration and hence, does not suffer from any disruption.  We repeat the same scale-out experiment as above and measure per-batch latency over time. Figure~\ref{fig:latency_comparison} shows the results (additional metrics in Appendix~\ref{appendix:full-remote-state-results}). \ours consistently achieves lower latency across all queries during the experiment, with $34\%$ lower latency on average.  More importantly, \ours's latency remains stable during reconfiguration, matching Remote's disruption-free behavior with external state storage.


\vspace{-2mm}
\subsection{\ours in action}\label{sec:protocol-generality}

We evaluate \ours under four reconfiguration scenarios: large-scale migration (\S\ref{sec:protocol-generality:operator-scalability}), repeated reconfigurations under fluctuating workload (\S\ref{sec:protocol-generality:repeated-operator-reconfiguration}), skew mitigation (\S\ref{sec:protocol-generality:skew-mitigation}), and task migration (\S\ref{sec:protocol-generality:task-migration}). We run all experiments in this section on \textbf{E1-AWS}, except large-scale migration, which we deploy on \textbf{E2-CloudLab}. We further evaluate concurrent multi-operator reconfiguration in Appendix~\ref{sec:protocol-generality:multi-operator-reconfiguration}.

\vspace{-2mm}


\subsubsection{Large-scale migration}\label{sec:protocol-generality:operator-scalability}
To demonstrate that \ours can accommodate production-scale workloads, our next experiment mirrors the setup of a real Alibaba use case~\cite{flink2} (Section 5.2). We use \textbf{Q6*} and initially set the target operator's parallelism to 16. We run a warm-up phase until the total state size reaches $\approx320GB$ and then double the input rate, before scaling out to 32 tasks.

Figure~\ref{fig:scalability} shows that \ours rapidly consumes backlog and restores throughput. In fact, its recovery duration is comparable to that reported in \S\ref{sec:exp-baselines}, even though the total state size is one order of magnitude larger and distributed across $4\times$ more tasks. Because \ours migrates active keys only, migration cost is independent of task count and total state size. 

\vspace{-1mm}

\subsubsection{Repeated reconfigurations}\label{sec:protocol-generality:repeated-operator-reconfiguration}
Repeated reconfigurations are common in production deployments due to fluctuating workloads. To evaluate such a scenario, we use \textbf{Q6*} and expand the cluster to 6 nodes to support higher parallelism. We perform two consecutive rate increases and two corresponding scale-out reconfigurations, from 4 to 8 to 12 tasks, followed by two load reductions and scale-in actions, from 12 to 9 to 6 tasks. The initial warm-up state size is as reported in Table~\ref{tab:state_size:total_state_accumulated}.

\begin{figure}[t]
    \centering
    \includegraphics[width=0.455\textwidth]{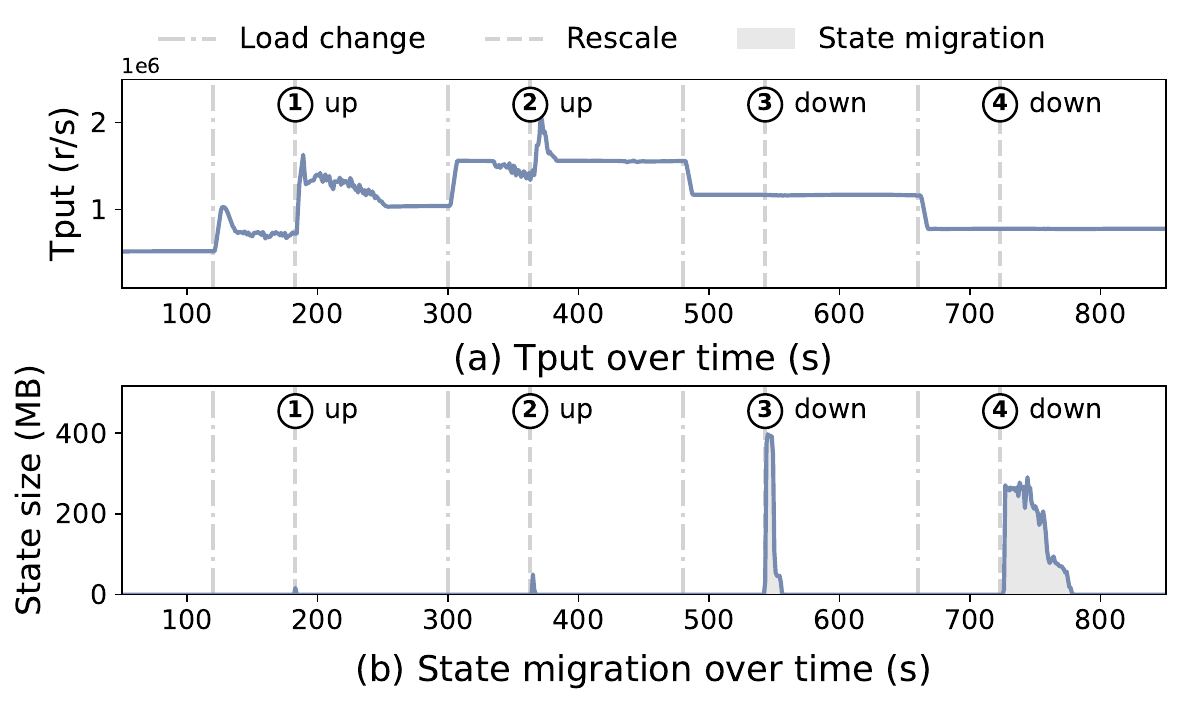}
    \caption{Repeated reconfigurations in response to fluctuating workload, with full state migration for scale-in scenarios.}
    \label{fig:repeated_operator_reconfiguration}
\end{figure}

Figure~\ref{fig:repeated_operator_reconfiguration} shows the throughput and state migration rate over the experiment duration. \ours maintains non-disruptive processing both when scaling out and when scaling in, demonstrating robustness under repeated reconfigurations. Notably, \ours transfers only $2.6GB$ of state at step \circled{3} and $8.9GB$ at step \circled{4}, significantly lower than the expected $10GB$ and $14GB$, respectively, if state were migrated fully after each reconfiguration. This is due to \ours's on-demand migration and one-hop state transfer property. Cold state is never migrated to new tasks during prior scale-outs, and therefore, need not be migrated back from canceling tasks.

\vspace{-2mm}


\begin{figure}[t]
    \centering
    \includegraphics[width=0.46\textwidth]{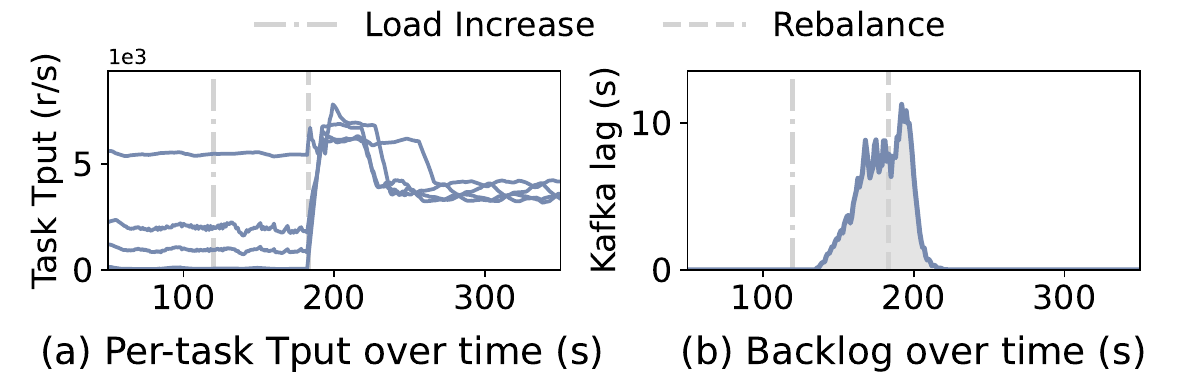}
    \caption{Rebalance for skew mitigation.}
    \vspace{-4mm}
    \label{fig:skew_mitigation}
\end{figure}

\begin{figure}[t]
    \centering
    \includegraphics[width=0.46\textwidth]{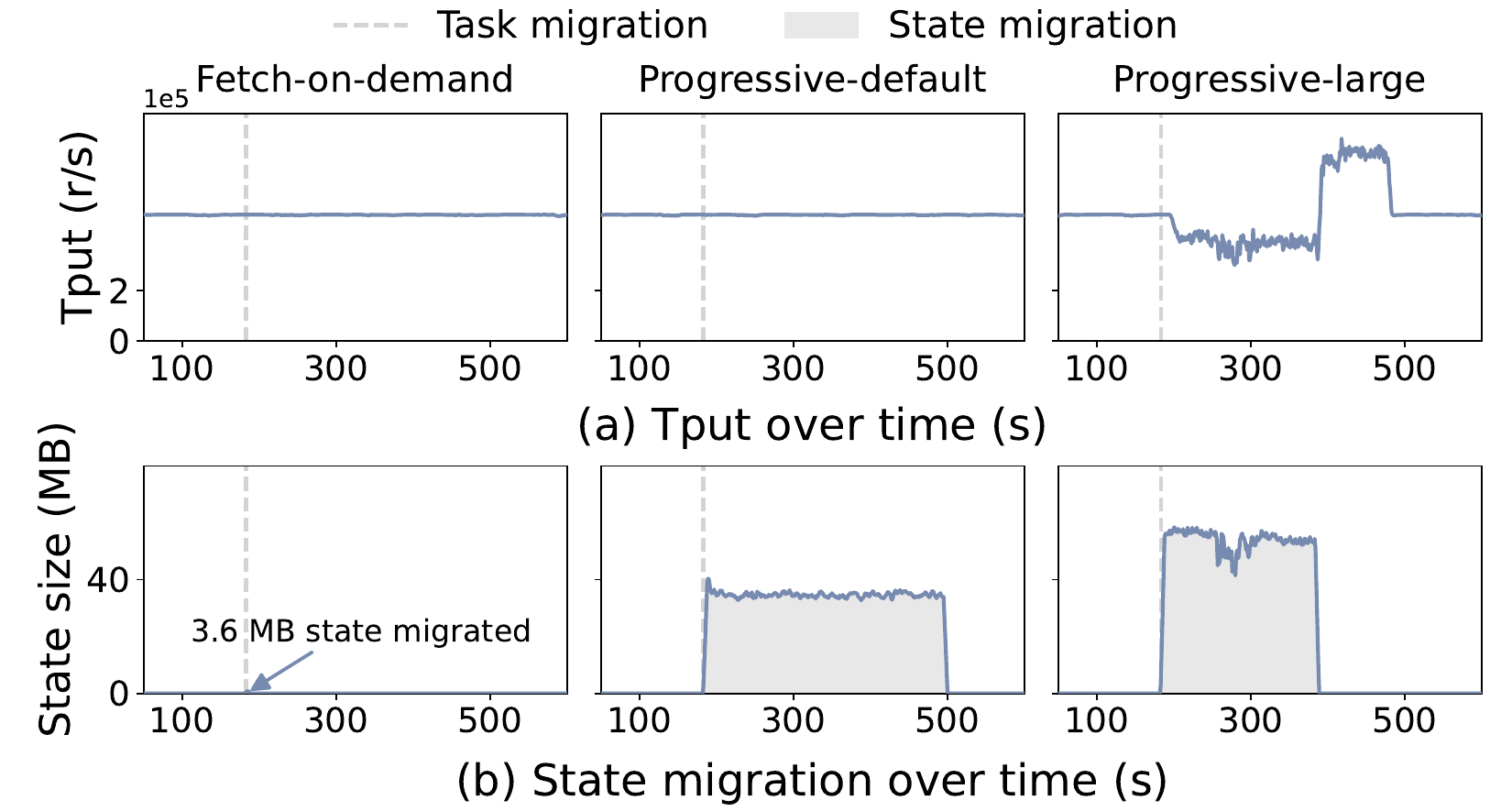}
    \caption{Task migration from one node to another.}
    \vspace{-4mm}
    \label{fig:task-migration}
\end{figure}

\subsubsection{Skew mitigation}\label{sec:protocol-generality:skew-mitigation}
We demonstrate a skew mitigation scenario using the \textbf{Taxi} query on the New York Taxi dataset~\cite{newyorktaxi}. We set the trip pickup area as the partition key, which yields a highly skewed distribution concentrated in busy city areas during peak hours.
The target operator is executed by four tasks, one of which is handling the majority of hot keys. Initially, the bottleneck task is operating close to capacity but without causing backlog. After 2 minutes, we increase the rate by 50\% and, one minute later, we trigger a reconfiguration that redistributes buckets across tasks in a skew-aware manner, balancing the load.

Figure~\ref{fig:skew_mitigation} shows the per-task throughput of the target operator (a) and Kafka lag (b) for the duration of the experiment. Before reconfiguration, the busy task is processing over 5000 records/s, while others are nearly idle. After rebalance, \ours quickly resolves the bottleneck, converging to a stable and balanced throughput.

\vspace{-2mm}






\subsubsection{Task migration}\label{sec:protocol-generality:task-migration}
We use \textbf{Q6*} to evaluate \ours's gradual migration strategy for a task migration scenario, where one of four tasks is moved to a different node. After warm-up, the task has accumulated $10 GB$ of state, which needs to be migrated in full.
As described in \S\ref{subsec-full}, gradual migration piggybacks additional keys on remote requests. We evaluate three strategies: (i) fetch-on-demand, the unchanged \ours protocol fetching only needed keys; (ii) progressive-default, fetching as many extra keys as keys in the batch ($\approx$25 for this query); and (iii) progressive-large, where we fetch $2\times$ more keys than progressive-default. Progressive-default is \ours's default policy for full state migration; users can tune $N$ according to their preference between non-disruptive processing and fast state migration.



Figure~\ref{fig:task-migration} shows throughput (a) and migrated state over time (b). Fetch-on-demand and progressive-default sustain the input rate without disruption, and the latter completes the migration in the background in about 5 minutes. 
Progressive-large halves the migration time relative to progressive-default at the cost of a slight throughput drop, illustrating the tunable trade-off between migration speed and disruption.

\vspace{-2mm}

\subsection{Microbenchmarks}\label{sec:exp-microbenchmarks}



We now examine \ours's sensitivity to varying state size and workload characteristics. We run these experiments on \textbf{E2-CloudLab} and use \textbf{Q6*}, as it is flexibly configurable through the Nexmark data source~\cite{beam-nexmark}. We ensure backlog-free conditions to isolate the effect of the variable under study.

\stitle{State size.} 
To understand the effect of state size on \ours, we run a series of experiments varying the target operator state size by controlling the key space, while fixing the accessed key size.
Specifically, we tune the warm-up duration to generate key space sizes of 5M, 10M, 20M, and 40M, resulting in total state sizes of 10, 20, 40, and 80 GB. We trigger a scale-out reconfiguration after 3 minutes, changing the parallelism of the target operator from 4 to 8.



\begin{figure}[t]
    \centering
    \includegraphics[width=0.47\textwidth]{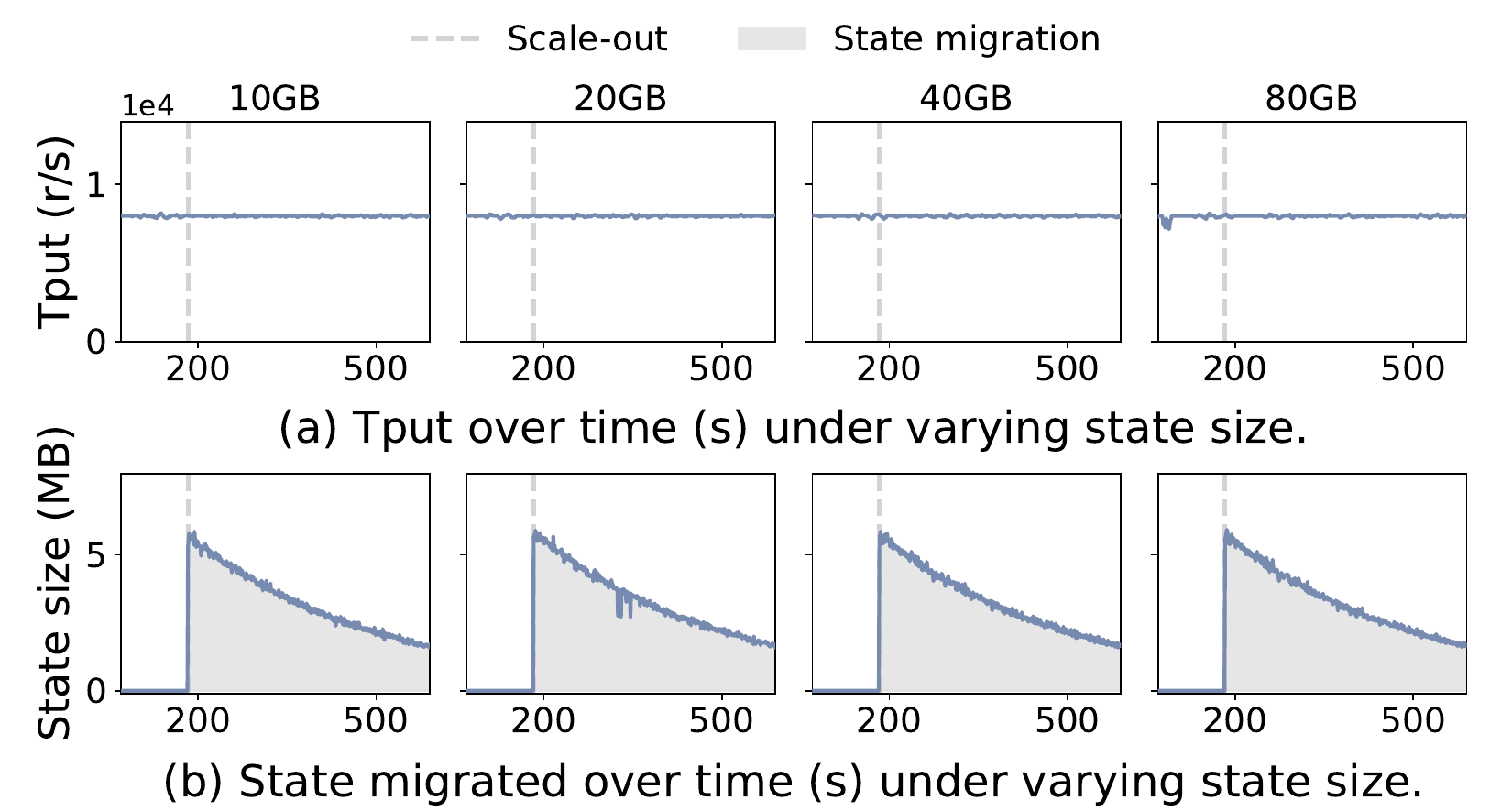}
    \caption{\ours's performance under varying state size.}
    \vspace{-2mm}
    \label{fig:micro-state-size}
\end{figure}

\begin{figure}[t]
    \centering
        \includegraphics[width=0.47\textwidth]{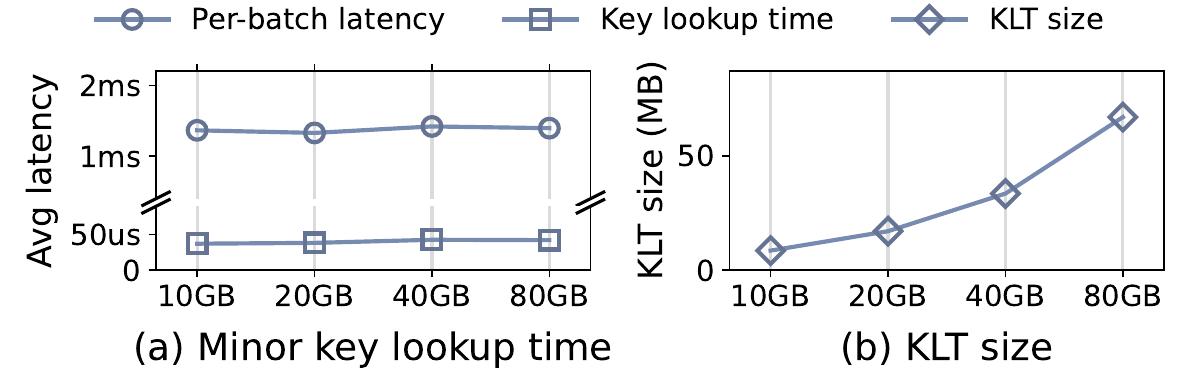}
        \caption{Key lookup overhead under varying state size.}
        \vspace{-3mm}
        \label{fig:key_lookup_overhead}
\end{figure}

Figure~\ref{fig:micro-state-size} shows throughput and state size transferred over time. \ours demonstrates stable behavior despite increasing state size. Across all runs, it migrates around $1.43 GB$, as it exploits the temporal key locality of the workload, relocating the active working set only. Figure~\ref{fig:key_lookup_overhead} shows the key lookup overhead and KLT memory footprint under varying state size. Key lookup takes less than 40$\mu$s per input batch across all runs, accounting for under 3\% of per-batch latency and adding negligible overhead to overall processing time (Figure~\ref{fig:key_lookup_overhead}a). Figure~\ref{fig:key_lookup_overhead}b shows the average per-task KLT memory usage after reconfiguration. While it grows with the key space, it remains compact, reaching at most 60 MB for 40M keys and 80 GB of total state. This is because each task only stores KLT entries for its owned buckets and the number of buckets per task decreases with increasing operator parallelism.

\begin{figure}[t]
    \centering
    \includegraphics[width=0.47\textwidth]{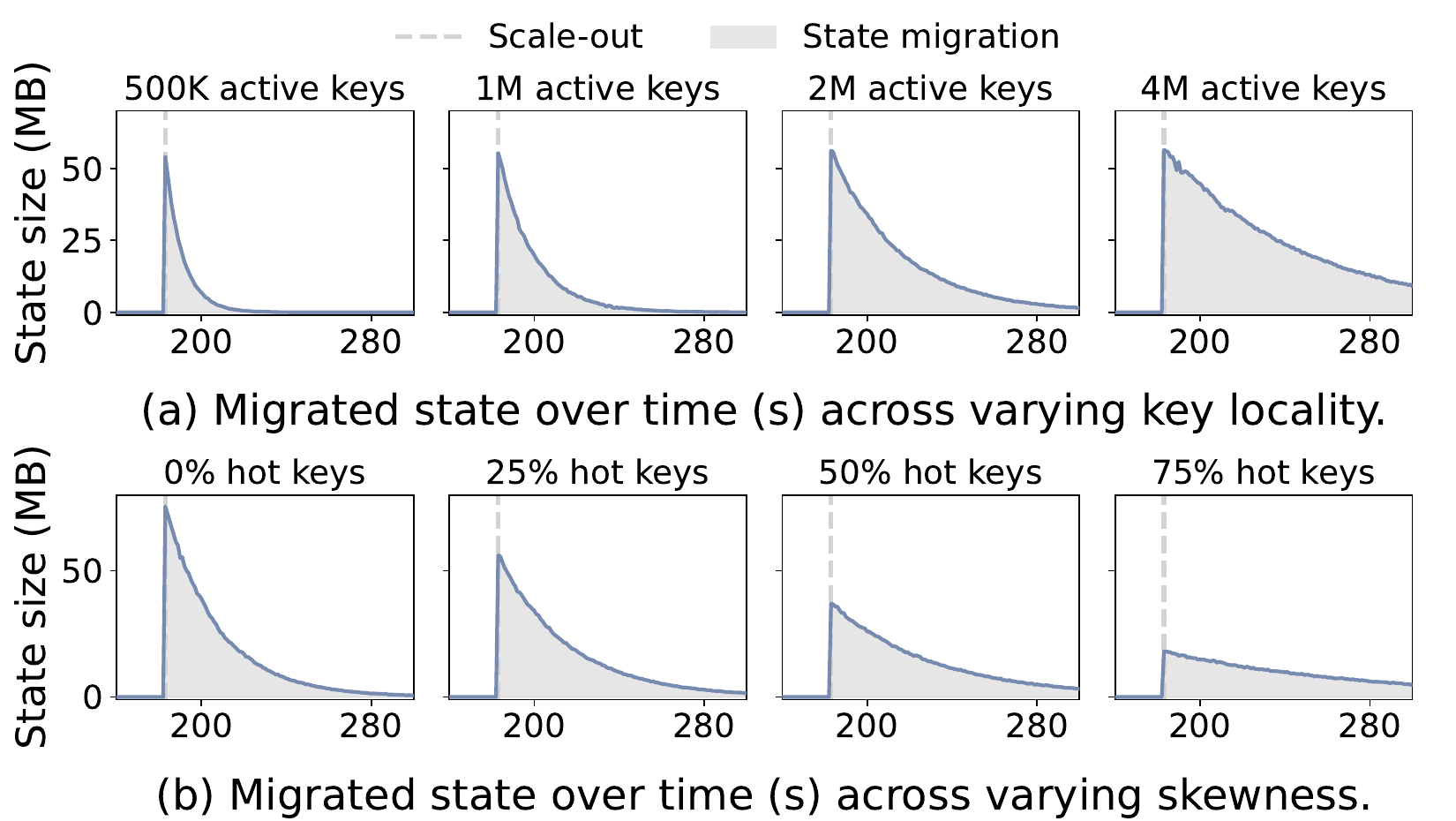}
    \caption{Effect of workload characteristics on \ours.}
    \vspace{-4mm}
    \label{fig:micro-data-distribution}
\end{figure}



\stitle{Workload characteristics.}
We evaluate the effect of (i) key temporal locality and (ii) skew on \ours. We fix the total key space at 20M keys with 40 GB total state size. To vary key locality, we configure the active key space at 500K, 1M, 2M, and 4M keys, with a fixed hot key ratio of 25\%---i.e., each input record has a 25\% probability of accessing a hot key randomly selected from a small predefined hot key set~\cite{beam-nexmark}. To vary skew, we fix the active key space at 2M keys and vary the hot key ratio across 0\%, 25\%, 50\%, and 75\%. We repeat the same $4 \rightarrow 8$ scale-out reconfiguration as above. We further evaluate the worst case for \ours---an active key space of all 20M keys with a hot key ratio of 0\%---in Appendix~\ref{appendix:uniform-access}.




As expected, Figure~\ref{fig:micro-data-distribution}a shows that migration duration depends on the size of the active key space, as more keys from the old tasks are accessed after reconfiguration, resulting in more state being migrated on demand.
Figure~\ref{fig:micro-data-distribution}b shows that higher skew leads to smoother reconfiguration, because (i) repeated accesses to hot keys incur no additional cost after initial fetch, and (ii) batch-based state access (\S\ref{sec:implementation}) deduplicates keys before issuing the migration request. 

\vspace{-1mm}

\section{Related work}\label{sec:related}
Since we discussed closely related work in \S\ref{sec:limitations}, here we only highlight important technical details. \ours shares ideas with prior work, such as fluid migration in Megaphone~\cite{megaphone}, fine-grained state transfer in Meces~\cite{Meces} and Spacker~\cite{spacker}, and record rerouting in DRRS~\cite{DRRS}. The key distinction, however, is that \ours migrates only active state, resulting in a significantly smaller migration footprint. Another important difference is that \ours never blocks record processing during reconfiguration. All prior protocols introduce at least partial blocking or buffering: Megaphone buffers records until frontiers advance, Meces, Spacker, and DRRS block processing until barrier alignment, and Falcon~\cite{Falcon} and Rhino~\cite{Rhino} block on markers for synchronization. \ours eliminates these barriers entirely. Unlike prior works, \ours's markers are not used for synchronization but only to signal that routing has changed and peer channels can be closed.

\ours primarily targets systems with long-running tasks, but it could also be integrated with dynamic scheduling systems~\cite{venkataraman2017drizzle,xu2021move,moritz2018ray,wang2021ownership}, as these face the same trade-off between keeping state remote or migrating it closer to compute.~Finally, \ours is policy-agnostic and complementary to works that focus on reconfiguration control~\cite{floratou2017dhalion,Mai2018Chi,bhardwaj2023cilantro,kalim2018henge,capsys,ds2}. These works determine when and how to trigger reconfigurations rather than how to execute them.

\section{Conclusion}
We presented \ours, a live reconfiguration protocol for distributed stateful dataflow systems that bridges the gap between the superior performance of embedded-state designs and the flexibility of remote-state approaches. \ours introduces a lazy state migration mechanism that decouples key partition assignment from state location, achieving truly non-disruptive fine-grained reconfiguration. We demonstrated \ours in a variety of use cases, including repeated rescaling reconfigurations, skew mitigation, and task migration.

\balance

\section{Acknowledgments}
We thank our shepherd and the anonymous reviewers for their thoughtful comments that substantially improved the paper, and the artifact evaluators for reproducing the results of this work.
This work was supported by the National Science Foundation under Grant No.~2440334.
This work used AWS through the CloudBank project~\cite{CloudBank}, which is supported by National Science Foundation grant \#1925001.
We also used computing resources provided by the CloudLab research testbed~\cite{Cloudlab}, which is supported by the National Science Foundation.

\bibliographystyle{plain}
\bibliography{reference}

@article{carbone2015apache,
  title={Apache flink: Stream and batch processing in a single engine},
  author={Carbone, Paris and Katsifodimos, Asterios and Ewen, Stephan and Markl, Volker and Haridi, Seif and Tzoumas, Kostas},
  journal={The Bulletin of the Technical Committee on Data Engineering},
  volume={38},
  number={4},
  year={2015},
  publisher={Institute of Electrical and Electronics Engineers (IEEE)}
}

@article{wang2023fries,
  title={Fries: fast and consistent runtime reconfiguration in dataflow systems with transactional guarantees},
  author={Wang, Zuozhi and Ni, Shengquan},
  journal={Proceedings of the VLDB Endowment},
  year={2023}
}

@article{flink2,
  author       = {Yuan Mei and
                  Zhaoqian Lan and
                  Lei Huang and
                  Yanfei Lei and
                  Han Yin and
                  Rui Xia and
                  Kaitian Hu and
                  Paris Carbone and
                  Vasiliki Kalavri and
                  Feng Wang},
  title        = {Disaggregated State Management in Apache Flink 2.0},
  journal      = {Proc. {VLDB} Endow.},
  volume       = {18},
  number       = {12},
  pages        = {4846--4859},
  year         = {2025},
  url          = {https://www.vldb.org/pvldb/vol18/p4846-mei.pdf},
  doi          = {10.14778/3750601.3750609},
  bibsource    = {dblp computer science bibliography, https://dblp.org}
}

@article{state-survey,
author = {To, Quoc-Cuong and Soto, Juan and Markl, Volker},
title = {A survey of state management in big data processing systems},
year = {2018},
issue_date = {December  2018},
publisher = {Springer-Verlag},
address = {Berlin, Heidelberg},
volume = {27},
number = {6},
issn = {1066-8888},
url = {https://doi.org/10.1007/s00778-018-0514-9},
doi = {10.1007/s00778-018-0514-9},
journal = {The VLDB Journal},
month = dec,
pages = {847–872},
numpages = {26}
}

@misc{Tikv,
  author = {TiKV},
  title  = {TiKV: A Distributed Transactional Key-Value Database},
  year   = {2024},
  howpublished    = {\url{https://tikv.org/}},
  urldate= {2024-06-01}
}

@misc{nexmark,
  title={NEXMark - A Benchmark for Queries over Data Streams [Technical Report]},
  author={Tucker, Pete and Tufte, Kristin and Papadimos, Vassilis and Maier, David},
  year={2007},
  publisher = {OGI School of Science & Engineering at OHSU},
}

@inproceedings{rappaz2021recommendation,
  title={Recommendation on live-streaming platforms: Dynamic availability and repeat consumption},
  author={Rappaz, J{\'e}r{\'e}mie and McAuley, Julian and Aberer, Karl},
  booktitle={Proceedings of the 15th ACM Conference on Recommender Systems},
  pages={390--399},
  year={2021}
}

@misc{GoogleClusterData,
  author = {Google},
  title  = {Google Cluster Data},
  year   = {2025},
  howpublished    = {\url{https://github.com/google/cluster-data}},
}

@misc{DEBS2015,
  author = {ACM DEBS 2015},
  title  = {DEBS2015 Grand Challenge},
  year   = {2015},
  howpublished    = {\url{http://www.debs2015.org/call-grand-challenge.html}},
}

@misc{newyorktaxi,
  author = {NYC Taxi \& Limousine Commission},
  title  = {NYC Taxi \& Limousine Commission - Trip Record Data},
  year   = {2025},
  howpublished    = {\url{https://www.nyc.gov/site/tlc/about/tlc-trip-record-data.page}},
}

@misc{AzureVMData,
  author = {Microsoft Azure},
  title  = {Azure Public Dataset},
  year   = {2025},
  howpublished    = {\url{https://github.com/Azure/AzurePublicDataset}},
  note   = {Last access: April 2026},
}

@inproceedings{sparkstructuredstreaming,
author = {Armbrust, Michael and Das, Tathagata and Torres, Joseph and Yavuz, Burak and Zhu, Shixiong and Xin, Reynold and Ghodsi, Ali and Stoica, Ion and Zaharia, Matei},
title = {Structured Streaming: A Declarative API for Real-Time Applications in Apache Spark},
year = {2018},
isbn = {9781450347037},
publisher = {Association for Computing Machinery},
address = {New York, NY, USA},
url = {https://doi.org/10.1145/3183713.3190664},
doi = {10.1145/3183713.3190664},
booktitle = {Proceedings of the 2018 International Conference on Management of Data},
pages = {601–613},
numpages = {13},
series = {SIGMOD '18}
}

@article{megaphone,
author = {Hoffmann, Moritz and Lattuada, Andrea and McSherry, Frank and Kalavri, Vasiliki and Liagouris, John and Roscoe, Timothy},
title = {Megaphone: latency-conscious state migration for distributed streaming dataflows},
year = {2019},
issue_date = {May 2019},
publisher = {VLDB Endowment},
volume = {12},
number = {9},
issn = {2150-8097},
url = {https://doi.org/10.14778/3329772.3329777},
doi = {10.14778/3329772.3329777},
journal = {Proc. VLDB Endow.},
month = may,
pages = {1002–1015},
numpages = {14}
}

@inproceedings{Rhino,
author = {Del Monte, Bonaventura and Zeuch, Steffen and Rabl, Tilmann and Markl, Volker},
title = {Rhino: Efficient Management of Very Large Distributed State for Stream Processing Engines},
year = {2020},
isbn = {9781450367356},
publisher = {Association for Computing Machinery},
address = {New York, NY, USA},
url = {https://doi.org/10.1145/3318464.3389723},
doi = {10.1145/3318464.3389723},
booktitle = {Proceedings of the 2020 ACM SIGMOD International Conference on Management of Data},
pages = {2471–2486},
numpages = {16},
location = {Portland, OR, USA},
series = {SIGMOD '20}
}

@inproceedings {Meces,
author = {Rong Gu and Han Yin and Weichang Zhong and Chunfeng Yuan and Yihua Huang},
title = {Meces: Latency-efficient Rescaling via Prioritized State Migration for Stateful Distributed Stream Processing Systems},
booktitle = {2022 USENIX Annual Technical Conference (USENIX ATC 22)},
year = {2022},
isbn = {978-1-939133-29-13},
address = {Carlsbad, CA},
pages = {539--556},
url = {https://www.usenix.org/conference/atc22/presentation/gu-rong},
publisher = {USENIX Association},
month = jul
}

@INPROCEEDINGS{Falcon,
  author={Mishra, Pritish and Bore, Nelson and Ramprasad, Brian and Thiessen, Myles and Gabel, Moshe and Da Silva Veith, Alexandre and Balmau, Oana and De Lara, Eyale},
  booktitle={2024 IEEE/ACM Symposium on Edge Computing (SEC)}, 
  title={Falcon: Live Reconfiguration for Stateful Stream Processing on the Edge}, 
  year={2024},
  volume={},
  number={},
  pages={234-248},
  doi={10.1109/SEC62691.2024.00026}}

@INPROCEEDINGS{DRRS,
  author={Qing, Yunfan and Zheng, Wenli},
  booktitle={2025 IEEE 41st International Conference on Data Engineering (ICDE)}, 
  title={Towards Fine-Grained Scalability for Stateful Stream Processing Systems}, 
  year={2025},
  volume={},
  number={},
  pages={3835-3848},
  doi={10.1109/ICDE65448.2025.00286}}

@article{MillWheel,
author = {Akidau, Tyler and Balikov, Alex and Bekiro\u{g}lu, Kaya and Chernyak, Slava and Haberman, Josh and Lax, Reuven and McVeety, Sam and Mills, Daniel and Nordstrom, Paul and Whittle, Sam},
title = {MillWheel: fault-tolerant stream processing at internet scale},
year = {2013},
issue_date = {August 2013},
publisher = {VLDB Endowment},
volume = {6},
number = {11},
issn = {2150-8097},
url = {https://doi.org/10.14778/2536222.2536229},
doi = {10.14778/2536222.2536229},
journal = {Proc. VLDB Endow.},
month = aug,
pages = {1033–1044},
numpages = {12}
}

@inproceedings {Sponge,
author = {Won Wook Song and Taegeon Um and Sameh Elnikety and Myeongjae Jeon and Byung-Gon Chun},
title = {Sponge: Fast Reactive Scaling for Stream Processing with Serverless Frameworks},
booktitle = {2023 USENIX Annual Technical Conference (USENIX ATC 23)},
year = {2023},
isbn = {978-1-939133-35-9},
address = {Boston, MA},
pages = {301--314},
url = {https://www.usenix.org/conference/atc23/presentation/song},
publisher = {USENIX Association},
month = jul
}

@article{akidau2021watermarks,
  title={Watermarks in stream processing systems: semantics and comparative analysis of Apache Flink and Google cloud dataflow},
  author={Akidau, Tyler and Begoli, Edmon and Chernyak, Slava and Hueske, Fabian and Knight, Kathryn and Knowles, Kenneth and Mills, Daniel and Sotolongo, Dan},
  journal={Proceedings of the VLDB Endowment},
  volume={14},
  number={12},
  pages={3135--3147},
  year={2021},
  publisher={VLDB Endowment}
}

@article{akidau2015dataflow,
  title={The dataflow model: a practical approach to balancing correctness, latency, and cost in massive-scale, unbounded, out-of-order data processing},
  author={Akidau, Tyler and Bradshaw, Robert and Chambers, Craig and Chernyak, Slava and Fern{\'a}ndez-Moctezuma, Rafael J and Lax, Reuven and McVeety, Sam and Mills, Daniel and Perry, Frances and Schmidt, Eric and others},
  journal={Proceedings of the VLDB Endowment},
  volume={8},
  number={12},
  pages={1792--1803},
  year={2015},
  publisher={VLDB Endowment}
}

@article{streamingsurvey_vldb24,
author = {Fragkoulis, Marios and Carbone, Paris and Kalavri, Vasiliki and Katsifodimos, Asterios},
title = {A survey on the evolution of stream processing systems},
year = {2023},
issue_date = {Mar 2024},
publisher = {Springer-Verlag},
address = {Berlin, Heidelberg},
volume = {33},
number = {2},
issn = {1066-8888},
url = {https://doi.org/10.1007/s00778-023-00819-8},
doi = {10.1007/s00778-023-00819-8},
journal = {The VLDB Journal},
month = nov,
pages = {507–541},
numpages = {35}
}

@inproceedings {ds2,
author = {Vasiliki Kalavri and John Liagouris and Moritz Hoffmann and Desislava Dimitrova and Matthew Forshaw and Timothy Roscoe},
title = {Three steps is all you need: fast, accurate, automatic scaling decisions for distributed streaming dataflows},
booktitle = {13th USENIX Symposium on Operating Systems Design and Implementation (OSDI 18)},
year = {2018},
isbn = {978-1-939133-08-3},
address = {Carlsbad, CA},
pages = {783--798},
url = {https://www.usenix.org/conference/osdi18/presentation/kalavri},
publisher = {USENIX Association},
month = oct
}

@inproceedings{capsys,
author = {Wang, Yuanli and Huang, Lei and Wang, Zikun and Kalavri, Vasiliki and Matta, Ibrahim},
title = {CAPSys: Contention-aware task placement for data stream processing},
year = {2025},
isbn = {9798400711961},
publisher = {Association for Computing Machinery},
address = {New York, NY, USA},
url = {https://doi.org/10.1145/3689031.3696085},
doi = {10.1145/3689031.3696085},
booktitle = {Proceedings of the Twentieth European Conference on Computer Systems},
pages = {654–670},
numpages = {17},
location = {Rotterdam, Netherlands},
series = {EuroSys '25}
}

@INPROCEEDINGS{spacker,
  author={Mao, Yancan and Zhang, Shuhao and Ma, Richard T. B.},
  booktitle={2025 IEEE 45th International Conference on Distributed Computing Systems (ICDCS)}, 
  title={Spacker: Unified State Migration for Distributed Streaming}, 
  year={2025},
  volume={},
  number={},
  pages={1000-1010},
  doi={10.1109/ICDCS63083.2025.00101}}

@inproceedings{kalavri2020support,
  title={In support of workload-aware streaming state management},
  author={Kalavri, Vasiliki and Liagouris, John},
  booktitle={12th USENIX Workshop on Hot Topics in Storage and File Systems (HotStorage 20)},
  year={2020}
}

@article{gulisano2012streamcloud,
author = {Gulisano, Vincenzo and Jim{\'{e}}nez-Peris, Ricardo and Pati{\~{n}}o-Mart{\'{i}}nez, Marta and Soriente, Claudio and Valduriez, Patrick},
doi = {10.1109/TPDS.2012.24},
issn = {10459219},
journal = {IEEE Transactions on Parallel and Distributed Systems},
title = {{StreamCloud: An elastic and scalable data streaming system}},
volume={23},
number={12},
pages={15},
year = {2012}
}

@article{Mai2018Chi,
  title={Chi: A scalable and programmable control plane for distributed stream processing systems},
  author={Mai, Luo and Zeng, Kai and Potharaju, Rahul and Xu, Le and Suh, Steve and Venkataraman, Shivaram and Costa, Paolo and Kim, Terry and Muthukrishnan, Saravanan and Kuppa, Vamsi and others},
  journal={Proceedings of the VLDB Endowment},
  volume={11},
  number={10},
  pages={1303--1316},
  year={2018},
  publisher={VLDB Endowment}
}

@inproceedings{castro2013integrating,
author = {Castro Fernandez, Raul and Migliavacca, Matteo and Kalyvianaki, Evangelia and Pietzuch, Peter},
title = {Integrating scale out and fault tolerance in stream processing using operator state management},
year = {2013},
isbn = {9781450320375},
publisher = {Association for Computing Machinery},
address = {New York, NY, USA},
url = {https://doi.org/10.1145/2463676.2465282},
doi = {10.1145/2463676.2465282},
numpages = {12},
location = {New York, New York, USA},
series = {SIGMOD '13},
booktitle    = {Proceedings of the {ACM} {SIGMOD} International Conference on Management of Data, {SIGMOD} 2013},
  pages        = {725--736},
}

@article{gedik2014elastic,
author = {Gedik, Buǧra and Schneider, Scott and Hirzel, Martin and Wu, Kun Lung},
doi = {10.1109/TPDS.2013.295},
isbn = {9781450327374},
issn = {10459219},
journal = {IEEE Transactions on Parallel and Distributed Systems},
title = {{Elastic scaling for data stream processing}},
volume={25},
number={6},
pages={17},
year = {2014}
}

@inproceedings{zhu2025impeller,
  title={Impeller: Stream Processing on Shared Logs},
  author={Zhu, Zhiting and Jia, Zhipeng and Ni, Newton and Tang, Dixin and Witchel, Emmett},
  booktitle={Proceedings of the Twentieth European Conference on Computer Systems},
  pages={637--653},
  year={2025}
}

@inproceedings{venkataraman2017drizzle,
  title={Drizzle: Fast and adaptable stream processing at scale},
  author={Venkataraman, Shivaram and Panda, Aurojit and Ousterhout, Kay and Armbrust, Michael and Ghodsi, Ali and Franklin, Michael J and Recht, Benjamin and Stoica, Ion},
  booktitle={Proceedings of the 26th Symposium on Operating Systems Principles},
  pages={374--389},
  year={2017}
}

@inproceedings{murray2013naiad,
  title={Naiad: a timely dataflow system},
  author={Murray, Derek G and McSherry, Frank and Isaacs, Rebecca and Isard, Michael and Barham, Paul and Abadi, Mart{\'\i}n},
  booktitle={Proceedings of the Twenty-Fourth ACM Symposium on Operating Systems Principles},
  pages={439--455},
  year={2013}
}

@inproceedings{xu2021move,
  title={Move fast and meet deadlines: Fine-grained real-time stream processing with cameo},
  author={Xu, Le and Venkataraman, Shivaram and Gupta, Indranil and Mai, Luo and Potharaju, Rahul},
  booktitle={18th USENIX Symposium on Networked Systems Design and Implementation (NSDI 21)},
  pages={389--405},
  year={2021}
}

@inproceedings{barber2021bladerunner,
  title={Bladerunner: Stream processing at scale for a live view of backend data mutations at the edge},
  author={Barber, Jeff and Yu, Ximing and Zamore, Laney Kuenzel and Lin, Jerry and Jazayeri, Vahid and Erlich, Shie and Savor, Tony and Stumm, Michael},
  booktitle={Proceedings of the ACM SIGOPS 28th Symposium on Operating Systems Principles},
  pages={708--723},
  year={2021}
}

@inproceedings{lin2016streamscope,
  title={$\{$StreamScope$\}$: Continuous Reliable Distributed Processing of Big Data Streams},
  author={Lin, Wei and Qian, Zhengping and Xu, Junwei and Yang, Sen and Zhou, Jingren and Zhou, Lidong},
  booktitle={13th USENIX Symposium on Networked Systems Design and Implementation (NSDI 16)},
  pages={439--453},
  year={2016}
}

@misc{pebble,
  author = {{Cockroach Labs}},
  title = {Pebble: RocksDB-Compatible Key-Value Store},
  year = {2020},
  howpublished = {\url{https://www.cockroachlabs.com/blog/pebble-rocksdb-kv-store/}},
  urldate = {2026-04-13}
}

@article{dynamodb,
author = {DeCandia, Giuseppe and Hastorun, Deniz and Jampani, Madan and Kakulapati, Gunavardhan and Lakshman, Avinash and Pilchin, Alex and Sivasubramanian, Swaminathan and Vosshall, Peter and Vogels, Werner},
title = {Dynamo: amazon's highly available key-value store},
year = {2007},
issue_date = {December 2007},
publisher = {Association for Computing Machinery},
address = {New York, NY, USA},
volume = {41},
number = {6},
issn = {0163-5980},
url = {https://doi.org/10.1145/1323293.1294281},
doi = {10.1145/1323293.1294281},
journal = {SIGOPS Oper. Syst. Rev.},
month = oct,
pages = {205–220},
numpages = {16}
}

@inproceedings{wang2021ownership,
  title={Ownership: A distributed futures system for $\{$Fine-Grained$\}$ tasks},
  author={Wang, Stephanie and Liang, Eric and Oakes, Edward and Hindman, Ben and Luan, Frank Sifei and Cheng, Audrey and Stoica, Ion},
  booktitle={18th USENIX Symposium on Networked Systems Design and Implementation (NSDI 21)},
  pages={671--686},
  year={2021}
}

@inproceedings{moritz2018ray,
  title={Ray: A distributed framework for emerging $\{$AI$\}$ applications},
  author={Moritz, Philipp and Nishihara, Robert and Wang, Stephanie and Tumanov, Alexey and Liaw, Richard and Liang, Eric and Elibol, Melih and Yang, Zongheng and others},
  booktitle={13th USENIX symposium on operating systems design and implementation (OSDI 18)},
  pages={561--577},
  year={2018}
}

@inproceedings{mei2020turbine,
  title={Turbine: Facebook’s service management platform for stream processing},
  author={Mei, Yuan and Cheng, Luwei and Talwar, Vanish and Levin, Michael Y and Jacques-Silva, Gabriela and Simha, Nikhil and Banerjee, Anirban and Smith, Brian and Williamson, Tim and Yilmaz, Serhat and others},
  booktitle={2020 IEEE 36th International Conference on Data Engineering (ICDE)},
  pages={1591--1602},
  year={2020},
  organization={IEEE}
}

@article{carbone2017state,
  title={State management in Apache Flink{\textregistered}: consistent stateful distributed stream processing},
  author={Carbone, Paris and Ewen, Stephan and F{\'o}ra, Gyula and Haridi, Seif and Richter, Stefan and Tzoumas, Kostas},
  journal={Proceedings of the VLDB Endowment},
  volume={10},
  number={12},
  pages={1718--1729},
  year={2017},
  publisher={VLDB Endowment}
}

@article{chandy1985distributed,
  title={Distributed snapshots: Determining global states of distributed systems},
  author={Chandy, K Mani and Lamport, Leslie},
  journal={ACM Transactions on Computer Systems (TOCS)},
  volume={3},
  number={1},
  pages={63--75},
  year={1985},
  publisher={ACM New York, NY, USA}
}

@misc{rappaz2021twitchdataset,
  author       = {J{\'e}r{\'e}mie Rappaz and Julian McAuley and Karl Aberer},
  title        = {Twitch Interaction Dataset for ``Recommendation on Live-Streaming Platforms: Dynamic Availability and Repeat Consumption''},
  year         = {2021},
  howpublished = {\url{https://drive.google.com/drive/folders/1BD8m7a8m7onaifZay05yYjaLxyVV40si?usp=sharing}},
  note         = {Dataset released by the authors via Google Drive}
}

@inproceedings{asyabi2022new,
  title={A new benchmark harness for systematic and robust evaluation of streaming state stores},
  author={Asyabi, Esmail and Wang, Yuanli and Liagouris, John and Kalavri, Vasiliki and Bestavros, Azer},
  booktitle={Proceedings of the Seventeenth European Conference on Computer Systems},
  pages={559--574},
  year={2022}
}

@misc{drrs_github,
  title        = {DRRS codebase},
  author={Yunfan Qing and Wenli Zheng},
  year={2025},
  howpublished = {\url{https://github.com/floudk/DRRS}},
}

@inproceedings{Cloudlab,
    title     = "The Design and Operation of {CloudLab}",
    author    = "Dmitry Duplyakin and Robert Ricci and Aleksander Maricq and Gary Wong and Jonathon Duerig and Eric Eide and Leigh Stoller and Mike Hibler and David Johnson and Kirk Webb and Aditya Akella and Kuangching Wang and Glenn Ricart and Larry Landweber and Chip Elliott and Michael Zink and Emmanuel Cecchet and Snigdhaswin Kar and Prabodh Mishra",
    booktitle = "Proceedings of the {USENIX} Annual Technical Conference (ATC)",
    pages     = "1--14",
    year      = 2019,
    month     = jul,
    url       = "https://www.flux.utah.edu/paper/duplyakin-atc19"
}

@inproceedings{CloudBank,
  title={{CloudBank}: Managed Services to Simplify Cloud Access for Computer Science Research and Education},
  author={Norman, Michael and Kellen, Vince and Smallen, Shava and DeMeulle, Brian and Strande, Shawn and Lazowska, Ed and Alterman, Naomi and Fatland, Rob and Stone, Sarah and Tan, Amanda and Yelick, Katherine and Van Dusen, Eric and Mitchell, James},
  booktitle={Practice and Experience in Advanced Research Computing (PEARC '21)},
  publisher={Association for Computing Machinery},
  address={New York, NY, USA},
  articleno={45},
  pages={1--4},
  year={2021},
  doi={10.1145/3437359.3465586}
}

@misc{beam-nexmark,
  title        = {Apache Beam {NEXMark} Benchmark Suite},
  author       = {{Apache Beam}},
  howpublished = {\url{https://beam.apache.org/documentation/sdks/java/testing/nexmark/}},
  note         = {Last access: April 2026},
}

@inproceedings{kalim2018henge,
  title={Henge: Intent-driven multi-tenant stream processing},
  author={Kalim, Faria and Xu, Le and Bathey, Sharanya and Meherwal, Richa and Gupta, Indranil},
  booktitle={Proceedings of the ACM Symposium on Cloud Computing},
  pages={249--262},
  year={2018}
}

@inproceedings{bhardwaj2023cilantro,
  title={Cilantro:$\{$Performance-Aware$\}$ resource allocation for general objectives via online feedback},
  author={Bhardwaj, Romil and Kandasamy, Kirthevasan and Biswal, Asim and Guo, Wenshuo and Hindman, Benjamin and Gonzalez, Joseph and Jordan, Michael and Stoica, Ion},
  booktitle={17th USENIX Symposium on Operating Systems Design and Implementation (OSDI 23)},
  pages={623--643},
  year={2023}
}

@article{floratou2017dhalion,
  title={Dhalion: Self-Regulating Stream Processing in Heron.},
  author={Floratou, Avrilia and Agrawal, Ashvin and Graham, Bill and Rao, Sriram and Ramasamy, Karthik},
  journal={Proc. VLDB Endow.},
  volume={10},
  number={12},
  pages={1825--1836},
  year={2017}
}

@inproceedings{mao2021trisk,
  title={Trisk: Task-centric data stream reconfiguration},
  author={Mao, Yancan and Huang, Yuan and Tian, Runxin and Wang, Xin and Ma, Richard TB},
  booktitle={Proceedings of the ACM Symposium on Cloud Computing},
  pages={214--228},
  year={2021}
}

\clearpage
\appendix
\section*{Appendix}

\section{Protocol analysis}\label{appendix:analysis}

\begin{definition}[Deployment]
\label{def:deployment}
A \emph{deployment} $\mathcal{D}$ of a dataflow DAG $G = (V,  E)$ is a tuple $\langle \Pi, M \rangle$, where $\Pi = \{\pi_v\}_{v \in V}$ assigns each operator $v$ a key partitioning $\pi_v = \{K_1^v, \ldots, K_{p_v}^v\}$, and $M\colon \bigcup_v \mathcal{T}_v \to \mathcal{W}$ maps each task $\tau_i^v \in \mathcal{T}_v$ to a physical worker in the set of available workers $\mathcal{W}$.
\end{definition}

A \emph{reconfiguration} is a transition from a running deployment $\mathcal{D}$ to a new deployment $\mathcal{D}'$. When the reconfiguration affects a stateful operator, the key assignment may change to reflect the new partitioning. We refer to the stateful operator whose deployment changes as the \emph{target operator}.

\begin{definition}[Reconfiguration]
\label{def:reconfiguration}
Given a running dataflow with deployment $\mathcal{D} = \langle \Pi, M \rangle$ and a target operator $v^*$, a \emph{reconfiguration} $\mathcal{R}(\mathcal{D},\, v^*) \to \mathcal{D}'$ transitions the system to a new deployment $\mathcal{D}' = \langle \Pi', M' \rangle$ that differs from $\mathcal{D}$ in the partitioning $\pi_{v^*}$, the worker mapping $M$, or both.
\end{definition}

\subsection{Correctness properties}\label{sec:problem}
We define a reconfiguration as \emph{correct} if it satisfies the following three properties.

\begin{property}[Per-key read/write isolation]
\label{prop:srw}
At any point in time, every key $k \in \mathcal{K}$ is assigned to exactly one task. 
No two tasks concurrently hold responsibility for the same key, and no two tasks 
concurrently read or write the state associated with the same key.
Formally, for any two distinct tasks $\tau_i^{v^*}$ and $\tau_j^{v^*}$ of the 
target operator, $K_i^{v^*} \cap K_j^{v^*} = \emptyset$ holds at all times, 
including during the transition from $\mathcal{D}$ to $\mathcal{D}'$.
\end{property}

\begin{property}[Exactly-once processing]
\label{prop:eo}
Every record that arrives at the target operator is processed exactly once: no record is dropped, and no record is processed more than once as a result of the reconfiguration.
\end{property}

\begin{property}[Watermark correctness]
\label{prop:wm}
For any watermark $W(t_w)$ propagated through the dataflow, two conditions must 
hold throughout the reconfiguration and under the new deployment $\mathcal{D}'$. 
First, watermarks are \emph{monotonically non-decreasing}: for any two watermarks 
$W(t_w)$ and $W(t_w')$ observed at an operator in that order, $t_w \leq t_w'$. 
Second, watermarks preserve  \emph{event-time semantics}: no record with event 
timestamp $t \leq t_w$ is delivered to any task after $W(t_w)$ has been 
propagated, including records that were in flight at the time of the 
reconfiguration.  
\end{property}

The above properties ensure that the output produced during and after reconfiguration is equivalent to the output that would have been produced had the application been deployed with the final configuration from the start.

\subsection{Proof sketches}\label{subsec-semantics}

In this section, we show that \ours preserves the three correctness properties from \S\ref{sec:problem}. 

\stitle{Property 1: Single-owner guarantee per key.}
The dataflow model~\cite{akidau2015dataflow, streamingsurvey_vldb24} requires that each key is processed by exactly one task at any point in time, ensuring consistent state access.
We show that at all times, including during the transition from $\mathcal{D}$ to $\mathcal{D}'$, every key $k \in \mathcal{K}$ is assigned to exactly one task of the target operator.

\begin{proof} (Sketch)
Recall that all tasks of the target operator $v$ are assigned disjoint partitions of the key space and the scheme $f$ generated by the coordinator respects this property; that is, $K_i^{v} \cap K_j^{v} = \emptyset~\forall i\neq j$ holds under both $\mathcal{D}$ and $\mathcal{D}'$, for all pairs of tasks $i,j$ of the target operator $v$. It remains to show that the property is preserved \emph{during} the transition.
Upon receiving a reconfiguration command, a sender task (i) completes processing the current input batch (if any), including pending updates to the state backend and KLT entries, (ii) transfers the reassigned KLT entries to the receiver over the peer channel, and (iii) redirects all subsequent records for those keys to the receiver. After step (ii), the sender can no longer access the transferred state; if the state still resides physically at the sender, the receiver must pull it locally before any update. Since steps (i) and (ii) are executed atomically (in practice, by the sender's main goroutine), no other task can acquire ownership between the hand-off and the forwarding.
In addition, the RT updates at upstream tasks are applied after steps (i)--(iii), ensuring that records for the reassigned keys are routed to the receiver only after the ownership transfer.
Therefore, there is no interval during which two distinct tasks concurrently hold responsibility for the same key. Hence the single writer per key invariant is maintained throughout the reconfiguration.
\end{proof}



\stitle{Property 2: Exactly-once processing.}
To guarantee exactly-once processing, we need to ensure that no records are missed and no duplicates are introduced due to the reconfiguration. 
\begin{proof} (Sketch)
We show that (i) no record arriving at the target operator is dropped and (ii) no record is processed more than once. Like all other protocols, \ours does not consider failures during reconfiguration; failures are handled by an external checkpointing mechanism that rolls back to the latest checkpoint and restarts the protocol  (\S\ref{sec:implementation}).

\emph{No drops.} A sender closes the peer channel only after receiving in-flight markers from all upstream tasks, confirming that no records for the reassigned keys remain in transit. Moreover, the RT updates at upstream tasks are applied only after all fast-forward processes have started and the receiver has obtained the reassigned KLT entries. Together, these two conditions ensure that every record for a reassigned key is either forwarded via the peer channel or delivered directly to the new owner following the RT update; no record can fall between the two delivery paths.

\emph{No duplicates.} A duplicate would require the same record to be delivered on both an upstream channel and a peer channel. This is ruled out because RT updates are applied atomically (in practice, using a mutex): once an upstream task's RT designates the receiver as the new owner, records for the reassigned keys are routed exclusively over the new path, so no record can appear on both routes simultaneously.

Combining the two arguments, every record arriving at the target operator is processed exactly once.
\end{proof}

\stitle{Property 3: Watermark correctness.}
Epoch-based event-time semantics requires that time-dependent operations (e.g., timers, windows) respect watermark boundaries (\S\ref{sec:background}): \emph{when a task advances its event time to $t$, all records destined for that task with timestamps $\leq t$ must have already been received}. 

Algorithm~\ref{alg:watermark} specifies how a receiver task advances its event time during reconfiguration in \ours. Upon receiving a watermark $w$ on input channel $c$ (upstream or peer channel), the task updates $c$'s recorded event time (line~\ref{line:update}), and then checks whether all expected in-flight markers have arrived (line~\ref{line:ready}). The expected count is determined by the \texttt{upstreams} parameter derived from the ownership scheme $f$.
Once all reconfiguration markers are present, the task computes the minimum event time across all input channels (lines~\ref{line:loop}--\ref{line:endloop}), skipping the progress update (line~\ref{line:noinit}) if any channel has not yet produced a watermark, denoted with $\perp$. If the minimum exceeds the task's current event time, the task advances to the new minimum (lines~\ref{line:advance}--\ref{line:advanceend}).

\begin{proof} (Sketch)
We verify both conditions: (i) the sequence of watermarks observed at any operator is monotonically non-decreasing, and (ii) no record with event timestamp $t \leq t_w$ is delivered to any task after $W(t_w)$ has been propagated. The argument rests on two invariants maintained by Algorithm~\ref{alg:watermark}.

\emph{Invariant 1 (Channel completeness).} The marker readiness check (line~\ref{line:ready}) ensures the task withholds any event-time advance until all expected in-flight markers have arrived. Receiving all markers implies that every upstream channel is established, since markers are injected only after peer channels are created. Hence the task has a complete view of all active input channels before any advance.

\emph{Invariant 2 (Minimum-time bound).} Once channel completeness is satisfied, the task advances its event time only to $\min_{c'} c'.\textit{event\_time}$ taken over all active input channels (lines~\ref{line:loop}--\ref{line:endloop}), provided every channel has produced at least one watermark (line~\ref{line:noinit}). Consequently, the task's event time is always upper bounded by the minimum event time of any input channel.

\emph{Condition (i)} follows from line~\ref{line:advance}: the task updates $\textit{task.event\_time}$ only when $\textit{min\_time} > \textit{task.event\_time}$, so the sequence of event-time values is non-decreasing by construction.

\emph{Condition (ii)} follows from Invariant 2: since the task's event time never exceeds the minimum over all input channels, no record with timestamp $t \leq t_w$ can remain in transit when the task advances to $t_w$. Hence no time-dependent operation fires before all relevant records have been delivered.

Finally, when a peer channel closes, the minimum is recomputed over the remaining channels (line~\ref{line:loop} with the updated $\textit{input\_channels}$), potentially allowing the receiver task to advance its event time if the closed channel was the one that prevented it from advancing. This recomputation is safe because the sender flushes all in-flight records before closing the channel; excluding it from the minimum cannot cause the task's event time to advance past any undelivered record's timestamp.
\end{proof}


\begin{algorithm}[t]
  \caption{Receiver task event-time advancement during reconfiguration.}
  \label{alg:watermark}
  \begin{algorithmic}[1]
  	\State // \emph{Input channel} $c \in \textit{upstream}~\cup~\textit{peer channels}$
    \Procedure{OnWatermark}{$w$, input channel $c$}\label{line:inputchan}
    \State $c.\textit{event\_time} \gets w.\textit{time}$ \label{line:update}
    \If{$|\textit{received\_markers}| < |\textit{expected\_upstreams}|$} \label{line:ready}
      \State \Return 
    \EndIf
    \State $\textit{min\_time} \gets \infty$
    \For{\textbf{each} channel $c'$ \textbf{in} $\textit{input\_channels}$} \label{line:loop}
      \If{$c'.\textit{event\_time} = \bot$} \label{line:noinit}
        \State \Return 
      \EndIf
      \State $\textit{min\_time} \gets \min(\textit{min\_time},\; c'.\textit{event\_time})$
    \EndFor \label{line:endloop}
    \If{$\textit{min\_time} > \textit{task.event\_time}$} \label{line:advance}
      \State $\textit{task.event\_time} \gets \textit{min\_time}$
      \State \Return 
    \EndIf\label{line:advanceend}
    \State \Return 
    \EndProcedure
  \end{algorithmic}
\end{algorithm}



\section{Comparison with DRRS on Flink}\label{sec:appendix-drrs}
We compare our implementation of DRRS (DRRS-ours) with the original DRRS implementation (DRRS-original) on Apache Flink~\cite{drrs_github} open-sourced by the DRRS authors~\cite{DRRS}. We run the \textbf{Twitch} query for evaluation, as it is also evaluated in DRRS-original with a Flink query implementation. We modify our \textbf{Twitch} implementation to include the retention history state field used by DRRS-original to accumulate state.

We run experiments on \textbf{E2-CloudLab}. We deploy DRRS-original with the \textit{even} task placement policy on Flink to balance the workload among available resources. We use the default configuration parameters of DRRS-original for both approaches, such as migration buffer size, key groups, and the number of subscales. Each experiment runs for 10 minutes, with reconfiguration triggered at 3 minutes, scaling out the target operator (ViewerLoyaltyCalculator) from 4 to 8 tasks.

\begin{figure}
\centering
\begin{subfigure}{\linewidth}
\centering
\includegraphics[width=\linewidth]{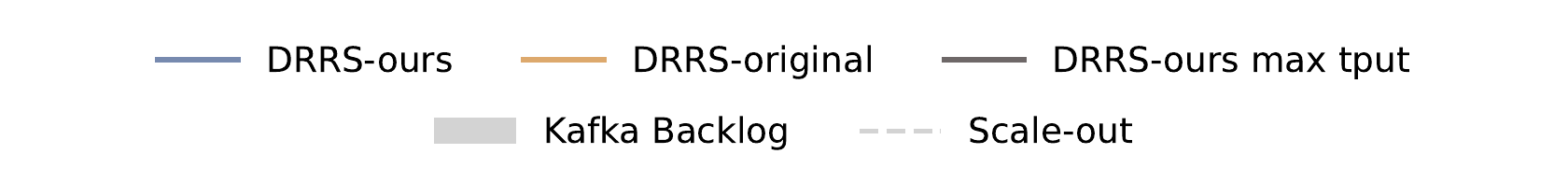}
\end{subfigure}
\hfill
\begin{subfigure}{\linewidth}
\centering
\includegraphics[width=\linewidth]{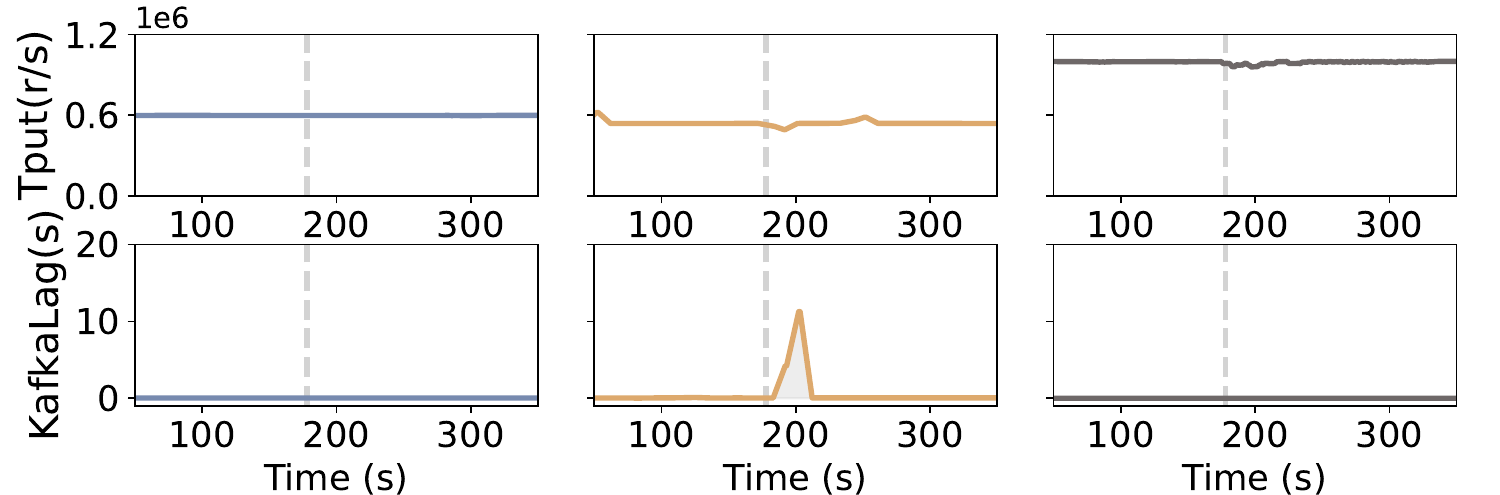}
\end{subfigure}
\caption{Comparison of DRRS-original on Flink with DRRS-ours on our runtime.}
\vspace{-4mm}
\label{fig:drrs_vs_flink}
\end{figure}

Figure~\ref{fig:drrs_vs_flink} shows the throughput and Kafka lag comparison between DRRS-ours and DRRS-original, with input rate set to Flink's maximum processing capacity. We also include DRRS-ours with input rate set to our runtime's maximum processing capacity.
DRRS-ours sustains the input rate without any disruption upon reconfiguration, while DRRS-original experiences a slight throughput drop with Kafka lag of up to 10 seconds, demonstrating that DRRS-ours outperforms DRRS-original under the default setting. In addition, we show that our runtime has larger processing capacity than Flink in steady state.

\vspace{-2mm}
\section{Workload details}\label{appendix:workload}
\subsection{Real-world queries}\label{appendix:real-world-queries}
We describe four real-world queries used in our evaluation, constructed from publicly available datasets.
\begin{enumerate}
    \item Twitch hot streamers (\textbf{Twitch}): 
    A live-streaming analytics workload~\cite{rappaz2021recommendation} that analyzes audience engagement events~\cite{rappaz2021twitchdataset} on the Twitch streaming platform. For each streamer, the query tracks the accumulated watch time, interaction counts, and engagement signals over the stream duration to compute a loyalty score reflecting the strength and consistency of audience engagement.
    \item NY taxi trends (\textbf{Taxi}): A ride analytics workload~\cite{DEBS2015} based on New York City taxi trip records~\cite{newyorktaxi}. For each origin-destination route, the query computes the median trip duration over recent trips, enabling continuous monitoring of travel time trends across the city.
    \item Azure VM CPU trends (\textbf{Azure}): A CPU utilization analytics workload based on the Microsoft Azure cloud operations trace~\cite{AzureVMData}. For each VM, the query computes the minimum, maximum, and median utilization over the recent CPU readings to support real-time identification of over- and under-provisioned instances.
    \item Borg task profiling (\textbf{Borg}): A resource consumption monitoring workload based on the Google cluster management trace~\cite{GoogleClusterData}. For each running task, the query computes the median CPU, RAM, and local disk utilization over the recent resource usage readings, providing operators with a live view of per-task resource demand.
\end{enumerate}

\vspace{-2mm}
\subsection{Nexmark queries}\label{appendix:nexmark-queries}
We describe the logic of Nexmark~\cite{nexmark} queries and two variants introduced in prior work~\cite{Meces} used in our evaluation.
\begin{enumerate}
\item Item Suggestion (\textbf{Q3}):
This query incrementally joins the person stream with the auction stream on seller ID to aggregate auctions by item category and seller location. The join operator is the target operator in our reconfiguration experiments, which maintains an ever-growing state size.

\item Average Price for a Category (\textbf{Q4}):
This query first calculates the closed auction event by window-joining the auction and bid streams, then groups those closed auctions by category to calculate the rolling average selling price. In this query, the window join is the target operator.

\item Hot Items (\textbf{Q5}):
This query groups the bid events by auction ID and counts the number of bids for each auction in the past window period (sliding window). The sliding window operator is the target operator in this query.

\item Average Selling Price by Seller (\textbf{Q6}):
This query, the same as \textbf{Q4}, first calculates the closed auction event by window-joining the auction and bid streams. It then groups those closed auctions by seller ID to calculate the average selling price for the recent ones. In this query, the window join is the target operator.

\item Highest Bid (\textbf{Q7}):
This query groups the bid events by bidder ID and applies a tumbling window to select the highest bid during the window. The target operator is the tumbling window in this experiment.

\item Monitor New Users (\textbf{Q8}):
This query performs a window join on person and auction streams to monitor whether a user has created an auction within a certain period after registering. In this query, the target operator is the window join operator.

\item Average Price for a Category Modified (\textbf{Q4*}):
This query, a variant of \textbf{Q4}, removes the window join operator and directly uses a closed auction stream as input to calculate the average selling price for each category. In this query, the target operator is the stateful mapper that calculates the rolling average selling price.

\item Average Selling Price by Seller Modified (\textbf{Q6*}):
Similar to \textbf{Q4*}, this query is a variant of \textbf{Q6} that removes the window join operator and directly uses a closed auction stream as input to calculate the average selling price for each seller. In this query, the target operator is the stateful mapper that calculates the average selling price for recent auctions.

\end{enumerate}

\vspace{-2mm}
\section{Full evaluation results}\label{appendix:full-results}

\subsection{Comparison with embedded state}\label{appendix:full-embedded-state-results}
\begin{figure*}
\centering
\begin{subfigure}{0.9\textwidth}
\centering
\includegraphics[width=\linewidth]{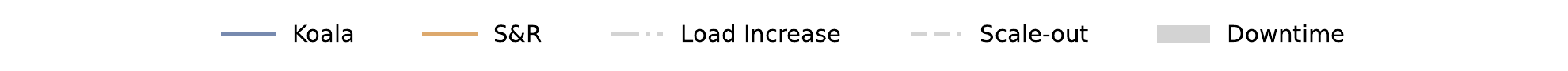}
\vspace{-7mm}
\end{subfigure}
\hfill
\begin{subfigure}{0.32\textwidth}
\centering
\includegraphics[width=\linewidth]{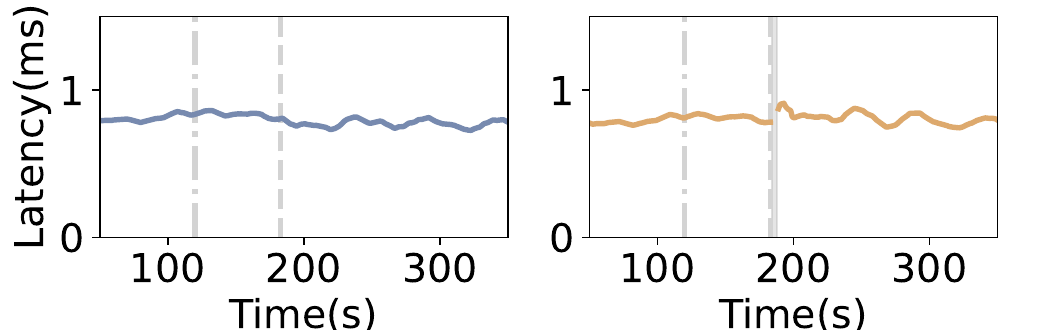}
\vspace{-6mm}
\caption{Twitch}
\end{subfigure}
\hfill
\begin{subfigure}{0.32\textwidth}
\centering
\includegraphics[width=\linewidth]{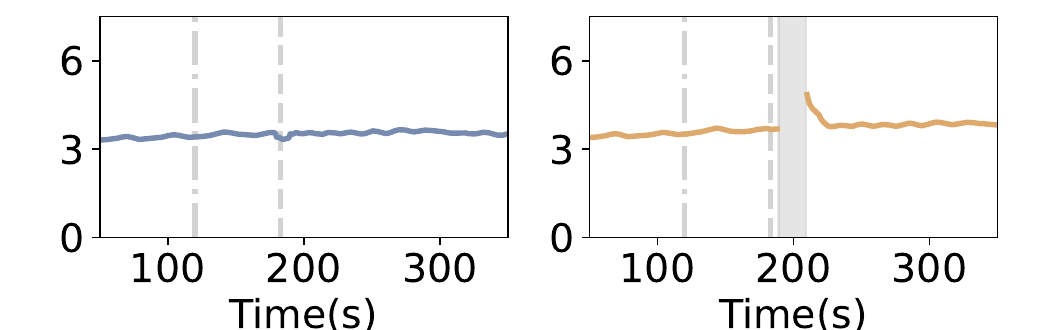}
\vspace{-6mm}
\caption{Taxi}
\end{subfigure}
\hfill
\begin{subfigure}{0.32\textwidth}
\centering
\includegraphics[width=\linewidth]{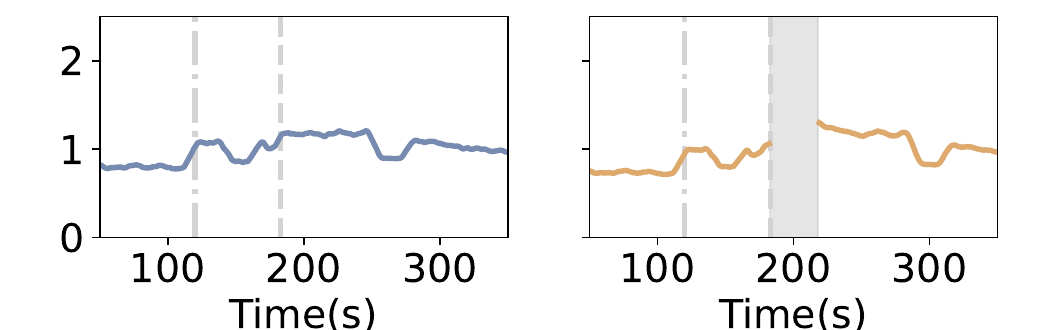}
\vspace{-6mm}
\caption{Borg}
\end{subfigure}
\hfill
\begin{subfigure}{0.32\textwidth}
\centering
\vspace{-1mm}
\includegraphics[width=\linewidth]{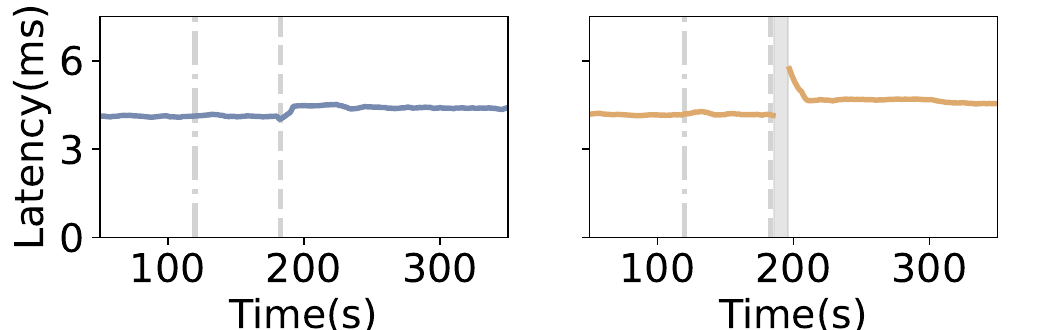}
\vspace{-6mm}
\caption{Azure}
\end{subfigure}
\hfill
\begin{subfigure}{0.32\textwidth}
\centering
\vspace{-1mm}
\includegraphics[width=\linewidth]{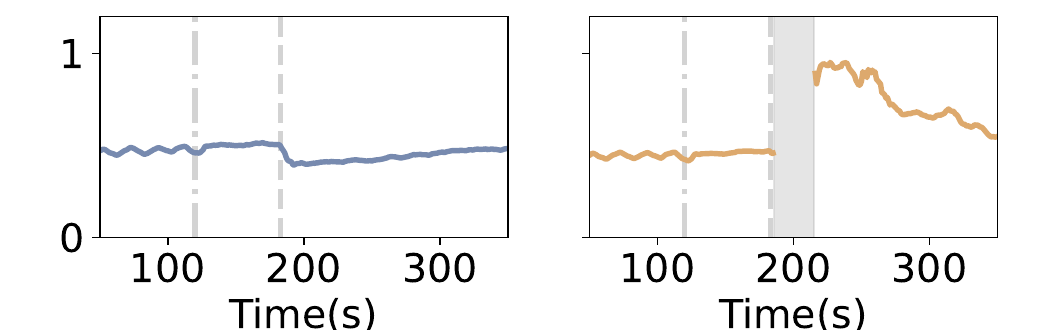}
\vspace{-6mm}
\caption{Q3}
\end{subfigure}
\hfill
\begin{subfigure}{0.32\textwidth}
\centering
\vspace{-1mm}
\includegraphics[width=\linewidth]{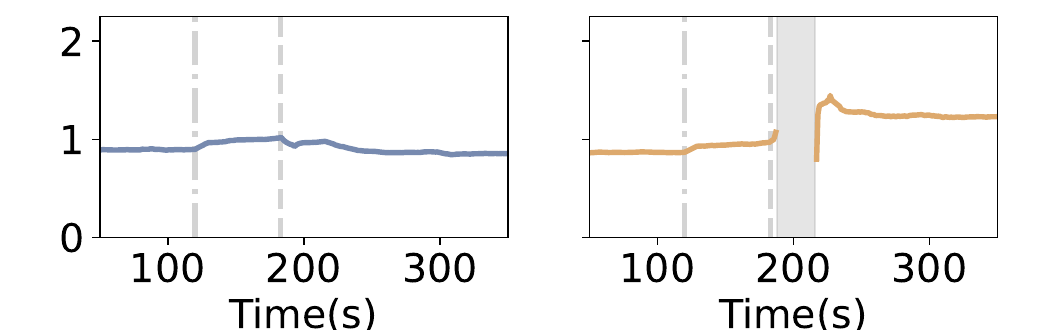}
\vspace{-6mm}
\caption{Q6*}
\end{subfigure}
\caption{Per-batch latency for Figure \ref{fig:baseline_comparison}. Per-batch latency is not available for DRRS as discussed in \S\ref{sec-eval-metrics}.}
\label{latency:LargeState}
\vspace{-3mm}
\end{figure*}
To complete the comparison results for embedded state experiments discussed in \S\ref{sec:exp-baselines}, Figure~\ref{latency:LargeState} shows the per-batch latency for \ours and the optimized stop-and-restart (S\&R) baseline. \ours achieves the same latency performance as S\&R in steady state, and shows no spikes upon reconfiguration. Note that DRRS is not included since per-batch latency is not available for DRRS as discussed in \S\ref{sec-eval-metrics}.


\vspace{-2mm}
\subsection{Comparison with remote state}\label{appendix:full-remote-state-results}
\begin{figure*}[t]
\centering
\begin{subfigure}{0.9\textwidth}
\centering
\includegraphics[width=\linewidth]{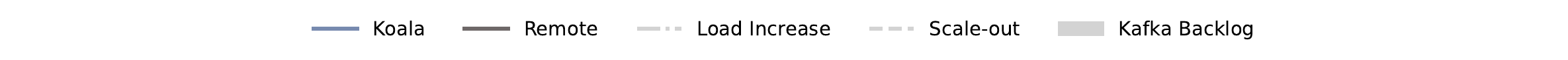}
\vspace{-6mm}
\end{subfigure}
\hfill
\begin{subfigure}{0.32\textwidth}
\centering
\includegraphics[width=\linewidth]{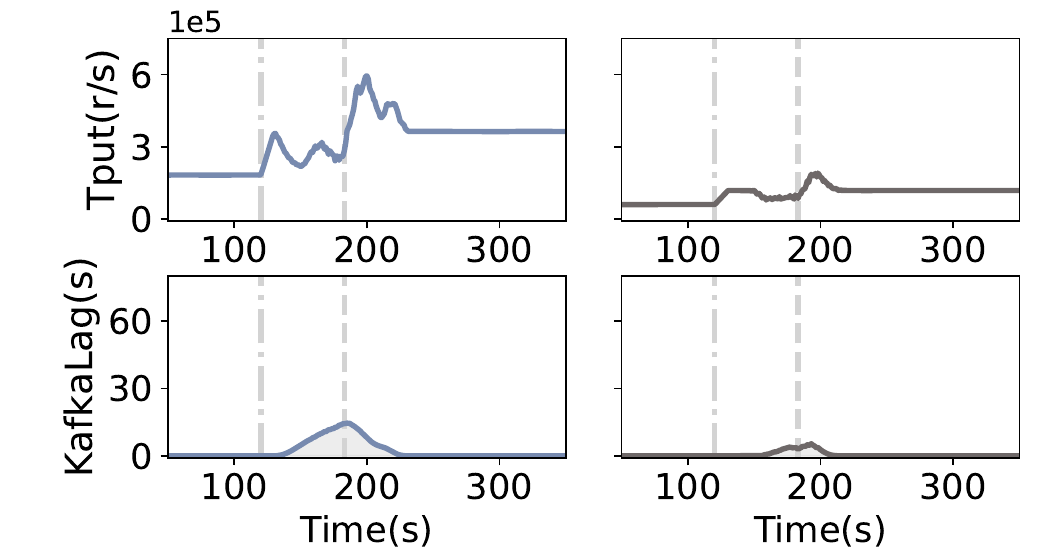}
\vspace{-6mm}
\caption{Twitch}
\end{subfigure}
\hfill
\begin{subfigure}{0.32\textwidth}
\centering
\includegraphics[width=\linewidth]{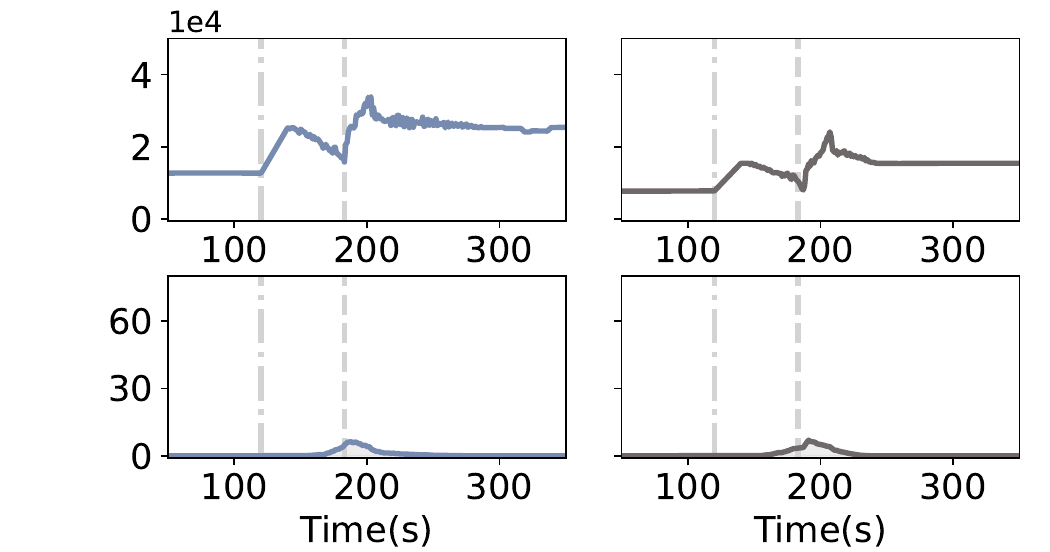}
\vspace{-6mm}
\caption{Taxi}
\end{subfigure}
\hfill
\begin{subfigure}{0.32\textwidth}
\centering
\includegraphics[width=\linewidth]{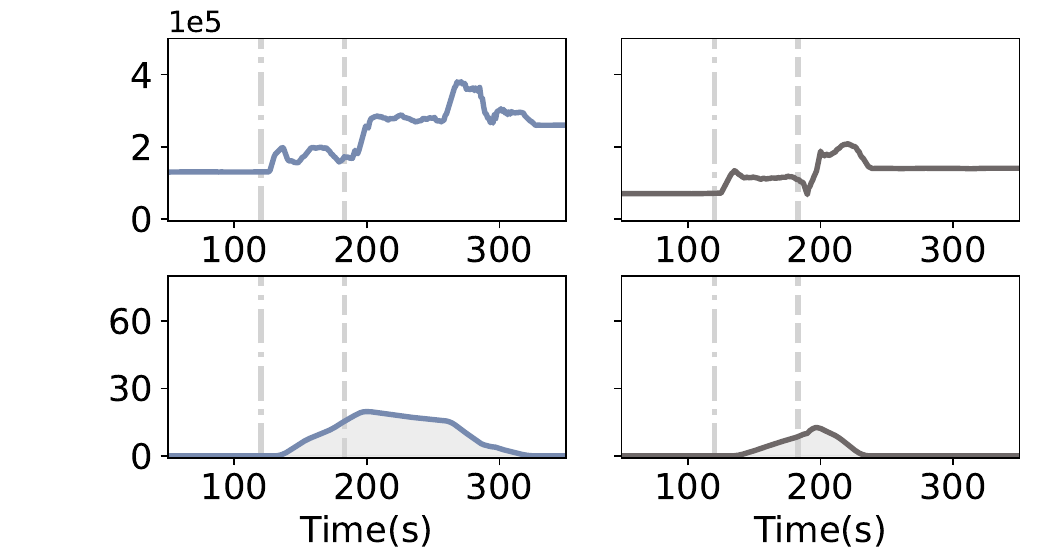}
\vspace{-6mm}
\caption{Borg}
\end{subfigure}
\hfill
\begin{subfigure}{0.32\textwidth}
\centering
\vspace{-1mm}
\includegraphics[width=\linewidth]{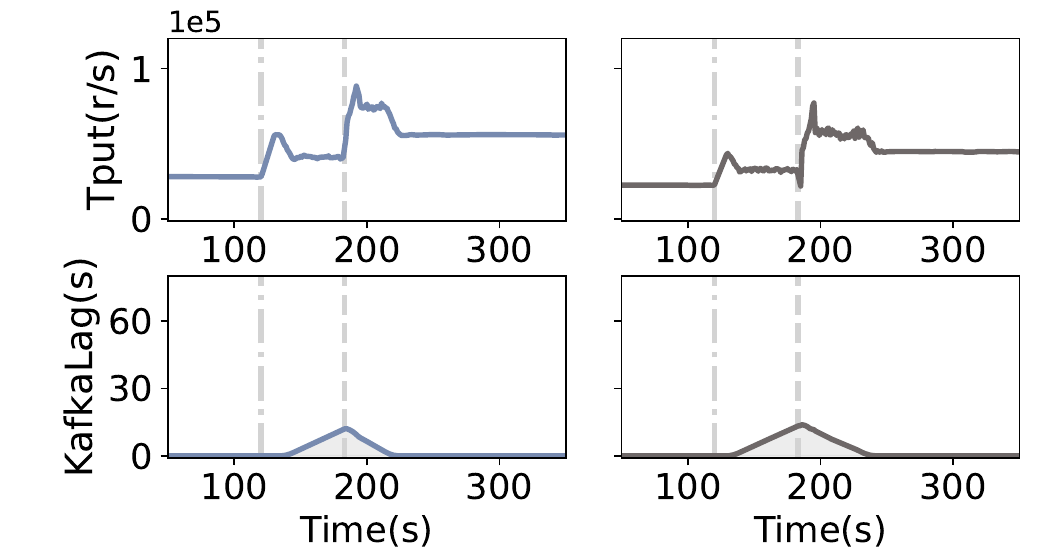}
\vspace{-6mm}
\caption{Azure}
\end{subfigure}
\hfill
\begin{subfigure}{0.32\textwidth}
\centering
\vspace{-1mm}
\includegraphics[width=\linewidth]{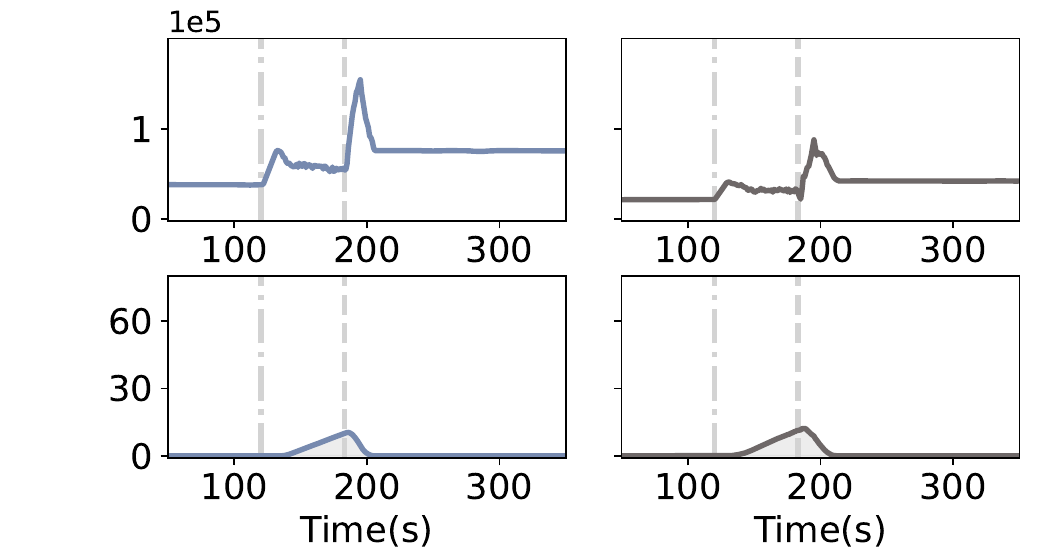}
\vspace{-6mm}
\caption{Q3}
\end{subfigure}
\hfill
\begin{subfigure}{0.32\textwidth}
\centering
\vspace{-1mm}
\includegraphics[width=\linewidth]{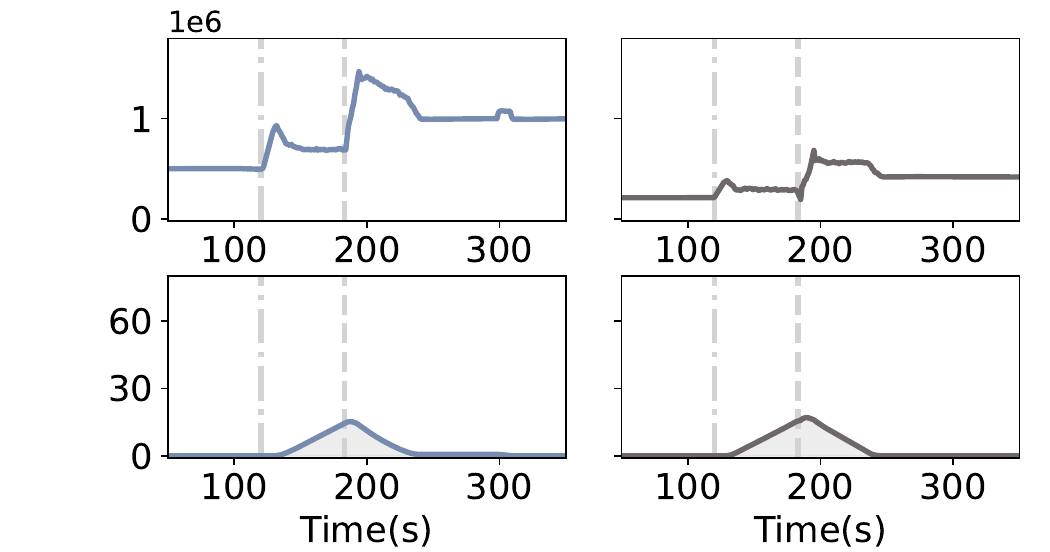}
\vspace{-6mm}
\caption{Q6*}
\end{subfigure}
\caption{Throughput and Kafka lag for Figure \ref{fig:latency_comparison}}
\label{throughput&kafkalag:remote}
\vspace{-3mm}
\end{figure*}
To complete the comparison results for remote state experiments discussed in \S\ref{sec:exp-baselines}, Figure~\ref{throughput&kafkalag:remote} shows the throughput and Kafka lag for \ours and the Remote baseline. \ours achieves similar non-disruptive performance to Remote (small Kafka lag), while sustaining higher throughput than Remote in all queries due to low-latency state access.


\vspace{-2mm}
\subsection{Full results on Nexmark benchmark}\label{appendix:full-nexmark-results}
\begin{figure*}
\centering
\begin{subfigure}{0.9\textwidth}
\centering
\includegraphics[width=\linewidth]{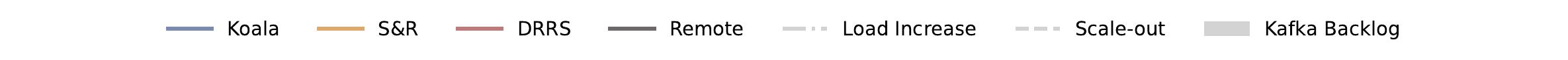}
\vspace{-6mm}
\end{subfigure}
\hfill
\begin{subfigure}{0.49\textwidth}
\centering
\includegraphics[width=\linewidth]{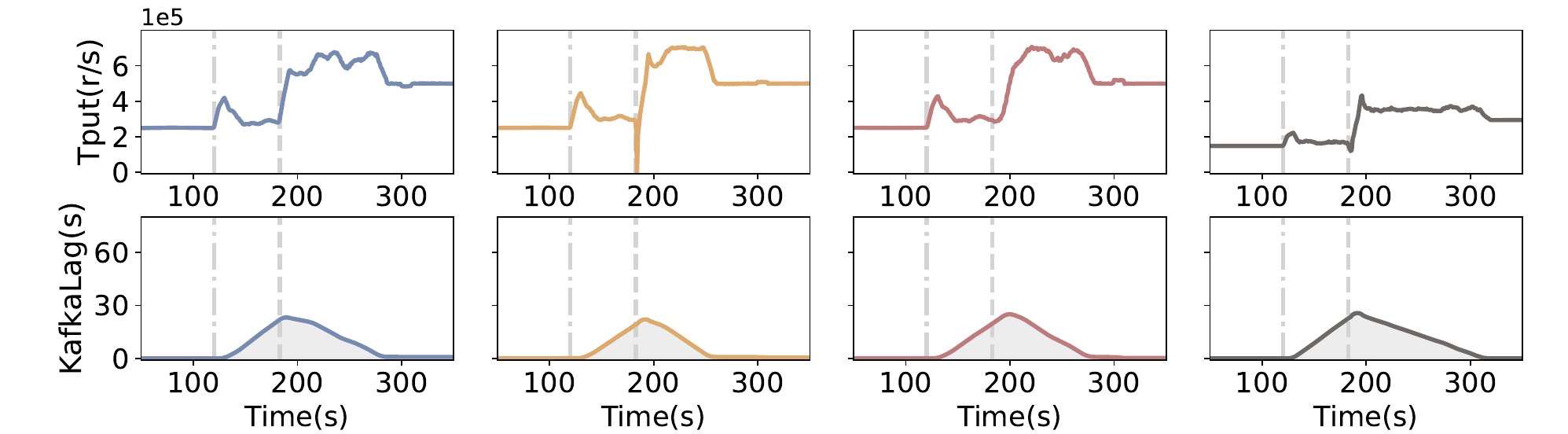}
\vspace{-6mm}
\caption{Q4}

\end{subfigure}
\hfill
\begin{subfigure}{0.49\textwidth}
\centering
\includegraphics[width=\linewidth]{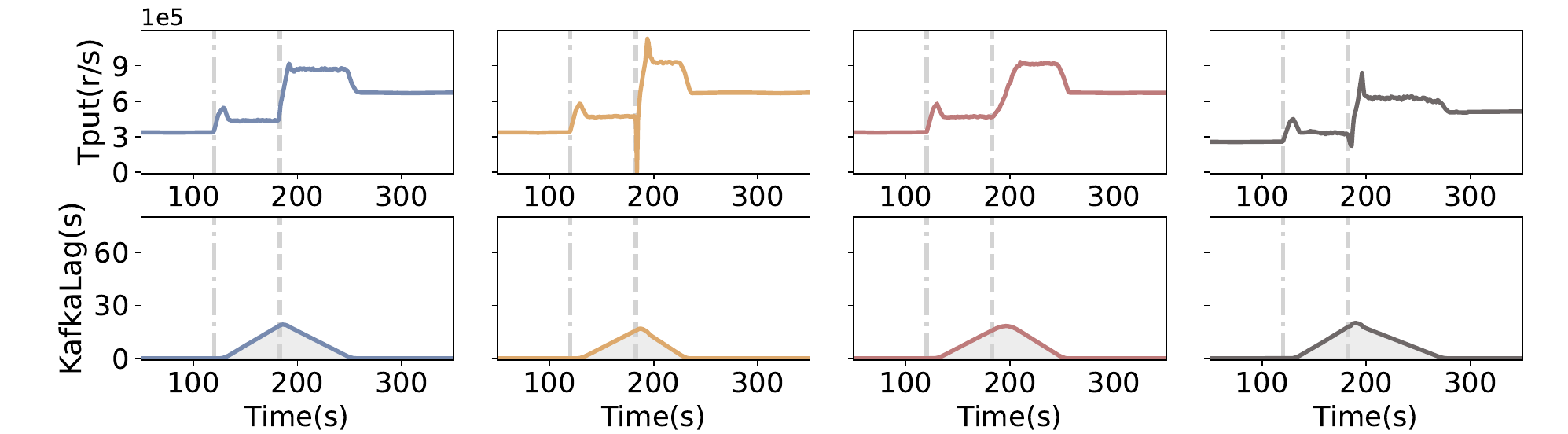}
\vspace{-6mm}
\caption{Q4*}
\end{subfigure}
\hfill
\begin{subfigure}{0.49\textwidth}
\centering
\includegraphics[width=\linewidth]{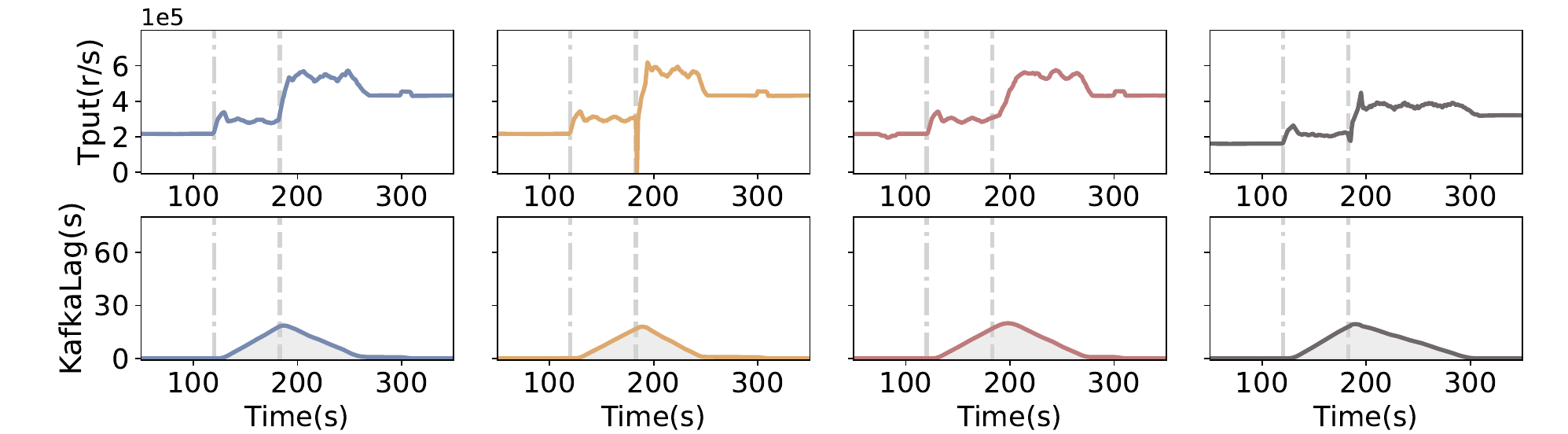}
\vspace{-6mm}
\caption{Q5}
\end{subfigure}
\hfill
\begin{subfigure}{0.49\textwidth}
\centering
\vspace{-1mm}
\includegraphics[width=\linewidth]{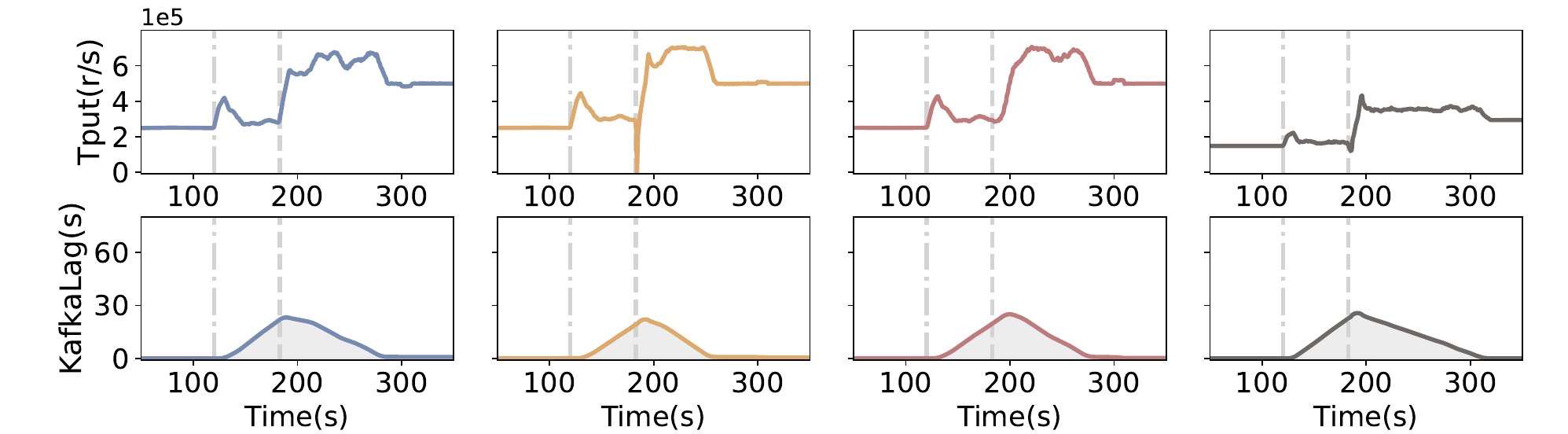}
\vspace{-6mm}
\caption{Q6}
\end{subfigure}
\hfill
\begin{subfigure}{0.49\textwidth}
\centering
\vspace{-1mm}
\includegraphics[width=\linewidth]{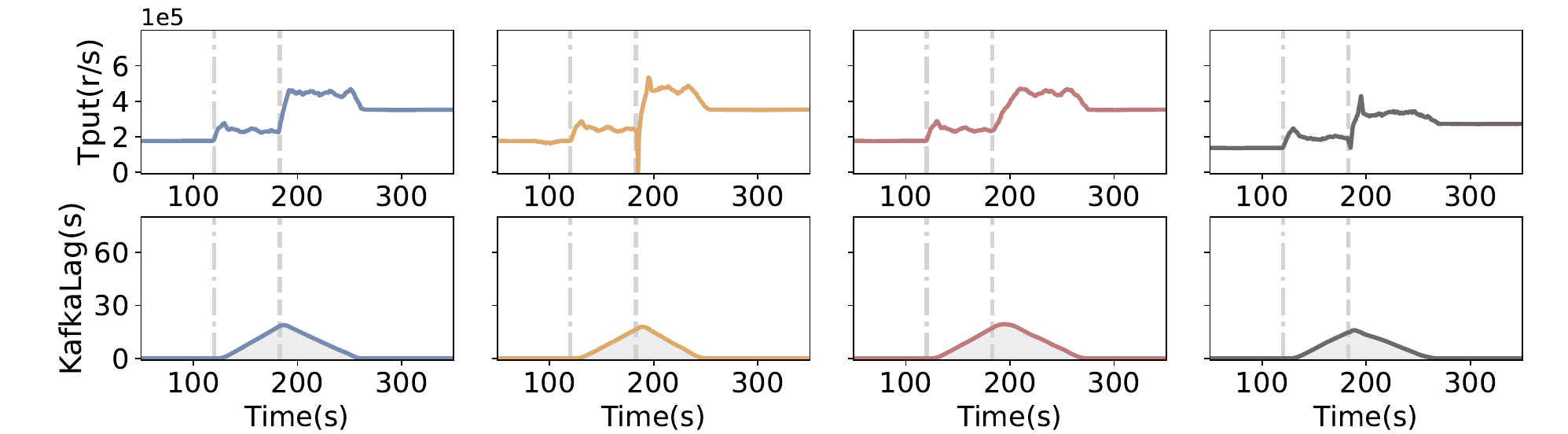}
\vspace{-6mm}
\caption{Q7}
\end{subfigure}
\hfill
\begin{subfigure}{0.49\textwidth}
\centering
\vspace{-1mm}
\includegraphics[width=\linewidth]{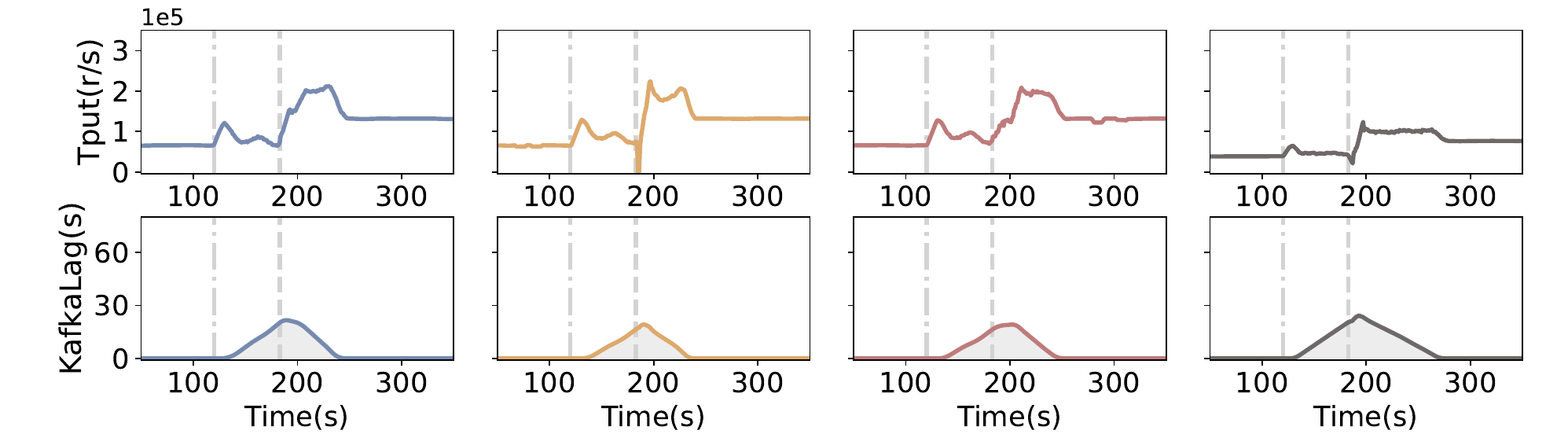}
\vspace{-6mm}
\caption{Q8}
\end{subfigure}
\caption{Throughput and Kafka lag for comparison with baselines on Nexmark small-state queries.}
\label{fig:small_state_throughput_kafkalag}
\vspace{-3mm}
\end{figure*}

\begin{figure*}
\centering
\begin{subfigure}{0.9\textwidth}
\centering
\includegraphics[width=\linewidth]{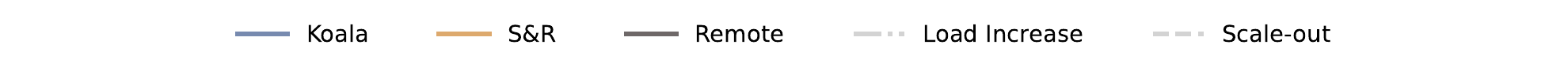}
\vspace{-7mm}
\end{subfigure}
\hfill
\begin{subfigure}{0.49\textwidth}
\centering
\includegraphics[width=\linewidth]{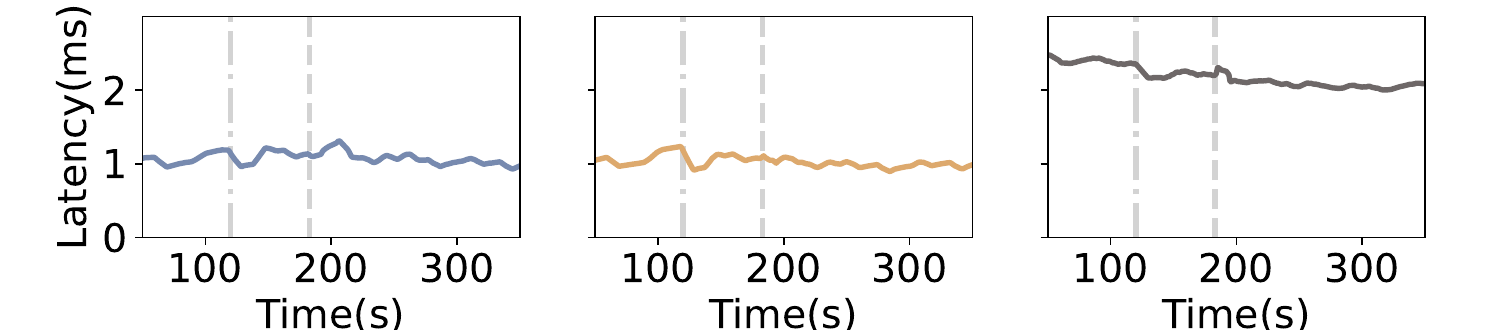}
\vspace{-6mm}
\caption{Q4}
\end{subfigure}
\hfill
\begin{subfigure}{0.49\textwidth}
\centering
\includegraphics[width=\linewidth]{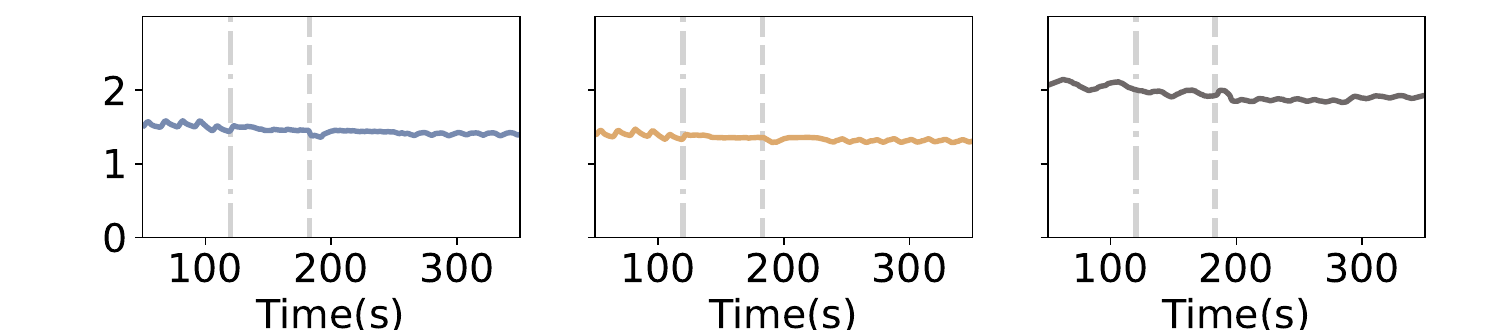}
\vspace{-6mm}
\caption{Q4*}
\end{subfigure}
\hfill
\begin{subfigure}{0.49\textwidth}
\centering
\includegraphics[width=\linewidth]{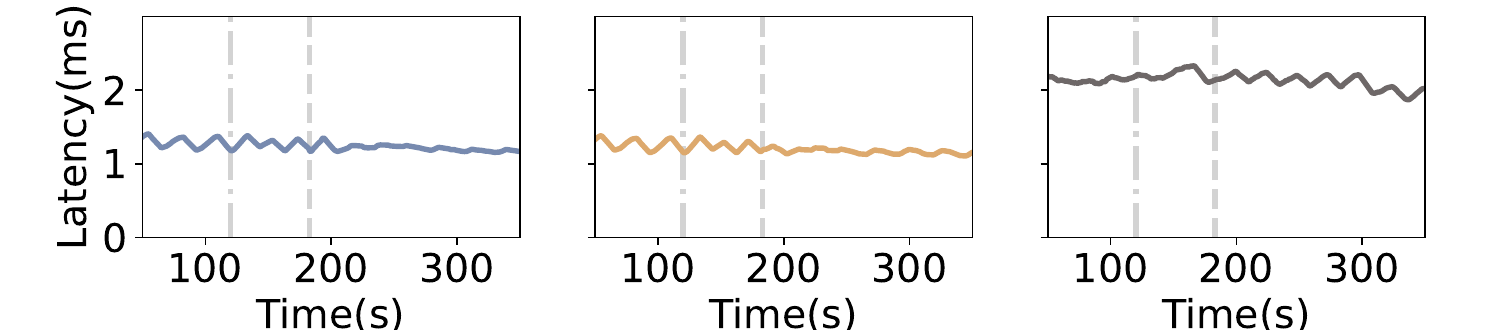}
\vspace{-6mm}
\caption{Q5}
\end{subfigure}
\hfill
\begin{subfigure}{0.49\textwidth}
\centering
\vspace{-1mm}
\includegraphics[width=\linewidth]{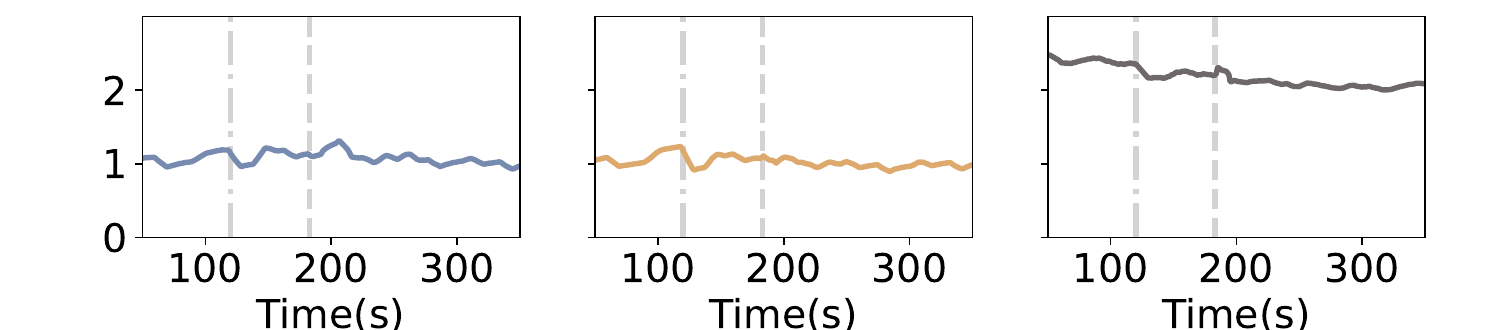}
\vspace{-6mm}
\caption{Q6}
\end{subfigure}
\hfill
\begin{subfigure}{0.49\textwidth}
\centering
\vspace{-1mm}
\includegraphics[width=\linewidth]{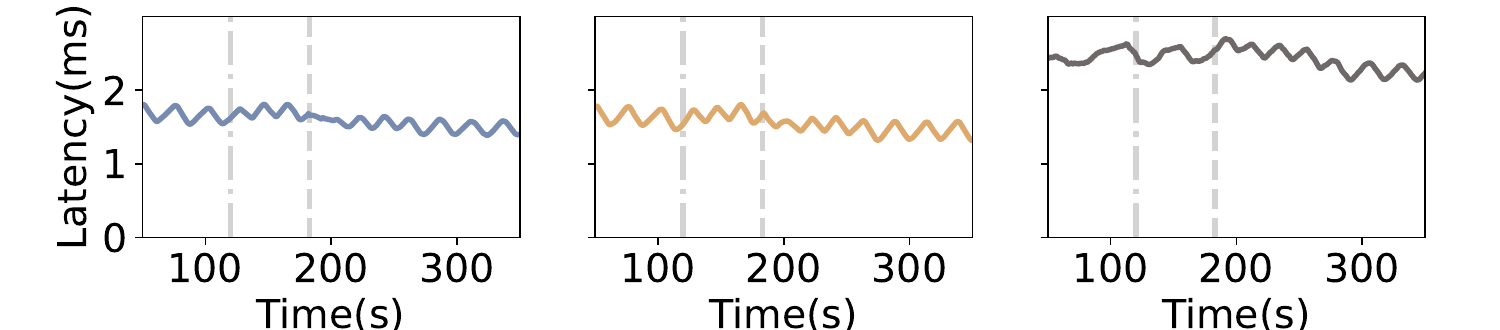}
\vspace{-6mm}
\caption{Q7}
\end{subfigure}
\hfill
\begin{subfigure}{0.49\textwidth}
\centering
\vspace{-1mm}
\includegraphics[width=\linewidth]{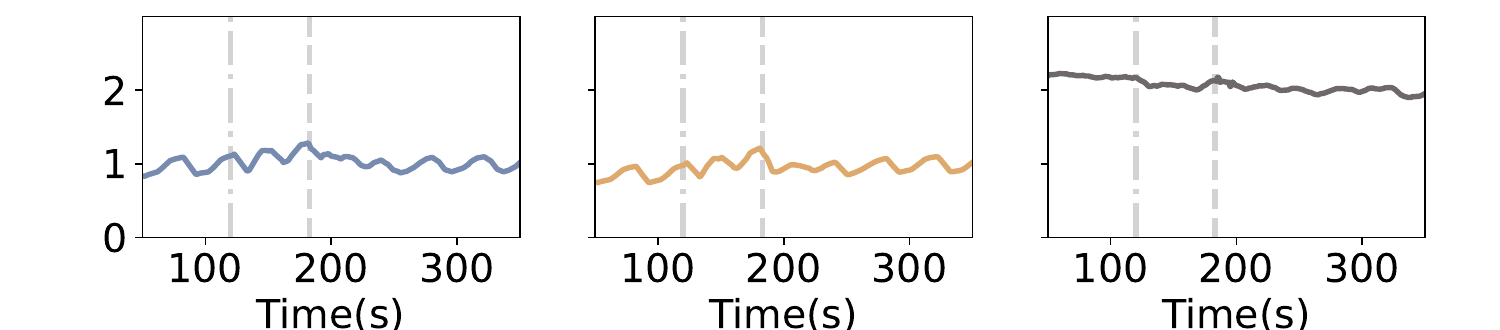}
\vspace{-6mm}
\caption{Q8}
\end{subfigure}
\caption{Per-batch latency for comparison with baselines on Nexmark small-state queries. DRRS is not included since per-batch latency is not available for DRRS as discussed in \S\ref{sec-eval-metrics}.}
\label{fig:small_state_latency}
\end{figure*}

Figure~\ref{fig:small_state_throughput_kafkalag} and Figure~\ref{fig:small_state_latency} demonstrate the full throughput, Kafka lag, and per-batch latency results for baseline comparisons over all Nexmark queries (\S\ref{appendix:nexmark-queries}). \textbf{Q3} and \textbf{Q6*} are excluded as we have discussed them in \S\ref{sec:exp-baselines}.


For the window-related queries (\textbf{Q4}, \textbf{Q5}, \textbf{Q6}, \textbf{Q7}, and \textbf{Q8}), expired state is cleaned up after window triggering, so the state size is bounded by the window size. Under the default key space and window duration~\cite{beam-nexmark}, total state size is small without inserting extra dummy state fields for accumulating state. 
\textbf{Q4*}, however, has a very small fixed key space, which also results in a small state size. Therefore, we include these small-state queries here for completeness.

\ours and all baselines achieve similar performance for these small-state queries, with low Kafka lag. S\&R exhibits a small throughput drop due to its blocking reconfiguration mechanism, but achieves the minimum backlog recovery time due to its efficient state transfer.




\clearpage
\nobalance\raggedbottom

\subsection{Multi-operator reconfiguration}\label{sec:protocol-generality:multi-operator-reconfiguration}
\begin{figure}[t]
    \centering
    \includegraphics[width=0.46\textwidth]{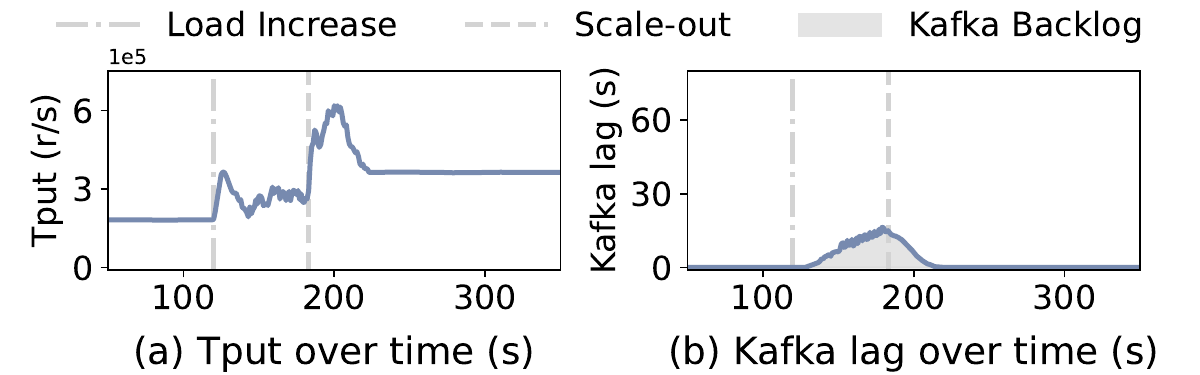}
    \vspace{1mm}
    \caption{Concurrent multi-operator reconfigurations.}
    \label{fig:multi_operator_reconfiguration}
\end{figure}

\ours can also support cases where multiple target operators need to be reconfigured at once. Multi-operator reconfiguration is realized by invoking the single-operator protocol for one target operator after another: the invocations proceed sequentially, while their on-demand state transfers run in parallel. Because each invocation is fast and non-blocking (\S\ref{subsec-procedure}), the back-to-back execution is effectively concurrent.
We choose the \textbf{Twitch} query to demonstrate this scenario, as it has two stateful operators which we can use as concurrent targets, and run the experiment on \textbf{E1-AWS}. After a $30min$ warm-up period, the operators have accumulated total state size of $11.2 GB$ and $5.6 GB$, respectively.
Figure~\ref{fig:multi_operator_reconfiguration} shows that \ours effectively overlaps the two reconfigurations, achieving the same backlog recovery time as in the single-operator reconfiguration case (Figure~\ref{fig:baseline_comparison}a).


\begin{figure}[t]
\centering
\begin{subfigure}{0.47\textwidth}
\centering
\includegraphics[trim=165 0 165 0, clip, width=\linewidth]{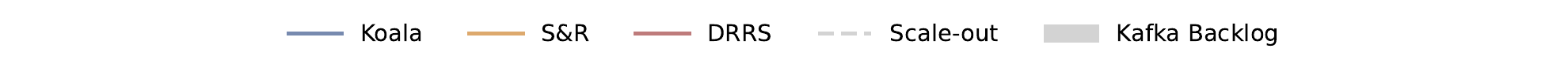}
\vspace{-6mm}
\end{subfigure}
\\
\begin{subfigure}{0.47\textwidth}
\centering
\includegraphics[width=\linewidth]{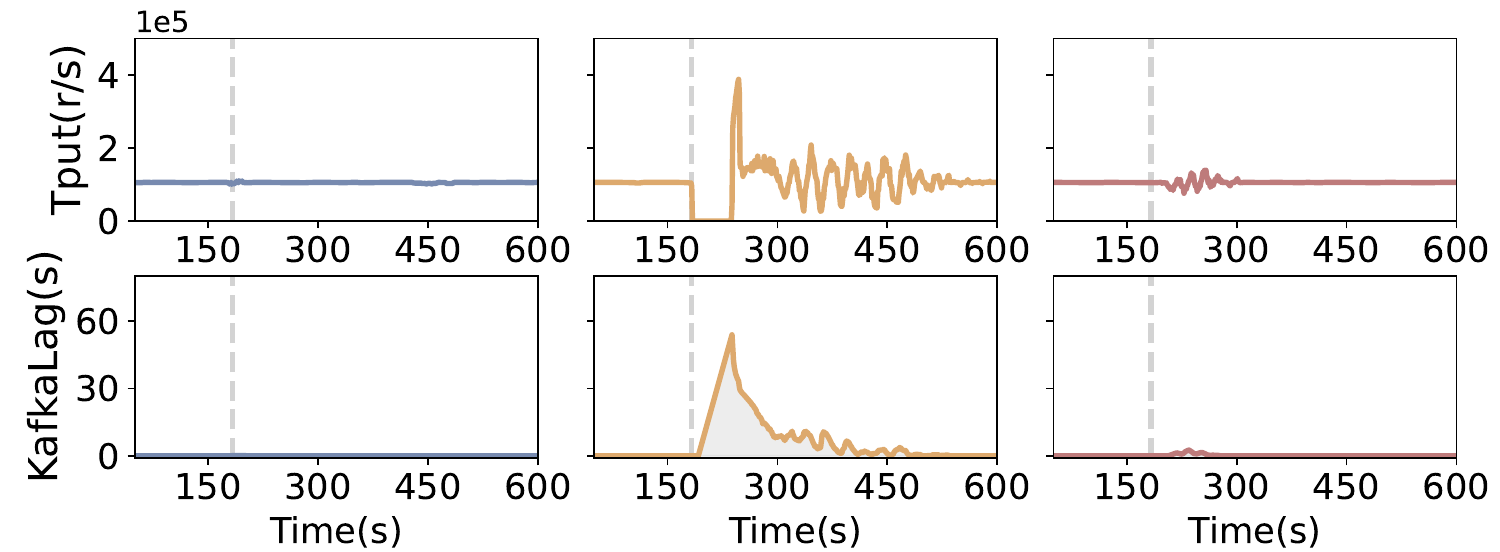}
\vspace{-6mm}
\caption{Throughput and Kafka lag}
\end{subfigure}
\\
\begin{subfigure}{0.47\textwidth}
\centering
\includegraphics[width=\linewidth]{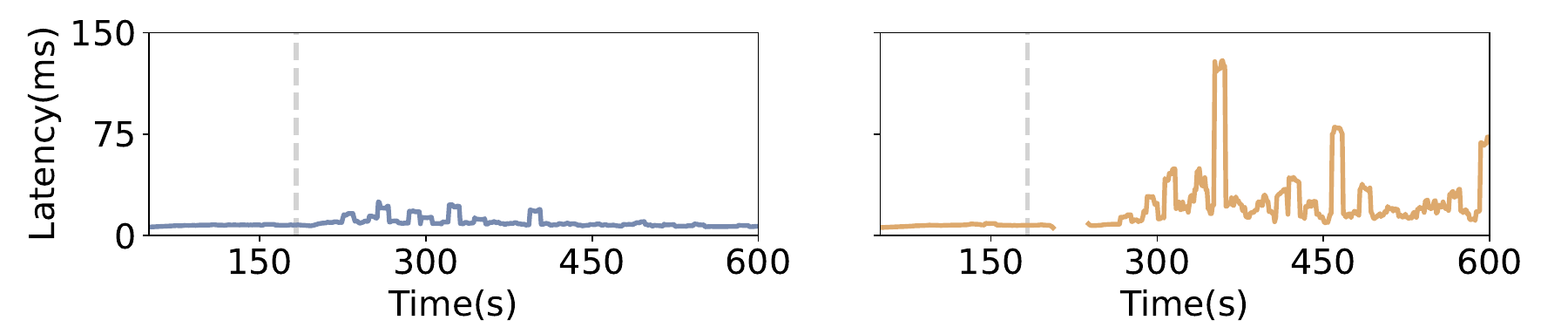}
\vspace{-6mm}
\caption{Per-batch latency}
\end{subfigure}
\caption{Comparison with baselines on \textbf{Q6*} under uniform state access over the entire key space. Per-batch latency is not available for DRRS, as discussed in \S\ref{sec-eval-metrics}.}
\label{fig:uniform_access_tput_latency}
\end{figure}

\subsection{Uniform state access}\label{appendix:uniform-access}
We now evaluate the worst case for \ours: state access is uniform over the entire key space, so there is no working set to exploit and nearly every reassigned key is eventually requested by its new owner. We ensure backlog-free conditions upon reconfiguration to isolate the effect of the variable under study.

\stitle{Setup.} We run this experiment on \textbf{E1-AWS} and use \textbf{Q6*}, as its access pattern is directly configurable through the Nexmark data source~\cite{beam-nexmark}. We set the active key space to the full key space of 20M keys and the hot key ratio to 0\%, so that each record accesses a key drawn uniformly at random. After warm-up, each of the four tasks of the target operator holds $10 GB$ of state. We trigger a scale-out from 4 to 8 tasks at $t=3$ min under a fixed sustainable input rate, and compare \ours against S\&R and DRRS.

Figure~\ref{fig:uniform_access_tput_latency} shows that \ours sustains the input rate with no visible throughput disruption; per-batch latency increases slightly while the reassigned keys are fetched remotely from their previous owners.
Due to the backlog-free conditions upon reconfiguration, the transient backpressure in \ours is absorbed by the empty intermediate buffers and resolves quickly, without affecting the source output rate that we report as throughput (\S\ref{sec-eval-metrics}).
S\&R, in contrast, migrates all reassigned state at once, blocking processing for about 45 seconds and requiring nearly 4 minutes to clear the accumulated backlog. DRRS keeps processing during reconfiguration, but its eager state transfer contends with record processing, causing throughput oscillations.


\end{document}